\documentclass{aa}

\usepackage{graphicx}
\usepackage{gensymb}
\usepackage{float}
\usepackage{hyperref}
\usepackage{subcaption}
\hypersetup{colorlinks=true,linkcolor=blue,citecolor=blue,filecolor=blue,urlcolor=blue,}
\usepackage{placeins}
\newcommand{\orcid}[1]{\protect\href{https://orcid.org/#1}{\protect\includegraphics[width=10pt]{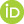}}}
\usepackage{txfonts}

\newcommand{\ha}{H$\alpha$\,} 
\newcommand{\EWha}{EW$_{H\alpha}$} 
\newcommand{\hb}{H$\beta$\,} 
\newcommand{\SBha}{$\Sigma_{H\alpha}$\,} 

\newcommand{\hii}{H\,{\footnotesize II}~\,}

\newcommand{\nii}{[N\,{\footnotesize II}]\,}
\newcommand{\sii}{[S\,{\footnotesize II}]\,}

\newcommand{\oi}{[O\,{\footnotesize I}]\,}

\newcommand{\oiii}{[O\,{\footnotesize III}]\,}
\newcommand{\mze}{{\small $\mathcal{M}$-Z-EW(H$\alpha$)}}
\defcitealias{2022A&A...659A..26B}{Bel22}
\defcitealias{2017MNRAS.466.3217Z}{Z17}
\defcitealias{2018MNRAS.474.3727L}{Lac18}
\defcitealias{2019A&A...623A..89C}{Com19}
\defcitealias{2024A&A...687A..20G}{BETIS I}

\begin{document}

   \title{Bidimensional
Exploration of the warm-Temperature Ionised gaS (BETIS)}

   \subtitle{II. Revisiting the ionisation mechanism of the extraplanar diffuse ionised gas}

   \author{R. Gonz\'alez-D\'iaz
          \inst{1,2,3}\orcid{0000-0002-0911-5141}
          \and
          F. F. Rosales-Ortega
          \inst{2}\orcid{0000-0002-3642-9146}
          \and
          L. Galbany
          \inst{1,3}\orcid{0000-0002-1296-6887}
          }

   \institute{Institute of Space Sciences (ICE-CSIC), Campus UAB, Carrer de Can Magrans, s/n, E-08193 Barcelona, Spain.\\ 
              \email{Raul.GonzalezD@autonoma.cat}
         \and
         Instituto Nacional de Astrof\'isica, \'Optica y Electr\'onica (INAOE-CONAHCyT),
              Luis E. Erro 1, 72840 Tonantzintla, Puebla, M\'exico.
         \and
            Institut d'Estudis Espacials de Catalunya (IEEC), 08860 Castelldefels, Barcelona, Spain.
         }

   \date{Received ... ; accepted ... }

 
\abstract{

The extraplanar diffuse ionised gas (eDIG) is a key component for understanding the feedback processes that connect galactic discs and their halos. In this paper, we present the second study of the Bidimensional Exploration of the warm-Temperature Ionised Gas (BETIS) project, which aims to explore the possible ionisation mechanisms and characteristics of the eDIG. We use a sample of eight edge-on galaxies observed with the Multi-Unit Spectroscopic Explorer (MUSE) integral field spectrograph (IFS) and apply the methodology for obtaining binned emission line maps developed in the first paper of the BETIS project.
We found that the vertical and radial profiles of the \nii/\ha, \sii/\ha, \oiii/\hb, and \oi/\ha ratios depict a complex ionisation structure within galactic halos, influenced by the spatial distribution of \hii regions across the galactic plane as observed from our line of sight, with Lyman continuum photon leakage from OB associations constituting the main ionisation source. 
Moreover, the electron temperature and S$^+$/S ionisation ratio also exhibit dependency on the distribution of \hii regions within the galactic discs. Our analysis excludes low-mass, hot, and evolved stars (HOLMES) as viable candidates for secondary ionisation sources to elucidate the unusual behaviour of the line ratios at greater distances from the galactic midplane. In contrast, we ascertain that shocks induced in the interstellar medium by star formation (SF) related feedback mechanisms represent a promising secondary ionisation source of the eDIG.
We present a suite of models integrating ionisation mechanisms arising from fast shocks and photoionisation associated with star formation. When applied to the classical Baldwin-Phillips-Terlevich (BPT) diagrams, these models reveal that the ionisation budget of the eDIG ranges from 20\% to 50\% across our sample, with local variations of up to 20\% within individual galaxy halos. This contribution correlates with the presence of filaments and other structural components observed within galaxy halos. The presence of shocks is additionally supported by the observation of high-density, high \oi/\ha ratios, characteristic of shock-compressed ionised gas, likely induced by feedback from regions of intense SF within the galactic disk. These results demonstrate consistency across all galaxies analysed in this sample.
}


   \maketitle
%

\section{Introduction}

To understand the physical mechanisms involved in feedback processes and the baryon cycle, which are fundamental to galaxy evolution it is essential to study the different gas phases present in the interstellar medium \citep[ISM,][]{2008IAUS..245...33C,2009ApJ...695..292C, 2014MNRAS.445..581H, 2016SAAS...43...85K, 2017MNRAS.466.1903G, 2017PhDT........59G,2023ApJ...944L..22B}. One of these gas phases is known as warm ionised medium (WIM, in the context of the Milky Way) or diffuse ionised gas (DIG, for extragalactic sources; \citealt{1963AuJPh..16....1H,1968Natur.217..709H,1971PhDT.........1R, 1973ApJ...179..651R, 1974IAUS...60...51G,1985ApJ...294..256R, 1989ApJ...345..811R, 1989ApJ...337..761K, 1996AJ....111.2265F}). In the Milky May, this gas was found to be warm ($0.6-1\cdot10^4$ K) and with low density ($\sim10^{-1}$cm$^{-3}$), representing one of the most prominent components of the ISM, constituting the 20\% of the ISM in volume and the 90\% of the ionising hydrogen in mass \citep{1997ApJ...483..666G,1997ApJ...491..114W,2000A&A...363....9Z,2001ApJ...560..207O,2003ApJS..146..407F, 2003ApJ...586..902H, RevModPhys.81.969}. However, the DIG is typically defined by its optical emission characteristics rather than its thermodynamics (e.g. see \citealt{2009RvMP...81..969H} or \citealt{2023A&A...678A..84D}).  In particular, the DIG component in extragalactic halos, also known as extraplanar diffuse ionised gas (eDIG), plays a crucial role in the physical processes driving the exchange of gas, metals, and energy between the galactic disk and halo, influencing the evolution of disk galaxies \citep{1988ARA&A..26..145T, 1990ApJ...352L...1R, 1990A&A...232L..15D, 2003ApJS..148..383M}. Studies of the eDIG through imaging encompass explorations using different wavelengths \citep{2024MNRAS.528.2145S}. In the optical range, the first detections of a diffuse component in the halo were made using \ha images \citep{1990A&A...232L..15D, 1990ApJ...352L...1R, 1994ApJ...427..160P, 1997ApJ...474..129R, 2000ApJ...537L..13R,2022MNRAS.513.2904T}, These studies found that in a wide sample of star-forming galaxies the eDIG \ha emission in the halo correlates with the star formation rate \citep{2003A&A...406..493R, 2003A&A...406..505R,2018ApJ...862...25J,2023MNRAS.519.6098L}. Additionally, sub-arcsecond resolution \ha imaging studies of individual galaxies \citep{2000AJ....119..644H,2004AJ....128..674R,2005ASPC..331..177R} reveal that the eDIG exhibit complex morphological structures such as filaments, superbubbles or chimneys (also detected in X-rays and radio-emission; \citealt{2000A&A...364L..36T,2006A&A...448...43T}). Alternatively, long-slit spectroscopy studies (in combination with photometric data) have enabled kinematic analyses and investigations into the potential ionisation mechanisms of the eDIG. The behaviour of the main ionisation species reveals that the  \nii/\ha, \sii/\ha, \oiii/\hb and \oi/\ha ratios tends to increase as a function of the distance from the midplane \citep{1990ApJ...352L...1R,1998LNP...506..527D, 1998ApJ...501..137R,2000ApJ...537L..13R, 2001ApJ...551...57C}. Additionally, this higher degree of ionisation of the collisional lines usually implies higher electron temperatures in the eDIG compared to the \hii regions \citep{1999ApJ...523..223H, 2009RvMP...81..969H, 2019ApJ...885..160B}. To explain these features and the increase in the emission of high ionisation species, several additional heating sources have been proposed (\citealt{2009RvMP...81..969H} and references therein). To infer the possible physical processes and the roles played by different morphological structures in the halo leading to the eDIG, the simultaneous spatial and spectral information provided by integral field spectrographs (IFS) has resulted in significant progress in this area. One of these studies, carried by the Sydney-AAO Multi-object Integral field spectrograph (SAMI; \citealt{2012MNRAS.421..872C,2015MNRAS.447.2857B}), used a sample of 40 low resolution edge-on galaxies (0.8 kpc < 2.8 kpc) to perform a systematical exploration of the eDIG \citep{2016MNRAS.457.1257H} noticing that the disc-halo interaction can influence the eDIG emission through starburst-driven winds. On the other hand, studies with the Mapping Nearby Galaxies at Apache Point (MaNGA; \citealt{2015ApJ...798....7B}) IFS using also low resolution large samples found an increment of the \nii/\ha indicating a temperature gradient in the halo, as well as a flattening in the \ha surface brightness (\SBha) comparable with the [O\,{\footnotesize II}] emission at higher distances ($\sim$6 kpc). 

The most recent studies with the Calar Alto Legacy Integral Field Area Survey (CALIFA; \citealt{2012A&A...538A...8S,2016RMxAA..52..171S}) studied the eDIG kinematically using a sample of 25 galaxies with spatial resolution of 0.8 kpc on average \citep{2019ApJ...882...84L}, finding a "lag" in the rotation velocity of the ionised gas, i.e., a decrement of the radial velocity with the increasing of the distance from the midplane (also reported by other authors, e.g., \citealt{2000ApJ...537L..13R, 2003ApJS..148..383M,2017ApJ...839...87B}). Nevertheless, they did not find correlations between the radial variation of the lag and the origin of the extraplanar gas. However, performing diagnoses of the main line ratios, they identified the primary ionisation sources of the eDIG the leaking of Lyman continuum (Lyc) photons from the OB stars of the \hii regions, with a low contribution of an additional source in the form of hot low mass evolved stars (HOLMES; \citealt{2011MNRAS.415.2182F,2022A&A...659A.153R}).

In the mentioned studies, despite the wide variety of large samples used, the spatial resolution of the data is typically insufficient to perform a comprehensive and in-depth bidimensional study of the eDIG. However, the most recent studies with IFS includes datasets of the Multi Unit Spectroscopic Explorer (MUSE, \citealt{2010SPIE.7735E..08B}), whose datasets exhibits spatial resolutions lower than 200 pc, enabling a more complete exploration of the line emission, structure and kinematics of the ionised gas in the halo of edge-on galaxies. For example, \citet{2022A&A...659A.153R} and \citet{2023A&A...678A..84D} used a sample of nearby edge-on galaxies from \citet[][\citetalias{2019A&A...623A..89C} henceforth]{2019A&A...623A..89C}. The resolution and data quality of this dataset (detailed in section \ref{sec:data}) enabled these authors to find that the \ha emission in the halo correlates with the star formation of the galaxy, revealing the possible existence of a feedback mechanism between the plane and the halo that ionised the latter. 

To advance the characterisation and exploration of the ionisation of extraplanar gas, it is crucial to study the impact of the various structures found in the halos of individual galaxies, such as filaments, plumes or bubbles. Previous studies using IFS  have presented large samples of low-resolution galaxies, with the exception of those conducted with the MUSE instrument, but focusing only on one galaxy or without inquiring in the connection between disc and halo that originates these structures and changes the ionisation conditions. 

In this paper we present the second study of the Bidimensional Exploration of the warm-Temperature Ionised gaS (BETIS) project, focused in the characterisation of the eDIG and the exploration of its possible ionisation mechanisms for a sample of edge-on galaxies. 
The paper is structured as follows. Section \ref{sec:data} presents and characterises the sample of edge-on galaxies, as well as describes the procedure followed for adaptive binning and the extraction of emission line features, based on the methodology developed in the first part of the series (\citealt{2024A&A...687A..20G}, \citetalias{2024A&A...687A..20G} henceforth). Section \ref{sec:sec_3} shows a description of the ionisation structure and features found in the halo of IC1553 and the connection of the eDIG with the disc for all the sample. Section \ref{sec:discusion} presents a set of star formation and fast shocks hybrid models to explain the ionisation structure found the halo. In section \ref{sec:results} we show the results of applying the methodology and models of section \ref{sec:discusion} for the rest of the sample and finally, section \ref{sec:conclusion} outlines the main findings in this paper.

The following notation is used throughout the paper: \nii$\equiv$\nii$\lambda6584$; \sii$\equiv$\sii$\lambda6717 +$\sii$\lambda6731$; \oiii$\equiv$\oiii$\lambda5007$; and \oi$\equiv$\oi$\lambda6300$. We adopt the standard $\Lambda$CDM cosmology with H$_0$ = 70 km/s/Mpc,\, $\Omega_{\lambda}$ = 0.7, \,$\Omega_M$ = 0.3.

\begin{table*}[ht!]
\centering
\caption{General characteristics of the eBETIS sample, in order of declination.}
\resizebox{\textwidth}{!}{%
\begin{tabular}{lrrlccrrccrcc}
\hline
\textbf{Galaxy} & \textbf{RA (J2000)} & \textbf{DEC (J2000)} & \textbf{Type} & \multicolumn{1}{l}{\textbf{z}} & \textbf{D} & \textbf{P.A.} & \textbf{FWHM} & \textbf{Incl.} & \textbf{log SB lim.} & \textbf{Bin size} & \textbf{L$_{eDIG}$(\ha)} & \textbf{log SFR$_{Disc}$} \\
 & (deg) & (deg) &  & \multicolumn{1}{l}{} & (Mpc) & (°) & (pc) & (°) & (erg/s/kpc$^2$) & (pc) & (erg/s) & (M$_\odot$/yr) \\ \hline
IC217 & 34.0435 & -11.9267 & Scd & 0.00630 & 24.32 & 35.7 & 117.91 & 82.6 & 35.6 & 94.3 & 39.11 & 0.063 \\
PGC28308 & 147.5582 & -12.0576 & Scd & 0.00907 & 45.22 & 125.2 & 153.46 & 85.5 & 35.2 & 175.4 & 39.12 & 0.084 \\
PGC30591 & 156.3603 & -15.3492 & Sd & 0.00676 & 35.22 & 169.2 & 187.83 & 86.6 & 36.0 & 136.6 & 39.13 & 0.055 \\
ESO544-27 & 33.2277 & -19.3168 & Sb & 0.00818 & 32.81 & 153.3 & 159.07 & 90.0 & 35.0 & 127.2 & 38.92 & 0.024 \\
IC1553 & 8.1671 & -25.6075 & Irr & 0.00979 & 32.25 & 15.0 & 171.99 & 78.6 & 36.5 & 125.1 & 39.80 & 0.229 \\
ESO443-21 & 194.9412 & -29.6002 & Scd & 0.00941 & 40.31 & 160.8 & 195.45 & 79.0 & 36.6 & 156.4 & 39.71 & 0.184 \\
ESO469-15 & 347.2317 & -30.8579 & Sb & 0.00545 & 19.81 & 149.2 & 86.44 & 90.0 & 35.9 & 76.9 & 38.90 & 0.029 \\
ESO157-49 & 69.9036 & -53.0126 & Sc & 0.00559 & 24.82 & 30.4 & 96.26 & 79.3 & 35.3 & 96.3 & 39.46 & 0.060 \\ \hline
\end{tabular}%
}
\tablefoot{The columns represents, from left to right: The designation of the galaxy, the RA and DEC in the J2000 epoch, the morphological Hubble type, the redshift, the distance in Mpc, the position angle (in deg), the PSF FWHM in pc, inclination with respect to the line of sight (in deg), the 2$\sigma$ \SBha limit and the average physical size of the bins in pc. The RA, DEC, the morphological type, the redshift and distances are obtained from NED. The position angles and inclinations are obtained directly from \citetalias{2019A&A...623A..89C} and Hyperleda. The last columns represent the integrated \ha luminosity within the eDIG and the integrated star formation rate (SFR) of the disc (see section \ref{sec:data}).}
\label{tab:sample}
\end{table*}

\begin{figure*}[ht!]
\centering
\includegraphics[width=0.32\textwidth]{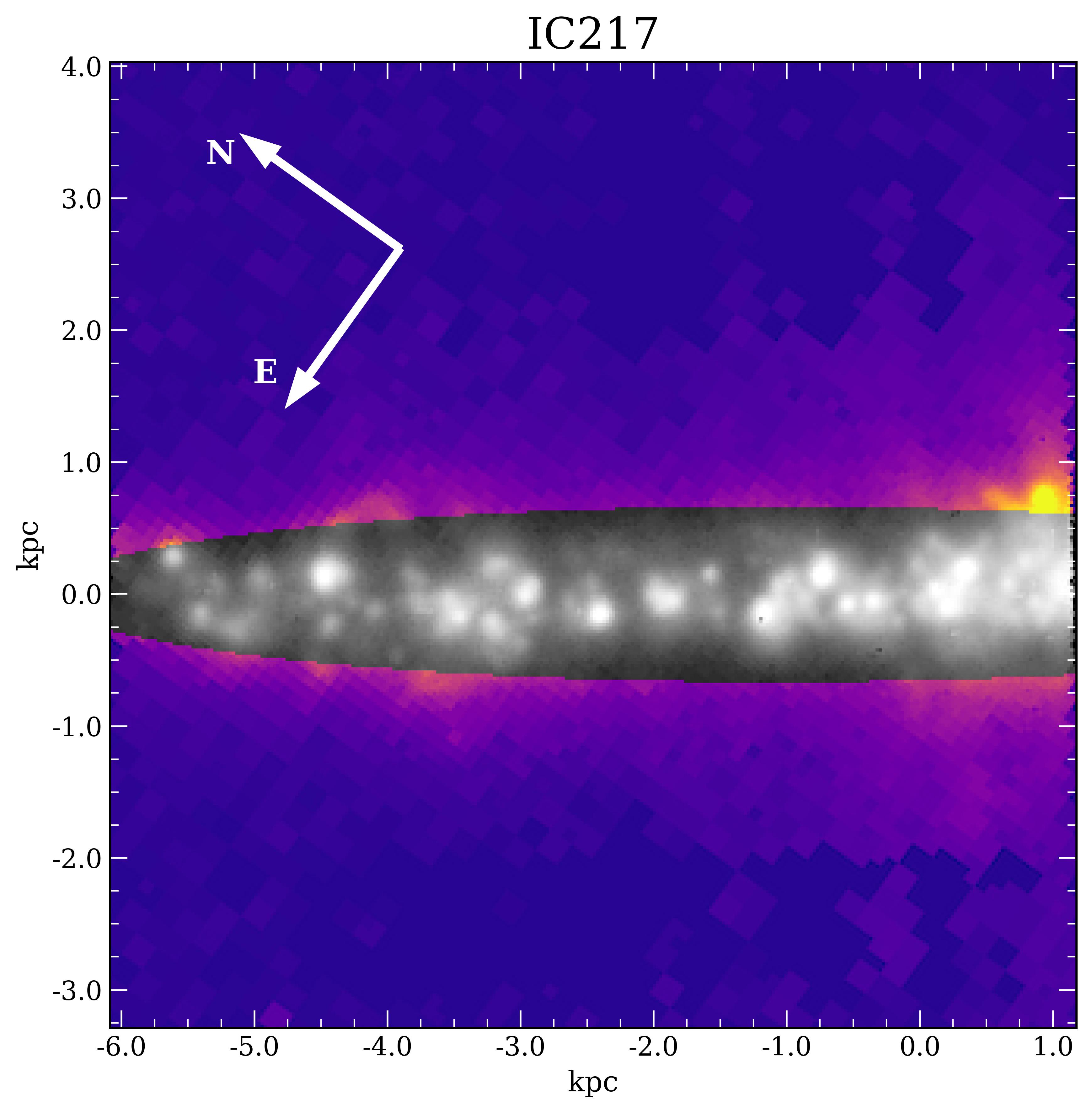}
\includegraphics[width=0.32\textwidth]{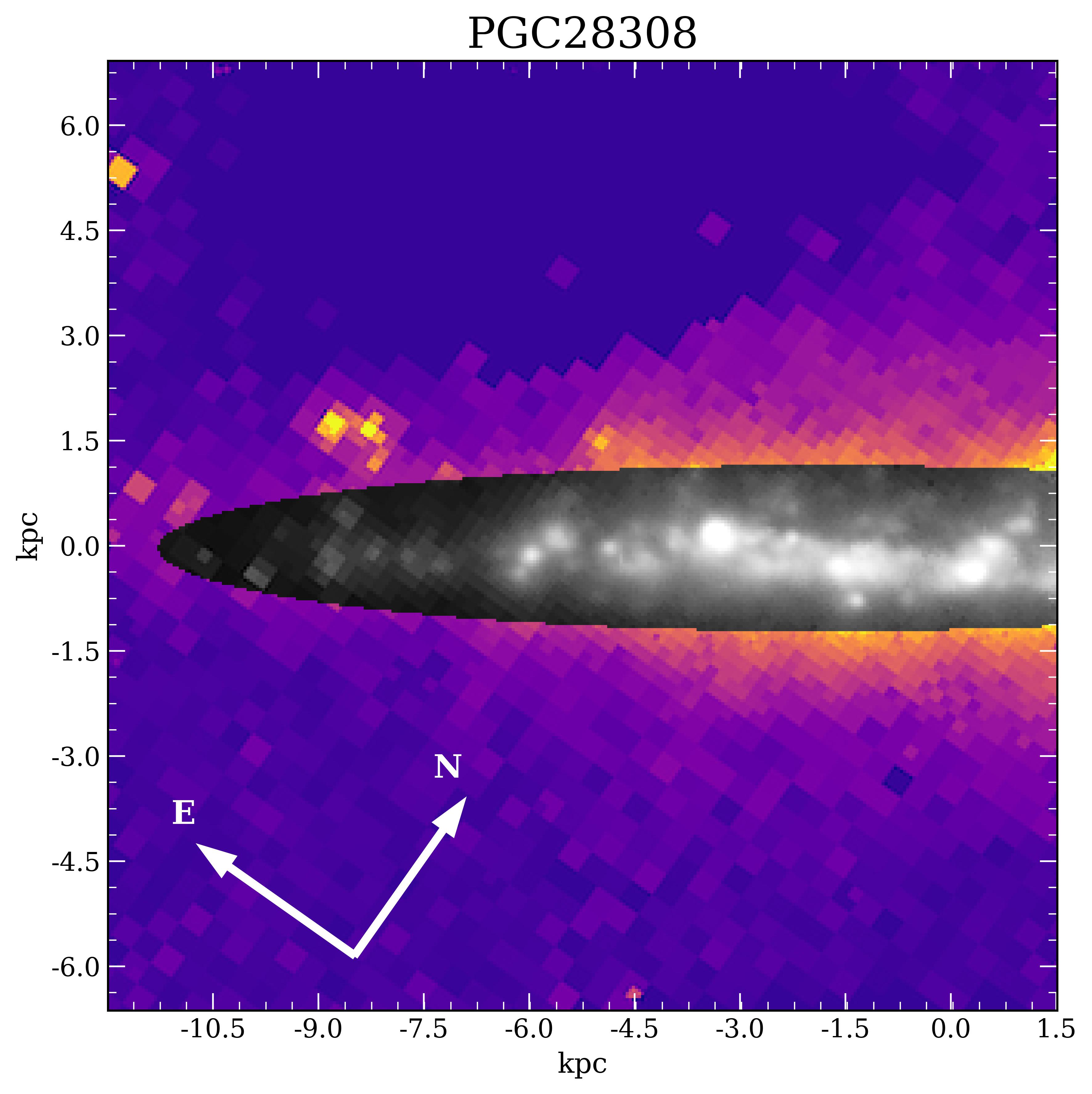}
\includegraphics[width=0.32\textwidth]{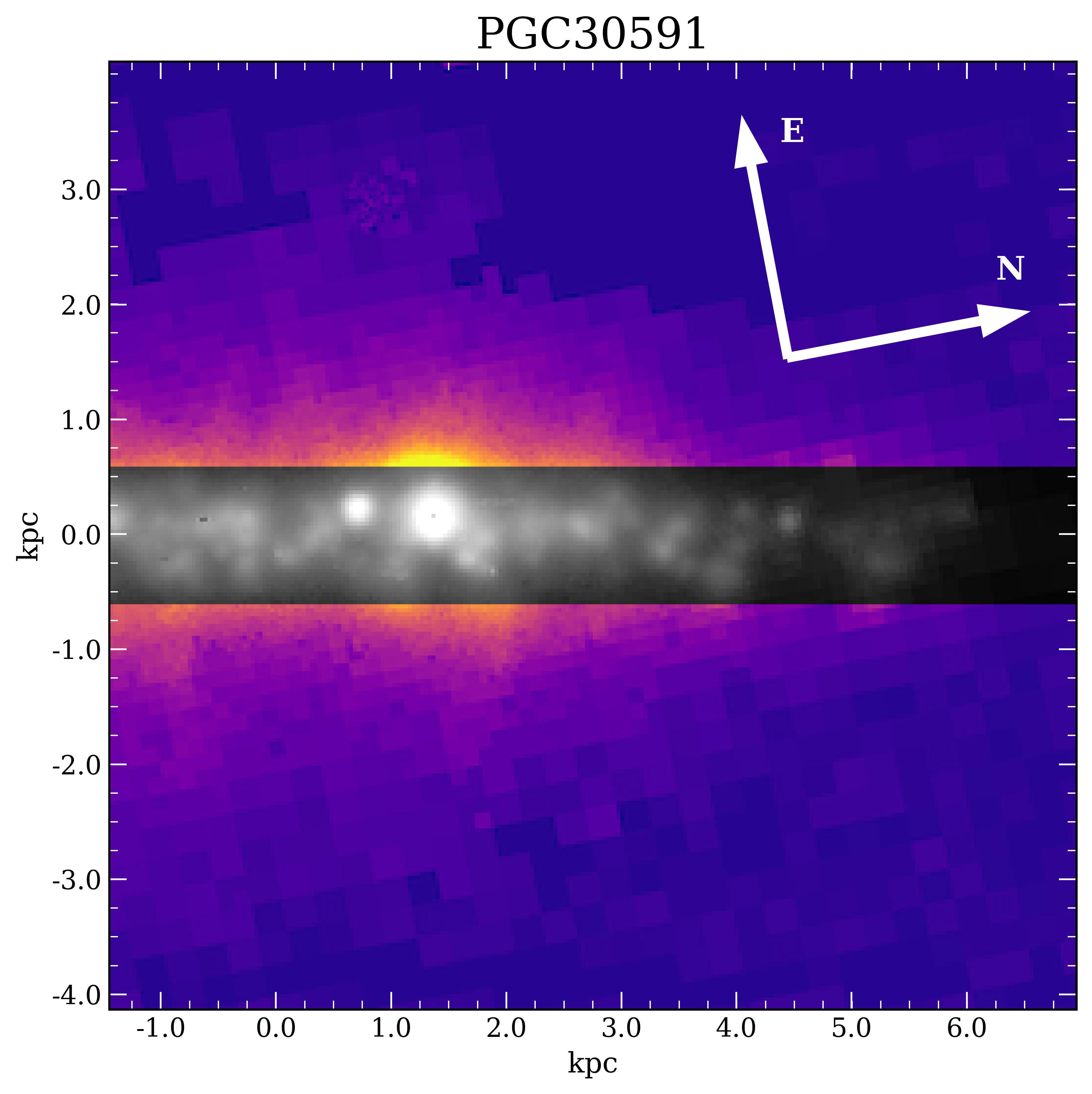}
\includegraphics[width=0.32\textwidth]{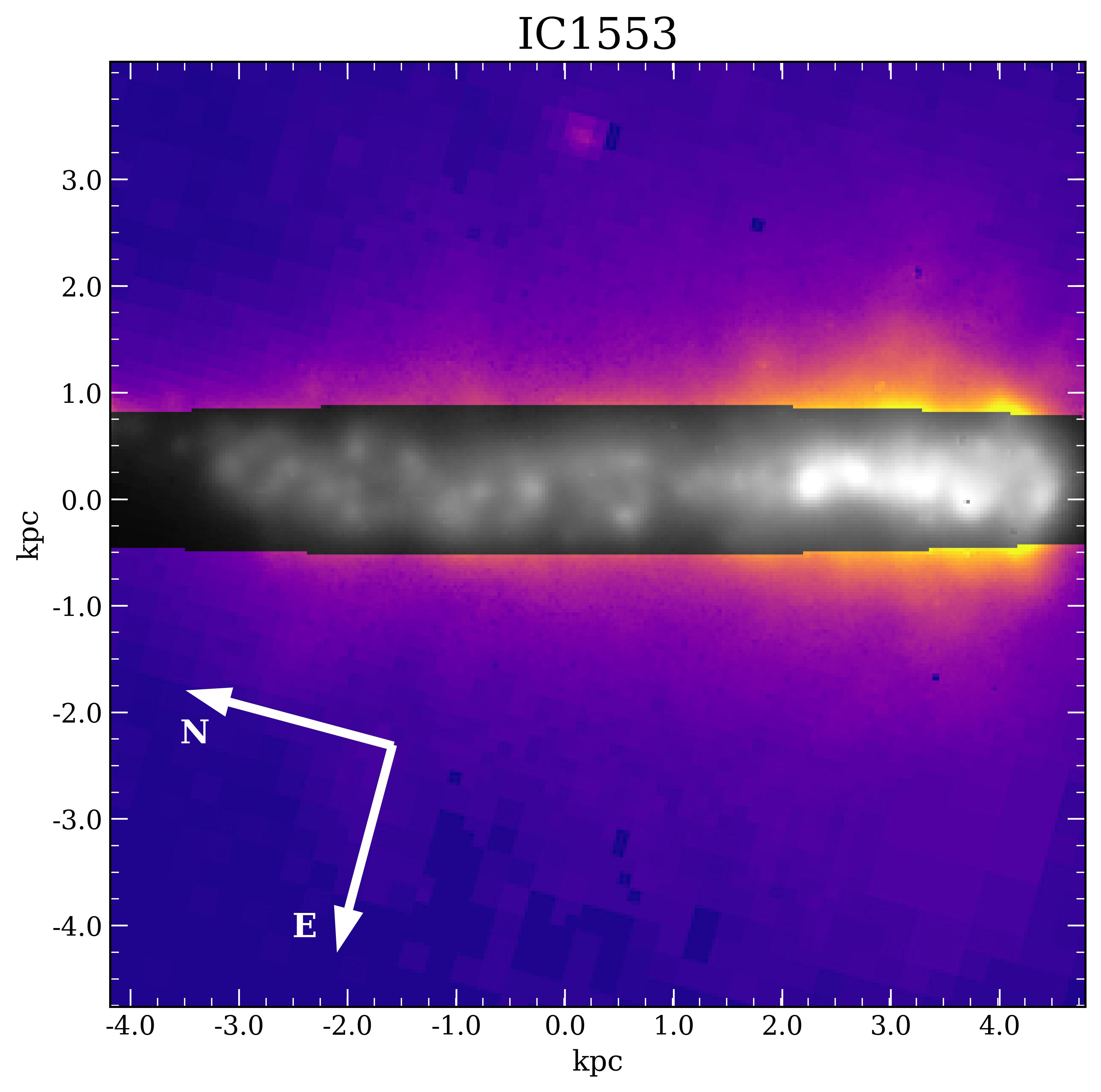}
\includegraphics[width=0.32\textwidth]{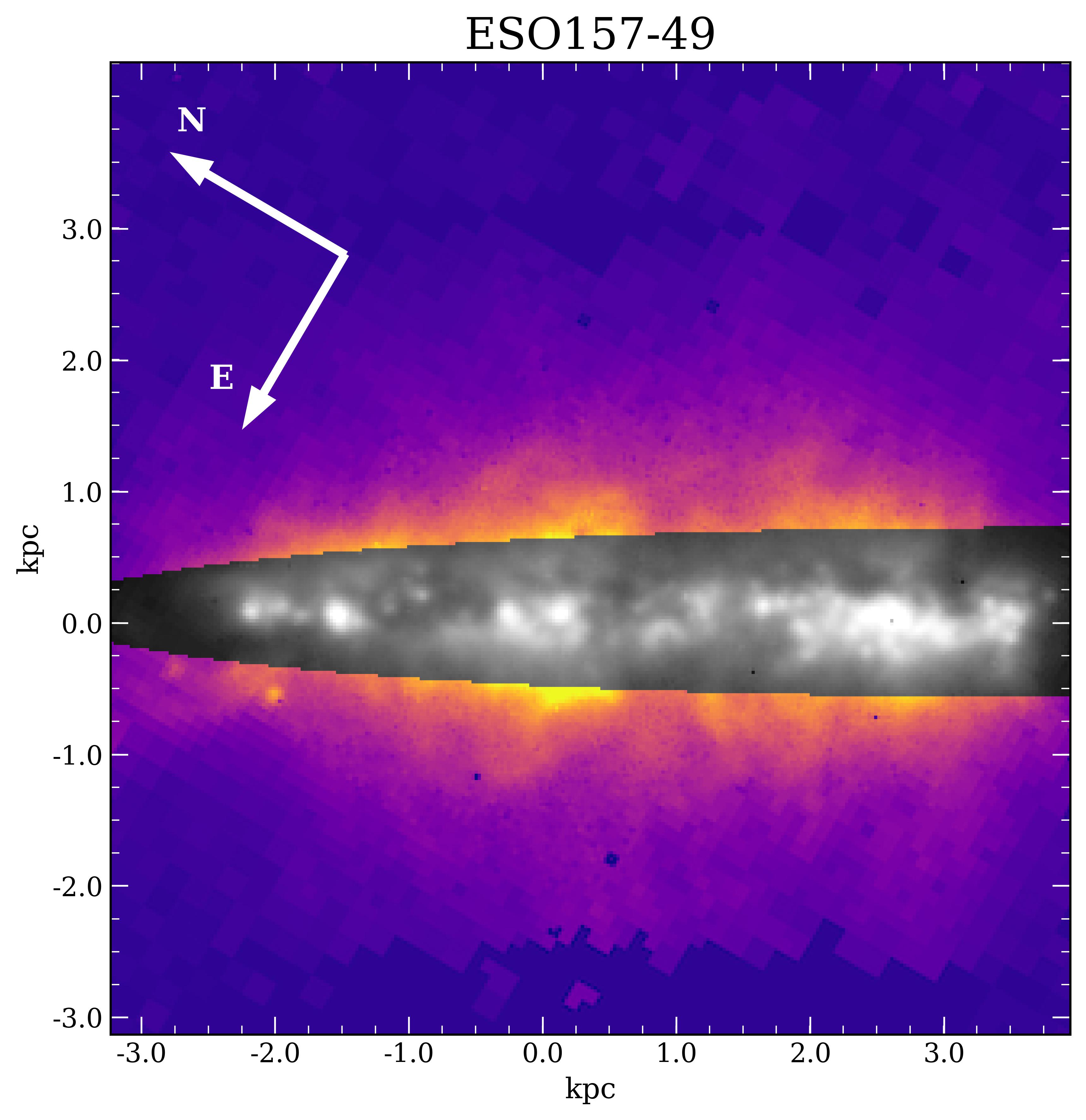}
\includegraphics[width=0.32\textwidth]{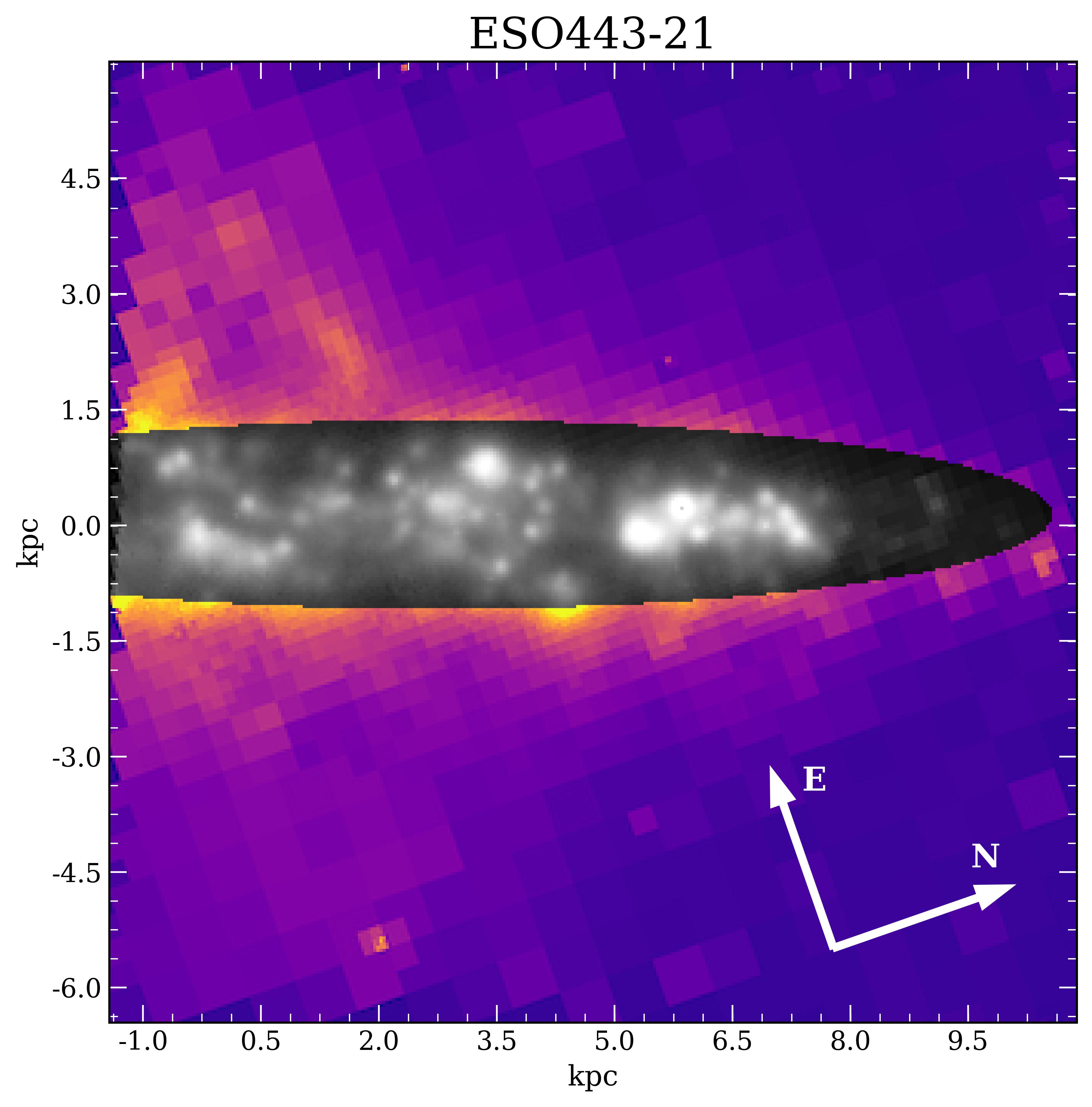}
\includegraphics[width=0.32\textwidth]{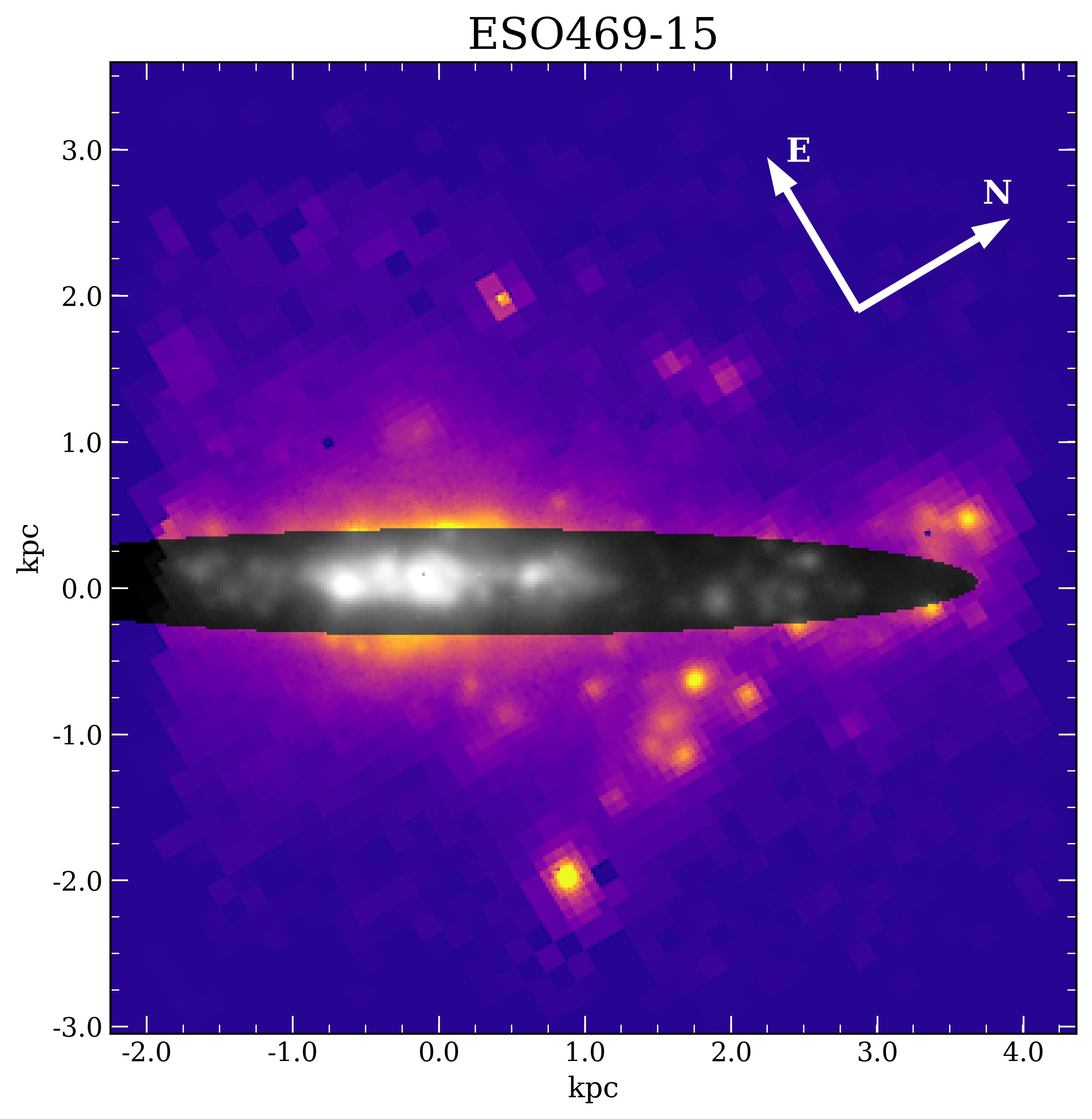}
\includegraphics[width=0.32\textwidth]{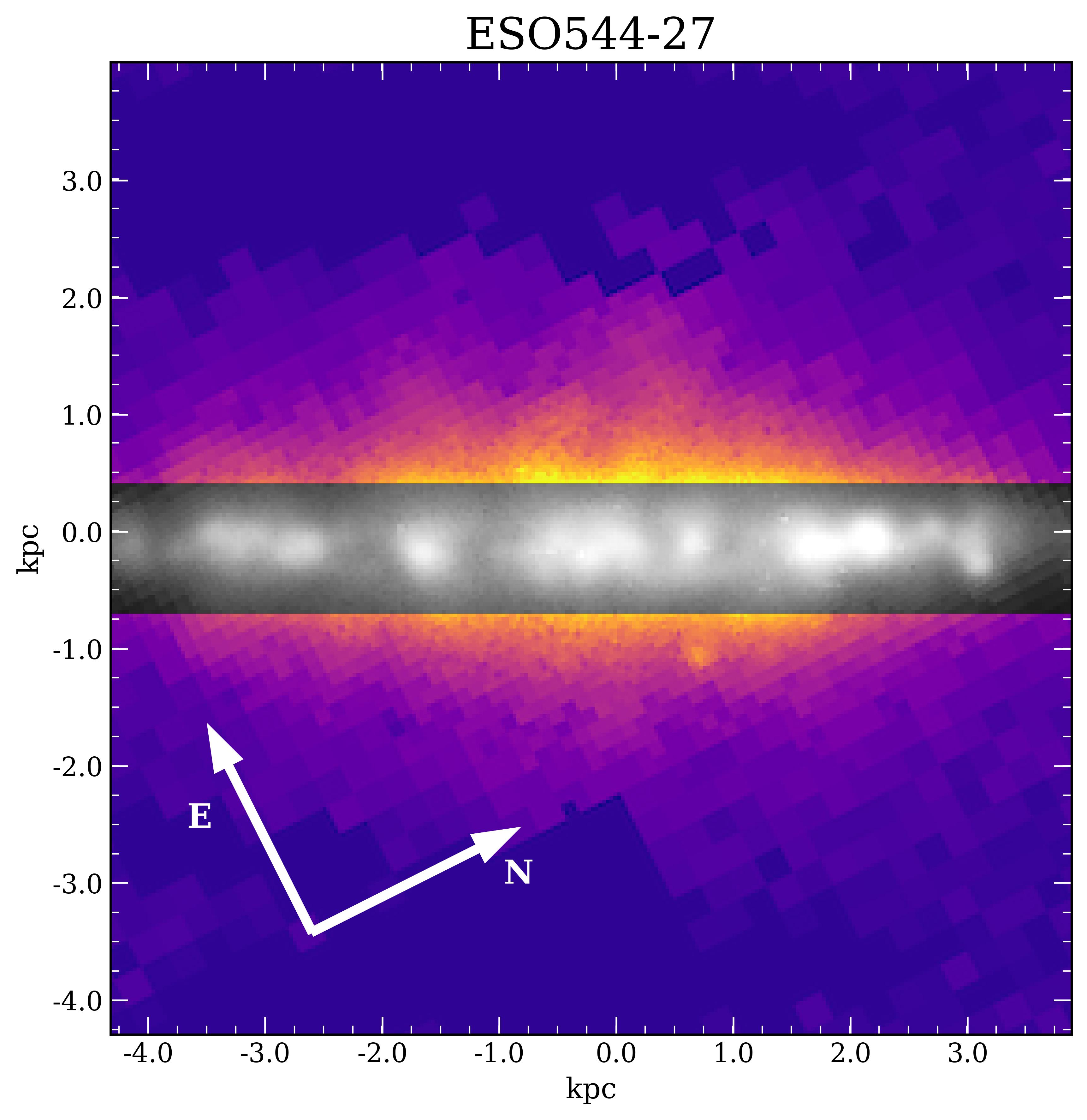}    
\caption{eBETIS sample. The images correspond to the binned \ha emission line maps obtained in Sec. \ref{sec:data}. Each image is rotated PA-90° with the white arrows indicating the direction of the celestial north and east in the original image. The ellipse fitting of galactic plane (see Sec. \ref{sec:data}) is coloured in grey scale. The contrast between the galactic plane and the extraplanar gas reveals the complexity of the ionised extraplanar gas, which comprises diffuse gas, filaments, and various structures.}
\label{fig:eBETIS_sample}     
\end{figure*}

\section{The data}\label{sec:data}

The characterisation of the eDIG and the exploration of its ionisation mechanisms, as for the case of the face-on galaxies, requires spatially resolved data and spectral information. This is particularly crucial for the eDIG, as the galactic halos may exhibit diverse mechanisms and structures that could contribute to the formation of the ionised extraplanar gas \citep{2019ApJ...882...84L}. These include phenomena such as plumes, fountains, filaments, gas accretion or gas outflows/inflows (\citealt{1996ApJ...462..712R,2003A&A...406..493R, 2003A&A...406..505R,2006MNRAS.370.1612K,2011MNRAS.415.1534M,2014ASPC..480..211C, 2019ApJ...882...84L}; \citetalias{2019A&A...623A..89C}).

For this reason, as in \citetalias{2024A&A...687A..20G}, we use MUSE IFS data to undertake this task. In particular, the sample of \citetalias{2019A&A...623A..89C} aligns well with the previous criteria. The sample, listed in table \ref{tab:sample}, consists on eight edge-on (i > 75°) nearby galaxies (z < 0.01, D < 46 Mpc) with high spatial resolution (full width at half maximum (FWHM) < 200 pc) from the 096.B-0054(A) and 097.B-0041(A) very-large telescope (VLT) programmes. Table \ref{tab:sample} shows the integrated star formation rate (SFR) in the disc spans between 0.024 M$_\odot$/yr < SFR < 0.229 M$_\odot$/yr. It was obtained using the \citet{2012ARA&A..50..531K} relation: log(SFR) = log($L_{H\alpha}$) - 41.27, with $L_{H\alpha}$ as the \ha luminosity, corrected for interstellar extinction assuming the Cardelli extinction law, assuming R$_V$ = 3.1 \citep{1989ApJ...345..245C} and a Balmer decrement \ha/\hb = 2.87.
The galaxies are then normal-disc galaxies with low star formation. This is consistent with the classification of these galaxies in the star formation main sequence, with specific SFR ranging between -10.5 < log(sSFR/yr)< -9.92, corresponding barely to star-forming galaxies (using MUSE data and 3.6 $\mu$m stellar masses; \citealt{2015ApJS..219....3M,2020MNRAS.499.1172Z,2022A&A...659A.153R}).

This sample, observed with MUSE, was selected since this type of investigation requires more constraints compared to the face-on case, since we are navigating within the galactic halo, where the signal-to-noise ratio of the observed emission line (S/N, \citealt{2012A&A...539A..73R};\citetalias{2024A&A...687A..20G}) is lower. Consequently, besides high resolution, we require data as deep as possible to observe low surface brightness lines such as \oi and \oiii, alongside data of sufficient quality to measure the \oi line without interference from sky lines. 

\noindent
The MUSE instrument brings 1 arcmin$^2$ of field of view (FoV), with a spatial and spectral sampling of $0.2\times0.2$ arcsec and 1.25 \r{A} respectively within a spectral range of $4650-9300$ \r{A}. 
Every datacube consists in four exposures (three for IC217) of 2624 seconds, and at least half of the galaxy is covered by one VLT pointing (from the centre to one of the edges of the galaxy, i.e., 0.5R$_{25}$ < 1 arcmin). The data was originally reduced using the ESO data-reduction phase 3 using the MUSE pipeline \citep{2020A&A...641A..28W}, and subsequently \citetalias{2019A&A...623A..89C} performed a sky subtraction using the {\sc ZAP} software (\citealt{2016MNRAS.458.3210S}, see \citetalias{2019A&A...623A..89C} for more details). 

\noindent
This dataset has previously been used to analyse the eDIG in prior studies. \citet{2022A&A...659A.153R} employed five of these galaxies, in conjunction with \ha and GALEX-FUV imaging, while \citet{2023A&A...678A..84D} exclusively utilised IC1553 for a more comprehensive examination of the eDIG ionisation mechanisms within this galaxy. In our paper, we use all eight galaxies from this dataset\footnote{We downloaded the reduced datacubes directly from the \href{https://archive.eso.org/scienceportal/home}{ESO archive science portal}.} (eBETIS sample henceforth) to perform an exhaustive spectral and spatial exploration of the eDIG structure and ionisation for every galaxy in particular. 

\noindent
To conduct this exploration, we first extract the emission line features of the eBETIS sample in the form of binned emission line maps, following the same procedure described in \citetalias{2024A&A...687A..20G}: we employ our own adaptive binning algorithm to the original datacubes, setting a S/N target of 10 based on \sii, since this line is one of its main features of the eDIG and DIG in general, as well as it typically exhibits low S/N in both planar DIG and eDIG. Subsequently, we perform a simple stellar population (SSP) synthesis to the binned datacube using {\sc STARLIGHT} \citep{2005MNRAS.358..363C} to the integrated spectra of each bin. This is followed by Gaussian fitting to every emission line of interest present in each resulting nebular spectrum (see \citetalias{2024A&A...687A..20G} for more details). This procedure finally results in the binned emission line maps for the \ha, \hb, \oiii$\lambda 5007$, \nii$\lambda\lambda 6548,83$, \sii$\lambda\lambda6717,31$ and \oi$\lambda 6300$ lines. The average sensitivity of the sample is 35.7 log(erg/s/kpc$^2$), obtained as the 2$\sigma$ of the distribution of \ha surface brightness (\SBha) of the bins of every \ha map. The average physical size of the bins is below 175 pc. Since the average point-spread function (PSF) FWHM of the sample is below 200 pc, the binning does not significantly restrict the resolution in our galaxies.

Figure \ref{fig:eBETIS_sample} shows the binned \ha maps of the eBETIS sample. For aesthetic purposes and to highlight the contrast between the galactic plane and the complex structure of the extraplanar gas, the plane of the galaxies is shown in greyscale. To define the galactic plane, we construct an elliptical mask by conducting a 2D Gaussian fitting to the binned \ha image using the {\sc python} routine {\sc scypy.opt.curve\_fit}
\footnote{\href{https://docs.scipy.org/doc/scipy/reference/generated/scipy.optimize.curve_fit.html}{{\sc scypy.opt.curve\_fit}} documentation.}, and centring the ellipse in the coordinates of the centre of the galaxy, which also coincides with the centre of the rotation curve of the galaxy. The midplane is then set at z = 0, being z the distance from the midplane, that corresponds to the centre of the minor axis of the ellipse fitted.
The integrated \ha luminosity of the eDIG for all the galaxies is listed in Table \ref{tab:sample}. It was obtained integrating the \ha flux outside the ellipses previously described, and we obtain an average value of 2.5$\cdot$10$^{39}$ erg/s, that is consistent with values obtained by previous authors \citep{2000ApJ...537L..13R,2003ApJS..148..383M, 2018ApJ...862...25J}.

In our data, for the subsequent analyses, we only include bins with S/N of the observed emission line greater than 2 and relative error lower than 40\% for all the lines of interest. The resulting binned maps exhibit an average physical bin size (listed in table \ref{tab:sample}) that remains below, or very similar to, the FWHM of the data, determined as the physical size of the seeing reported in the datacube header.

\section{The complex morphology of the eDIG} \label{sec:sec_3}

The \ha maps of Figure \ref{fig:eBETIS_sample} reveal the complex structure of the extraplanar gas in normal disc galaxies. The most discernible morphological structures observed in the outer regions of the galaxy are the filaments. In galaxies such as ESO157-49 and IC1553, these filaments are especially visible, emerging from the galactic plane and extending more than 2 kpc from the midplane. Other structures can be observed, for instance, in ESO443-21, which features a thick knot-shaped filament around the galactic centre that reaches a distance of $\sim$4.5 kpc from the midplane.

\begin{figure*}[ht!]
\centering
    \includegraphics[width=\textwidth]{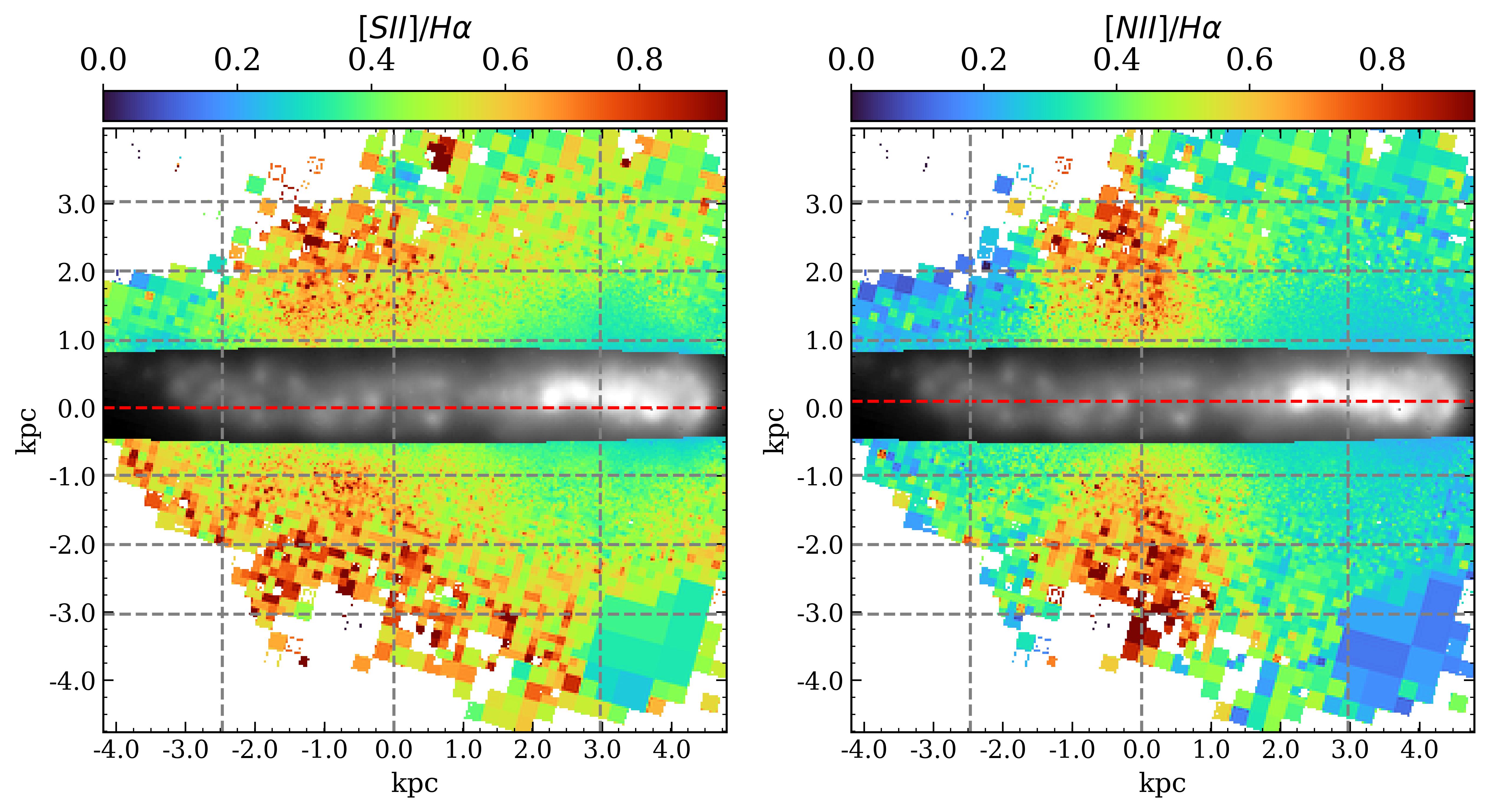}
    \includegraphics[width=\textwidth]{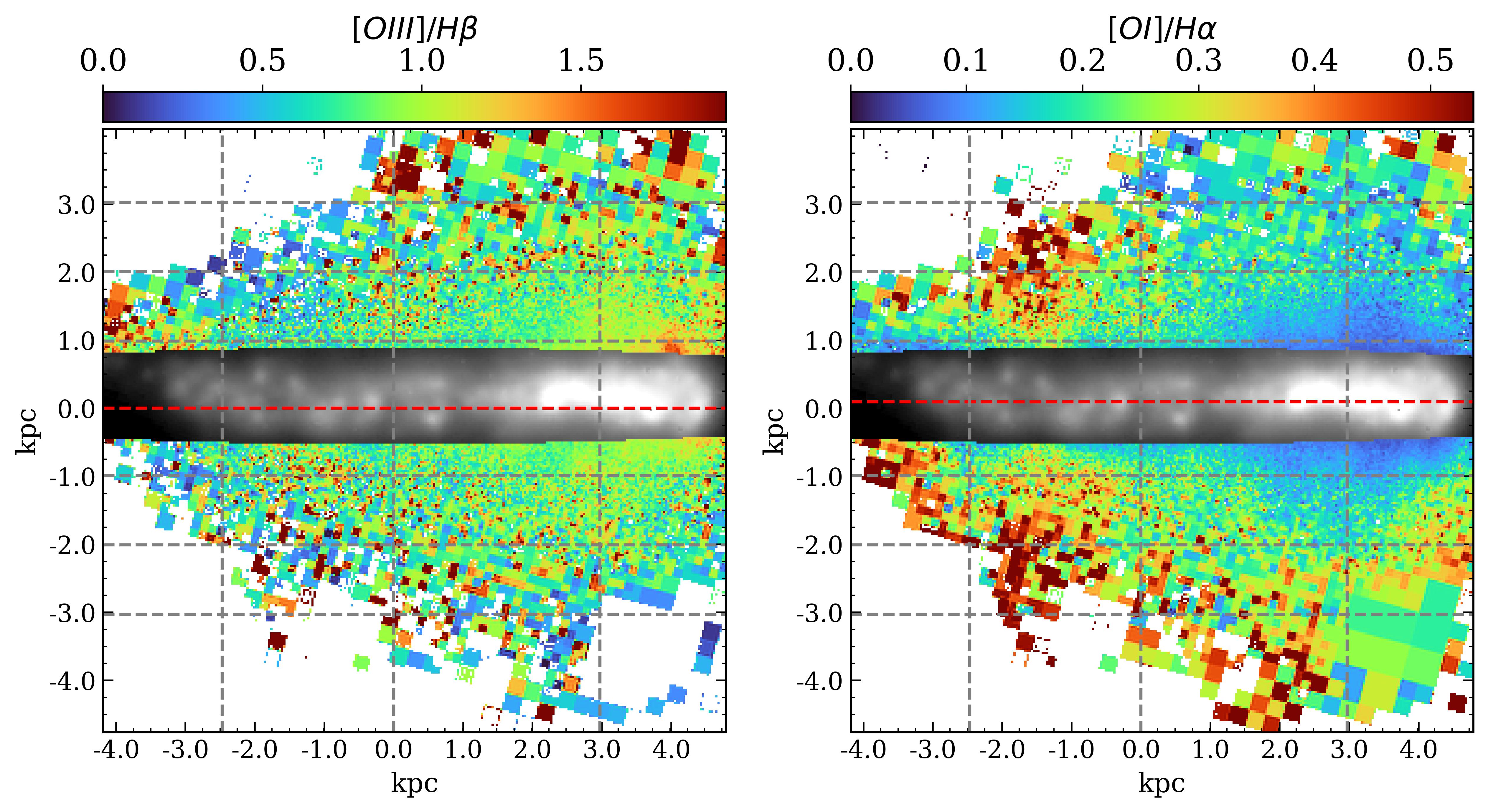} 
\caption{IC1553 line ratio maps. Every map is masked, excluding those bins with S/N < 2 or relative error greater than 40\% for the respective lines. We overlay the same \ha map of the galactic plane masked in Figure \ref{fig:eBETIS_sample}. The dashed red line indicates the midplane, located at z = 0. The y-axis is re-scaled in function of the galaxy inclination.In this case, the galactic plane is defined as a rectangle with the minor side equal to the minor axis of the fitted ellipse and the major side extending the full length of the x-axis. The grey dashed lines indicates the heights with respect the midplane z = $\pm$1, $\pm$2, $\pm$3 kpc and major axis distances MAD = -2.5, 0, 3 kpc.}
\label{fig:IC1553_maps}     
\end{figure*}

The presence of structures in the halo becomes more evident when examining the maps of the \nii/\ha, \sii/\ha, \oi/\ha, and \oiii/\hb line ratios. These maps reveal new ionised features that were initially invisible when only examining the \hii images. For instance, Figure \ref{fig:IC1553_maps} shows the binned maps of the mentioned lines ratios of the IC1553 galaxy.

The images show the intricate and asymmetrical structure of the eDIG. Having differences of 0.4 dex in the \sii/\ha ratio along the major axis and over 0.8 dex in \nii/\ha. Additionally, the line ratios increase with respect to the distance from the midplane, a phenomenon that is well-documented in the literature \citep{1998ApJ...501..137R, 2019ApJ...882...84L}. However, the 2D spectroscopic analysis of the eDIG allows to recognise that the variation of the line ratios are dependent not only on height, but along the major axis distance (MAD) of the galaxy. 

Between 1.5 kpc $\lesssim$ MAD $\lesssim$ 4.5 kpc, the \nii/\ha ratio presents significantly lower values than the range between -1 kpc $\lesssim$ MAD $\lesssim$ 1 kpc, showing a a conical-shaped structure that is slightly off-centred with respect to the galaxy. These features are also reported by other authors \citep{2022A&A...659A.153R, 2023A&A...678A..84D}. However, with the adaptive binning method applied \citepalias{2024A&A...687A..20G}, we recover data with S/N > 2 at z > 3, sampling almost the entire MUSE FoV in this galaxy and also in the rest of the galaxies. Besides, this apparent conical structure is shifted at MAD = -0.5 kpc in the \sii/\ha ratio, and at MAD = -1.5 kpc at \oi/\ha ratio, being not present in the \oiii/\hb ratio.
This asymmetry in the halo coincides with the asymmetry observed in the galactic plane, where most of the \ha emission from the \hii regions is concentrated between 1 kpc $\lesssim$ MAD $\lesssim$ 3.5 kpc. At this radial range, the line ratios found in the eDIG resemble more closely the line ratios of the \hii regions rather than the typical DIG line ratios (below 0.3 and 0.4 in the \nii/\ha and \sii/\ha, and below 0.1 in \oi/\ha; \citetalias{2024A&A...687A..20G}).

In the appendices from \ref{sec:Ap_A} to \ref{sec:Ap_F}, we present the line ratio maps for the rest of the sample, where the complex bidimensional structure of the eDIG and its apparent connection between the line-of-sight \hii regions of the galactic plane is evident for all galaxies.

\begin{figure}[t!]
\centering
\includegraphics[width=\columnwidth]{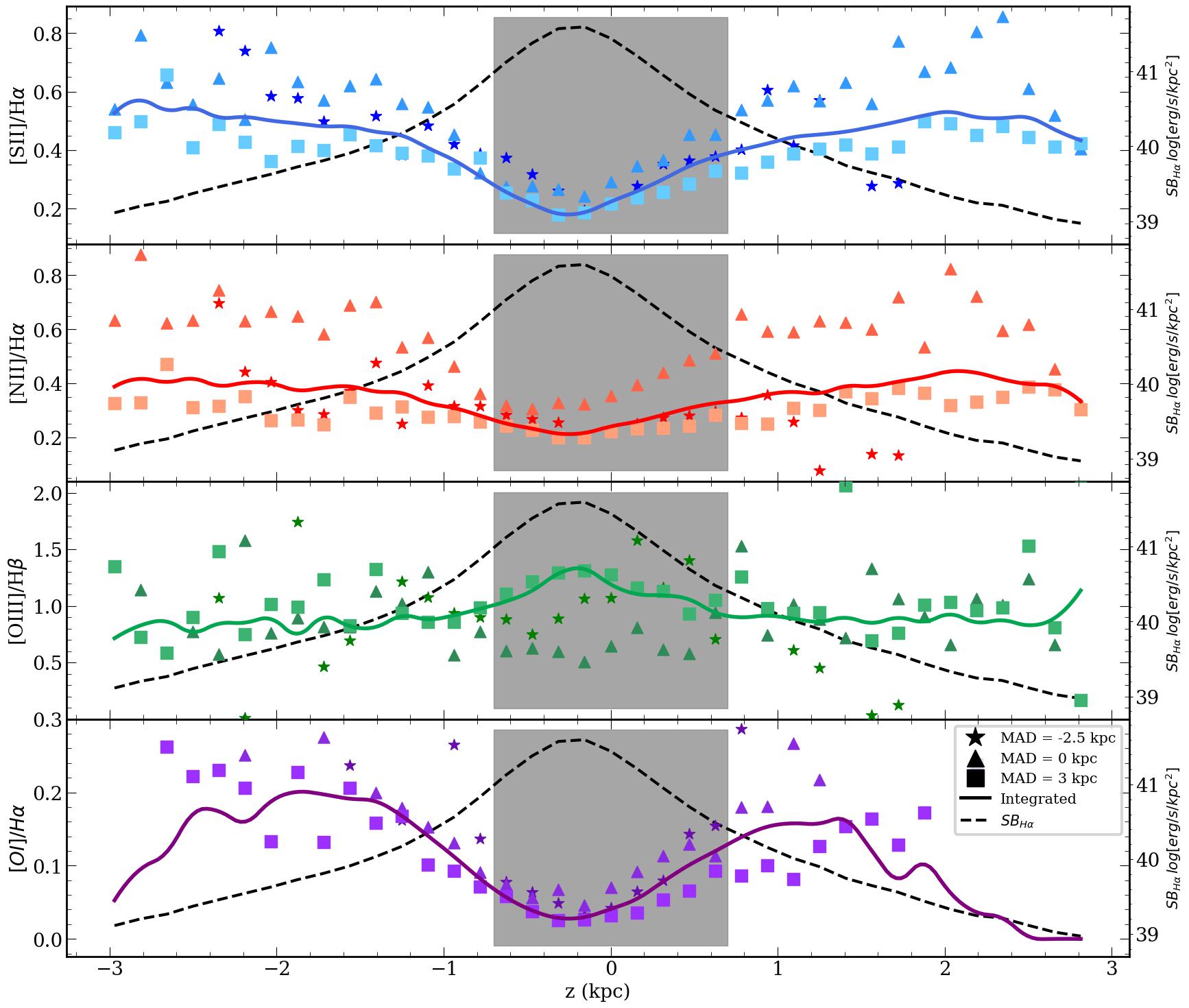}
\caption{Distribution of the line ratios with respect the distance from the midplane for the IC1553 galaxy. 
The stars, triangular and square markers represent the line ratio at that z for MAD = -2.5, 0 and 3 kpc respectively, analogous to a long-slit perpendicular to the galactic
plane at a given MAD. The solid lines represents the integrated values along the major axis distance. The dashed black line represents the \SBha height distribution of the galaxy. The grey background between z = - 0.7 kpc and 0.7 kpc represents the galactic plane.}
\label{fig:IC1553_lines}     
\end{figure}


Figure \ref{fig:IC1553_lines} shows the distribution of the eDIG characteristic line ratios with respect the distance from the midplane for IC1553. The stars, triangles, and squares represent the values of the line ratio at that height for MAD equal to -2.5 kpc, 0 kpc, and 3 kpc, respectively. Analogously, this corresponds to the line ratio height distribution in a long-slit located at those MADs and perpendicular to the galactic plane. The solid lines are constructed then integrating the fluxes along the major axis. The dashed lines represent the \SBha distribution, constructed similarly to the solid lines by integrating the \ha flux along the major axis, and the grey background represents the location of the galactic plane. Since the \hii regions distribution is not symmetrical with respect z = 0, the minimum of the \SBha 
and line distributions are shifted with respect z = 0.

The integrated distribution of the \nii/\ha, \sii/\ha and \oi/\ha line ratios tends to increase in height, reaching a minimum inside the galactic plane. The \nii/\ha and \sii/\ha line ratios increase up to 0.4 and 0.6, respectively, from z = 0 to 2 kpc, and then remain constant. \oi also increases from z = 0 to 2 kpc, reaching up to 0.2, but then decrease dramatically. This line ratios seems anti-correlated with the \SBha vertical distribution. In contrast, for the \oiii/\hb ratio, at z = 0 it reaches its maximum of 1.3 and then decreases slowly to 0.7.

This trend is consistent with findings from previous studies in other edge-on galaxies \citep{1998ApJ...501..137R, 2000ApJ...537L..13R, 2001ApJ...551...57C,2009RvMP...81..969H,2016MNRAS.457.1257H, 2023A&A...678A..84D}. However, in addition to this, when restricting the height distributions to specific MADs, the overall behaviour of the distributions remains consistent, but the line ratios exhibit notable differences. At MAD = 3 kpc (represented by squares in Figure \ref{fig:IC1553_lines}, where the most prominent \hii regions emission is observed in Figure \ref{fig:IC1553_maps}), the \nii/\ha, \sii/\ha and \oi/\ha line distributions follow the same trend as their respective integrated distributions, but consistently reach lower values. Similarly, at MAD = 0 (represented by triangles in Figure \ref{fig:IC1553_lines}, where the \hii regions emission observed in Figure \ref{fig:IC1553_maps} is lower), the line distributions also follow the same trend, but consistently reach higher values. This behaviour in the line ratio distributions is consistent in all the galaxies, as shown in the appendices from \ref{sec:Ap_A} to \ref{sec:Ap_F}.

\begin{figure}[t!]
        \centering
        \includegraphics[width=\columnwidth]{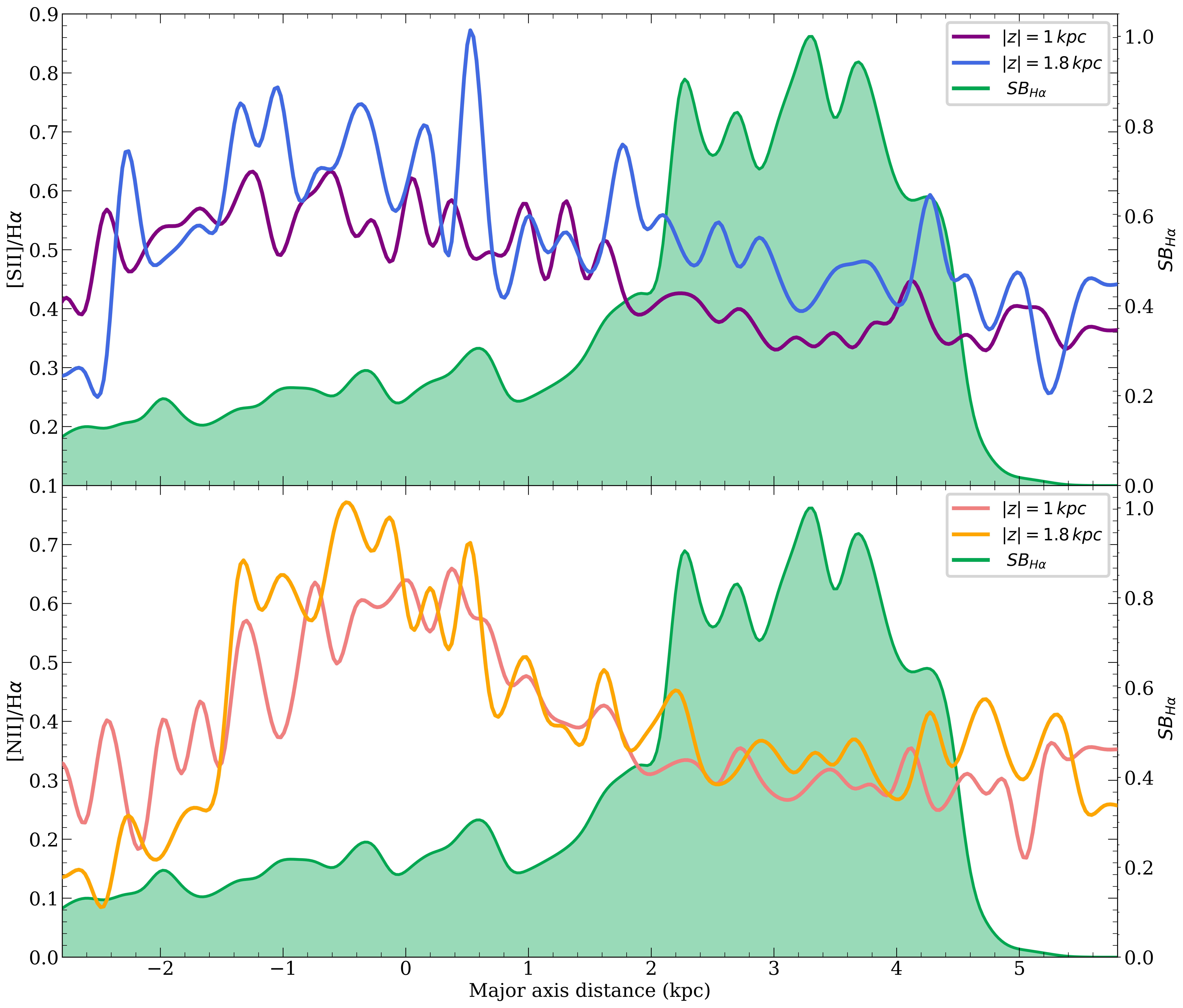}
        \caption{Major axis distance (MAD) distribution of the \nii/\ha and \sii/\ha lines for IC1553 at z = 1 and 1.8 kpc. In green, the normalised \SBha MAD distribution, constructed integrating the \ha flux within the ellipse fitted in section \ref{sec:data}  throughout the major axis. Both line distributions show a correspondence with the \SBha distribution, decreasing and showing more difference between both heights as the \SBha increase. This tendency is consistent for all galaxies.}
        \label{fig:IC1553_MAD}
\end{figure}

Figure \ref{fig:IC1553_MAD} shows the MAD distribution of the \nii/\ha and \sii/\ha lines for z = 1 and 1.8 kpc, as well as the normalised \SBha MAD distribution of IC1553.
Two effects are noticeable in this figure: Firstly, all the line ratio distributions tend to be lower between 2 kpc and 4.5 kpc, where the \SBha is higher, and increase while the \SBha decreases, particularly for the \nii/\ha ratio. Secondly, the effect of the increment of the line ratio with height, as seen in Figure \ref{fig:IC1553_lines}, remains present along the major axis, with the difference being clearer in regions of high \SBha, especially for the \sii/\ha ratio, where there is a constant difference of 0.1 between both distributions. These effects are not exclusive to IC1553, the MAD distributions present the same behaviour for all galaxies in the sample (see appendix). These figures evidences the high complexity that the eDIG presents, suggesting that the physical conditions in the eDIG are influenced by the distribution of the \hii regions emission along the galactic plane.

\section{Discussion}
\label{sec:discusion}

\subsection{The linkage between the galactic plane and the eDIG}

The correlation between the morphological distribution of star-forming regions in the disks of edge-on galaxies and the shapes and morphology of the halos presents the strongest evidence supporting the interpretation that energy sources from star formation (SF) within the galaxy disks drive the observed disk-halo interactions \citep{1989ApJ...345..372N,1997PASP..109.1298D}.
Over the past decade, there has been a unanimous consensus in the literature that the primary ionisation mechanism for the WIM (DIG) and the eDIG is attributed to star formation in the discs of late-type galaxies (\citealt{2016MNRAS.457.1257H,2019ApJ...885..160B,2019ApJ...882...84L,2019MNRAS.483.2382W,2022A&A...659A..26B,2022A&A...660A..77D,2022A&A...666A..29D,2022A&A...659A.153R,2022MNRAS.513.2904T,2023A&A...678A..84D,2023MNRAS.519.6098L,2024MNRAS.528.2145S}, etc.). Specifically, Lyman continuum (LyC) photons leaking from OB associations in the galactic discs escape into the ISM in the halos through transparent pathways, ionising layers of low-density gas that can be traced out to distances of up to 5 kpc or more from the midplane of the disks \citep{2009RvMP...81..969H,2019A&A...622A.115W}.

The specifics of the underlying mechanisms governing the disc-halo connection are complex and widely debated. From an energetic perspective, several external sources have been proposed to account for the origin of the eDIG. These include gas accretion from the intergalactic medium \citep[IGM, e.g.][]{Binney2005,Putman2017} and the circumgalactic medium \citep[CGM, e.g.][]{2019ApJ...882...84L,Bizyaev2022}, the latter also being suggested as a probable origin of the UV halo \citep[e.g.][]{2016ApJ...833...58H,ShinnSeon2015} or a hot halo \citep{Marasco2012,Li2023}. However, in a recent study, \citet{2024MNRAS.528.2145S} argue, using a sample of nearby highly inclined ($i \geq 80\degree$), isolated galaxies, that the eDIG phenomenon is uncorrelated with the galaxy environment.
Consequently, the primary energy sources driving the disc-halo interactions should reside within the galaxy discs, as no other pervasive external energy sources are known that could account for the observational data \citep{1997PASP..109.1298D}. Therefore, the halo gas expelled or heated by these energy sources must also originate from the discs.

Several heating mechanisms have been proposed to account for the energetics observed in the halos of galaxies: AGNs and high-level star formation, or a combination of both. In the case of AGNs, many spirals exhibit very characteristic outflow cones associated with a nuclear engine, which can be traced using X-rays, radio continuum emission, and optical emission-line maps (e.g. \citealt{1994AJ....107.1227W}; \citealt{1996ApJ...467..551C}; \citealt{2019MNRAS.482.4032L}). However, considering that only 10–15\% of galaxies in the local universe ($z < 0.1$) exhibit AGN activity, and that disk-halo interactions are observed over extensive regions of the discs, AGNs cannot be the common cause for the observed disk-halo interactions. Indeed, no AGNs are found in the eBETIS sample (Sec. \ref{sec:data}) or in recent eDIG studies of normal late-type galaxies \citep{2016MNRAS.457.1257H,2019ApJ...882...84L,2023MNRAS.519.6098L}.

On the other hand, high-level SF can occur in various locations within galaxy discs, either in circumnuclear starbursts or in giant \hii regions dispersed across the discs. Several scenarios for the disk-halo interaction due to SF have been proposed: the superbubble breakout model \citep[e.g.][]{MacLowMcCray1988}, the galactic fountain model \citep{1976ApJ...205..762S}, the champagne model \citep{Tenorio-Tagle1979}, the chimney model \citep{1989ApJ...345..372N}, and the galactic wind model \citep[][and reference therein]{Veilleux2005}. For the discussion purpose, we exclude phenomena related to nuclear activity in galaxies or incoming gas from the IGM/CGM, and we focus on those models in which ionising radiation originating from SF must not only escape the \hii regions but also traverse the dense clouds of neutral hydrogen near the midplane.

In the galactic fountain model, the outflow velocities are expected to be lower than the escape velocity of the gas ($v < v_{esc}$), indicating that all material will ultimately return to the disc, resulting in a dynamic circulation of matter as the primary mechanism of the disc-halo interaction. 
In the chimneys model, OB associations are born within giant molecular clouds, and the ionising ultraviolet flux and stellar winds from successive generations of OB stars provide significant momentum input, creating free cavities in the surrounding ISM. The sequential explosions of supernovae (SNe) result in the formation of a dense, cool wall of neutral gas and a central flow of hot gas, which constitutes a chimney. The total energy output from stellar winds is comparable to that of SNe, albeit released over a more extended period. For a simple instantaneous starburst model with a canonical IMF and a total mass of $10^6$ M$_{\odot}$, the total energy is approximately $10^{55}$ ergs \citep{1992ApJ...401..596L}. The hot gas ascends to the halo through the chimneys, reaching distances of several kpc. In this model, the chimneys create vertical channels that facilitate the rapid propagation of SN-driven shocks, with gas velocities reaching $v \geq v_{esc}$, thereby enabling the ejection of matter into intergalactic space. The gas rises to $\gtrsim 3$ kpc and returns to the disk after a dynamical time of $\sim10^7$ years with an infall velocity of $< 10^2$ km s$^{-1}$, resulting in an ongoing exchange of mass, energy, and momentum. For canonical Galactic parameters, the chimney phase is associated with a mass flow rate of $0.3-3$ M$_{\odot}$ yr$^{-1}$ and a global power input of $10^{40-42}$ ergs s$^{-1}$ \citep{1989ApJ...345..372N}.

This scenario aligns with the observed multiphase nature of the halo ISM (\citealt{2003A&A...406..493R};\citealt{2010ApJ...721..505R}). The chimney model can be described as an intermediate state between a two-phase model \citep{1989HiA.....8..567F}—relevant in galaxies with higher mean ambient density—and a three-phase model \citep{1977ApJ...218..148M}, in which the third hot phase consists of chimneys with a filling factor of about $10\%$ \citep{1989ApJ...345..372N}. In both models, disk-halo interactions occur, but in the fountain mode, the energy input over the disk is distributed across extensive areas, whereas in the chimney mode, the energy injection into the ISM is more localised.

Independent of the various outflow scenarios, they provide absorption-free pathways for H-ionising photons into the halo, as well as channels for the vertical transport of hot processed material and cosmic rays into the halo, where a hot gas phase with temperatures of up to $\sim 10^7$ K has been observed. This X-ray emitting plasma offers a further compelling argument favouring SF-related processes as the heating sources of the halo, given that the cooling time for such gas is very short, requiring continuous heating. Indeed, several authors have identified stellar winds and SNe as the principal energy sources driving this process (\citealt{1977STIN...7826045W}; \citealt{2019ApJ...887..161G}; \citealt{2022ApJ...940...44S}).

However, real galaxies exhibit greater complexity than the models depicted. For instance, the Type II supernova rate and the average gas density at the disc can vary with galactocentric distance, resulting in different processes and phases occurring at various radii within the same galaxy. Additionally, these parameters can fluctuate over time, especially in galaxies experiencing bursts of star formation throughout their lifetimes. The total SFR, the distribution of \hii regions, and the local level of star formation in different regions across a galactic disk are not constant over time. Galaxies may undergo cycles of star formation, implying they might temporarily be in states of high, intermediate, or quiescent phases of star formation \citep{1988A&A...192...57R}. The natural extension of the disk ISM into the halo via star-formation-driven outflows also indicates that the halo ISM is dynamic rather than static. The multiphase ISM exists under the influence of both the galaxy's gravitational potential and the energy input from heating sources, leading to interactions between different phases and constituents of the ISM. Consequently, in the presence of ionising UV radiation from hot massive stars, stellar winds, and supernova shock waves, the ISM in galaxy halos is never in local thermodynamic equilibrium.

\begin{figure}[ht!]
    \centering
    \includegraphics[width=\columnwidth]{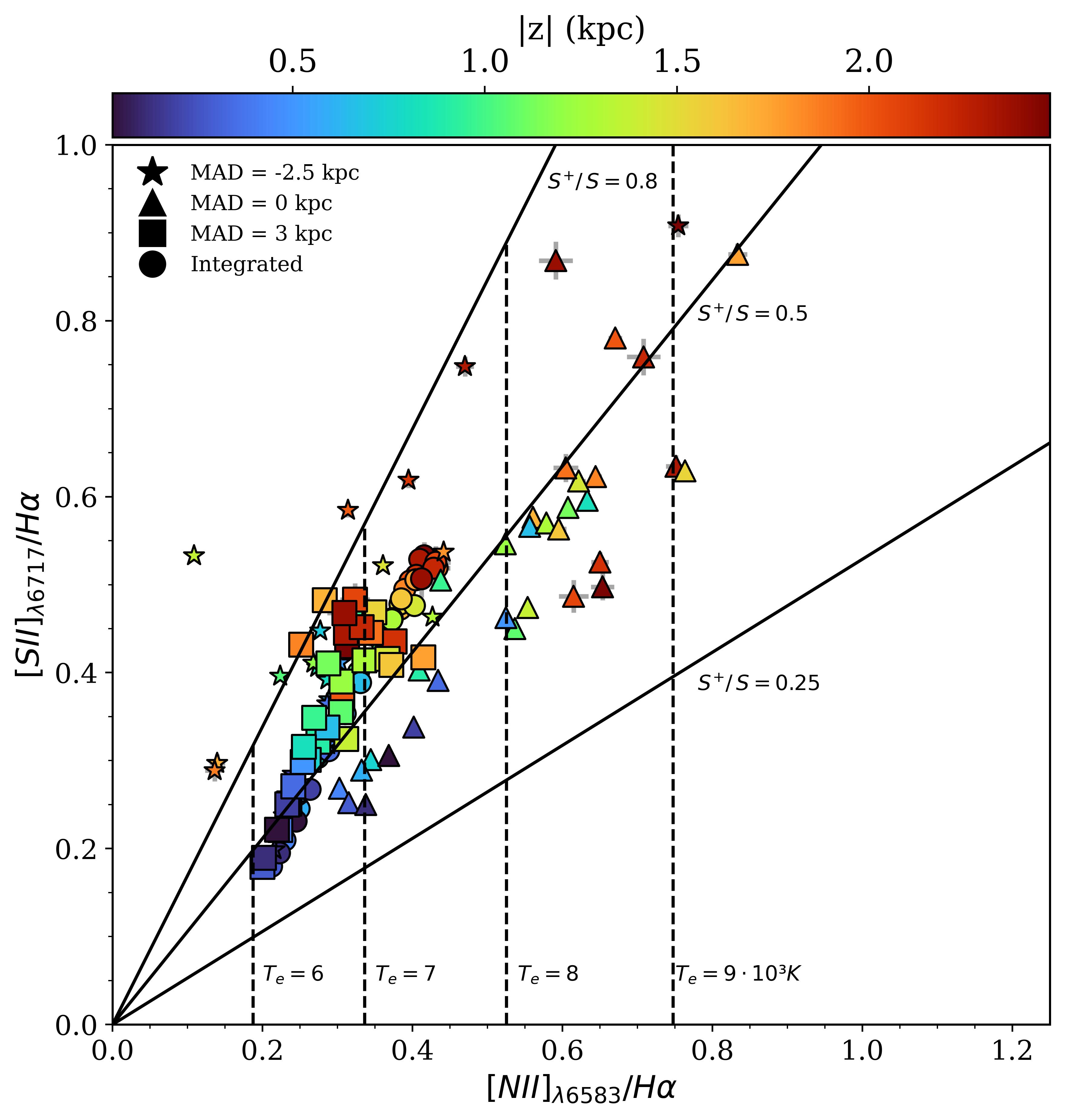}
    \caption{\nii$_{\lambda\,6583}$/\ha vs. \sii$_{\lambda\,6717}$/\ha diagnostic diagram for IC1553. Colours represent height with respect the midplane, equally from above or below the plane. Vertical dashed lines are the theoretical ratios obtained from \ref{eq:N_H} at a constant T$_e$ (6, 7, 8 and 9 in units of 10$^3$ K respectively). Solid lines represent the theoretical ratios from equations \ref{eq:N_H} and \ref{eq:S_H} for at a fixed S$^+$/S (0.25, 0.5, 0.8 respectively). The circles indicate the line ratios obtained by integrating the fluxes along the major axis at a certain z. Stars, triangles and squares indicate the line ratio at a certain z for MAD = -2.5, 0 and 3 kpc respectively.}
    \label{fig:IC1553_T_S+S}   
\end{figure}

The morphology and properties of gaseous halos are influenced by the level of SF activity in the underlying disk \citep{2003A&A...406..493R}. Galaxies require a sufficient level of SF to sustain a widespread extraplanar eDIG layer \citep{1995ApJ...444..119D}. Here, itt is essential to distinguish between pure starburst galaxies (typically with circumnuclear SF, e.g. M82) and normal galaxies with varying levels of widespread SF in their disks. For the latter, it is known that the most actively star-forming galaxies possess the most prominent halos \citep{1997PASP..109.1298D}. In galaxies with intermediate SF levels, such as those in the eBETIS sample, disk-halo interactions occur only in regions with high SFR. Theoretical calculations indicate that only those regions surpassing a threshold of $10^{-2}$ ergs s$^{-1}$  cm$^{-2}$ are capable of initiating disk-halo interaction \citep{1995ApJ...444..119D}.

Indeed, the connection between the \SBha distribution and the ionisation structure of the eDIG in the eBETIS galaxies (see Sec. \ref{sec:sec_3}) demonstrates a clear correlation between the morphology of halo emission and the distribution of SF regions and complexes in the underlying galaxy disks. Furthermore, the detection of extended eDIG in the halos of the eBETIS sample indicates that even low-mass galaxies with modest SFR (Tables \ref{tab:sample} and \ref{tab:eDIG_char}) are capable of producing a pervasive layer of extraplanar diffuse ionised emission. This finding contradicts previous assumptions that halos are a rare phenomenon among normal spirals \citep{1991A&A...248...23H}, a bias likely caused by the insufficient sensitivity of earlier observations.

In the following, we discuss the observational evidence supporting this scenario as derived from the eBETIS sample.

\subsection{The ionisation mechanism of the eDIG} 

In the scenario delineated above, the eDIG emission originates in situ within low-density gas environments, induced by ionising photons propagated from the galactic disk via transparent conduits formed by superbubbles or chimneys. However, the precise underlying physics explaining the large distance between the extraplanar gas and ionising stars in the midplane, as well as the fine-tuned details of the ionisation mechanism of the eDIG, remain a topic of ongoing debate.

Long-standing issues regarding the specifics of OB-star-driven ionisation remain prevalent in the literature. A notable characteristic of the DIG that proves difficult to explain through OB star photoionisation is its optical emission-line spectrum. Should the eDIG be ionised by radiation escaping from the midplane H II regions, the spectrum of the ionising radiation would be expected to resemble that of the OB star (or the SED dominated by such stars). Consequently, one would anticipate similar emission-line spectra in the eDIG as in the H II regions. Nonetheless, certain observed optical emission line ratios are not explicable by pure photoionisation from OB stars. The distinctive emission-line ratios observed in the eDIG, namely \nii$_{\lambda6584}$/\ha, \sii$_{\lambda6717}$/\ha, \oiii$_{\lambda5007}$/\hb and \oi$_{\lambda6300}$/\ha, are elevated compared to those observed in \hii regions \citep{1997ApJ...474..129R,1998ApJ...501..137R,2001ApJ...551...57C,2009RvMP...81..969H,2009ApJ...703.1159S}, as illustrated in Fig. \ref{fig:IC1553_lines} and elaborated upon in Sec. \ref{sec:sec_3} for the eBETIS sample, indicating the relative significance of additional heating processes.

Some of the differences between the DIG and \hii region emission-line spectra (i.e., enhanced \nii/\ha and \sii/\ha) can be explained by differences in temperatures. In a model of photoionised gas, the temperature is determined by the equilibrium between heating and cooling mechanisms. Heating is facilitated through the thermalisation of the excess kinetic energy of electrons during the photoionisation-recombination process, whereas cooling predominantly arises from collisional excitation followed by radiative decay of metastable states of trace ions \citep{2006agna.book.....O}.


Relative temperatures can be explore by analysing the emission line ratios of the forbidden lines with respect to the H-recombination emission. The line intensities of \nii and \sii relative to \ha probe the ionisation fractions ($X^{+}/X$) and elemental abundances of ionised the gas, showing a strong dependence on the electron temperature (T$_e$). These parameters relate with the line intensity ratios as follows \citep{1999ApJ...523..223H, 2006agna.book.....O}:

\begin{equation}\label{eq:N_H}
    \frac{I{([N\,{II}]_{\lambda 6583}})}{I(H{\alpha})} = 1.63 \cdot 10^5 \left(\frac{N^+}{N}\right) \left(\frac{N}{H}\right) \left(\frac{H^+}{H}\right)^{-1} T_{4}^{0.426}e^{-2.18/T_4} ,
\end{equation}

\begin{equation}\label{eq:S_H}
    \frac{I{([S\,{II}]_{\lambda 6717})}}{I(H{\alpha})} = 7.64 \cdot 10^5 \left(\frac{S^+}{S}\right) \left(\frac{S}{H}\right) \left(\frac{H^+}{H}\right)^{-1} T_{4}^{0.307}e^{-2.14/T_4} ,
\end{equation}

\noindent
with T$_{4}$ being the electron temperature in units of 10$^4$ K.

We assume that usually the 100\% of the hydrogen and the 80\% of the nitrogen are ionised in the DIG, varying little within the DIG and basically tracing the T$_{e}$ \citep{2000ApJ...528..310S,2002ApJ...572..823O,2009RvMP...81..969H}. We also assume solar abundances: N/H = 6.8$\cdot10^{-5}$ and S/H = 1.3$\cdot10^{-5}$ \citep{2021A&A...653A.141A}.

\citet{1999ApJ...523..223H} and \citet{2006ApJ...652..401M} used these relationships to build diagnostic diagrams to estimate both T$_e$ and S$^{+}$/S from observations of \nii/\ha and \sii/\ha. Given that H and N have similar first ionisation potentials (13.56 eV and 14.5 eV, respectively) and a weak charge-exchange reaction, and assuming that N is not doubly ionised, then N$^+$/N$^0$ $\sim$ H$^+$/H$^0$ vary little within the DIG, so that variations in \nii/\ha essentially traces variations in T$_e$ \citep{2000ApJ...528..310S,2002ApJ...572..823O,2009RvMP...81..969H}.  
Figure \ref{fig:IC1553_T_S+S} shows the distribution of \nii$_{\lambda\,6583}$/\ha vs. \sii$_{\lambda\,6717}$/\ha line ratios in function of |z| for IC1553. Circles are the integrated along the major axis for a certain z. Stars, triangles and squares are bins at MAD = -2.5, 0 and 3 kpc respectively. We use the equations \ref{eq:N_H} and \ref{eq:S_H} to show the predicted ratios for a constant T$_e$ (6, 7, 8 and 9 in units of 10$^3$ K) and S$^+$/S (0.25, 0.5 and 0.8).


The general behaviour indicates a linear, constant increment of T$_e$ with increasing distance from the galactic plane. This trend is consistent with previous studies of the WIM in the Galaxy \citep{2009RvMP...81..969H, 2006ApJ...652..401M} and with long-slit observations perpendicular to other edge-on galaxies \citep{2001ApJ...551...57C,2001ApJ...560..207O,2003ApJ...586..902H,2019ApJ...885..160B}. However, the spatial coverage of the eBETIS galaxies permits the study of the dependence of the eDIG line ratios on temperature at different MADs. Figure \ref{fig:IC1553_T_S+S} illustrates significant variations in T$_e$ and S$^{+}$/S with respect to the line of sight. For instance, at the galaxy minor-axis (MAD = 0), the temperature increases from 7 to 9 $\times 10^3$ K and the S$^{+}$/S ratio rises from 0.3 to 0.5, whereas at MAD $= -2.5$ kpc, the S$^{+}$/S ratio attains high values of approximately 0.8 at all vertical distances. At MAD = 3 kpc, the temperature of the eDIG is confined to a narrow range between $6-7.5 \times 10^3$ K, with S$^{+}$/S $\approx 0.5-0.8$. To comprehend the variations in S$^{+}$/S, we could examine the energetics of the eDIG trace ions. The ionisation energies of N and S are 14.5 eV and 10.4 eV, respectively. If N is mostly ionised, then S should also be mostly ionised. However, the energy required to double-ionised S is 23.3 eV (S$^+$ $\rightarrow$ S$^{++}$), while for N is 29.6 eV (N$^+$ $\rightarrow$ N$^{++}$). Therefore, S$^+$/S may vary while N$^+$/N remains constant if the ionising energy does not reach 29.6 eV. Consequently, we anticipate greater variation in S$^+$/S under eDIG conditions.

The variations in the temperature and ionisation structure of the eDIG as a function of spatial position within a single galaxy, as illustrated in Fig. \ref{fig:IC1553_T_S+S}, provide more compelling evidence of the intricate dynamical heating structure of the ISM in the halo of late-type galaxies. A similar trend is observed across the entire eBETIS sample, as depicted in the associated figures in the appendices, indicating that the eDIG tends towards a lower ionisation state (higher S$^+$/S $\sim 0.5-0.8$) compared to classical \hii regions, which would be situated in the lower-left corner of these diagrams (e.g. \nii/\ha $\approx 0.25$, \sii/\ha $\approx 0.1$, \citealt{2006ApJ...652..401M}). The anomalous \nii/\ha and \sii/\ha eDIG line ratios can be attributed to an increase in the T$_e$ temperature in the halo, resulting from photoionisation with a low ionisation parameter $q$, i.e., the ratio of ionising photon density to electron density, which measures the diluteness of the radiation field \citep{1994ApJ...428..647D}. Nonetheless, while the elevated \nii/\ha and \sii/\ha line ratios can be explained by low $q$ photoionisation, other eDIG line ratios, such as \oiii/\hb and \oi/\ha, remain unexplained \citep{2009RvMP...81..969H}.

What remains puzzling in this scenario is the increase in T$_e$ with respect to the midplane distance |z|, as supplementary heating sources may be necessary to account for this phenomenon. The literature has extensively debated the origin and characteristics of an additional source of ionisation to explain the observational features of the (e)DIG. The observed temperature behaviour of the eDIG, as depicted in Fig. \ref{fig:IC1553_T_S+S}, could be attributed to the spectrum of the ionising radiation being reprocessed as it traverses the ISM, an effect known as radiation hardening: a preferential absorption of photons near the ionisation thresholds within the sources' environment, resulting in (on average) more energetic photons escaping from the \hii regions. This results in higher temperatures in the (e)DIG than in \hii regions, which, in turn, leads to stronger emission of collisionally excited optical lines \citep{2003ApJ...586..902H,2004MNRAS.353.1126W,2005ApJ...633..295W,2009RvMP...81..969H}. This model explains the WIM in our Galaxy but fails to explain an increasing \oiii/\hb ratio at large distances from the galactic plane \citep{2019A&A...622A.115W}. Therefore, the rise in the \oiii ratio with H-recombination lines with |z| needs an additional source of heating and/or ionisation.

Several alternative heating mechanisms that may raise the electron temperature of the eDIG have been proposed, including differential absorption and/or scattering of radiation due to dust \citep{1993ASPC...35..540S,1996ApJ...467L..69F}, dissipation of turbulence \citep{1997ApJ...485..182M}, magnetic reconnection \citep{1999ApJ...525L..21R,2012A&A...544A..57H}, photoelectric heating from small grains \citep{2001ApJ...558L.101R}, cosmic rays heating \citep{2013ApJ...767...87W,2019ApJ...885..160B}, etc. However, models still struggle to fully reproduce the (e)DIG emission-line spectrum.

A potential heating source that has gained prominence in the literature in recent years is the photoionisation by a hard spectrum produced by the high-temperature end of the white-dwarf distribution and the central stars of planetary nebulae. These low mass, hot, and evolved (post-AGB) stars, known as HOLMES, has been proposed as an ionisation mechanism for the DIG in general (e.g., \citealt{1991PASP..103..911S, 2018MNRAS.474.3727L, 2022A&A...659A..26B, 2024MNRAS.tmp..356L}) and for the eDIG in particular (e.g., \citealt{2011MNRAS.415.2182F, 2019A&A...622A.115W, 2022A&A...659A.153R}).
HOLMES are abundant and smoothly distributed in the thick disc and lower halos of the galaxies, with scale heights greater than those of the OB stars. They are a significant source of ultraviolet emission, capable of reproducing the increment in the line ratios found in the (e)DIG, and specifically for the high ionisation species such as \oiii \citep{2011MNRAS.415.2182F}. 
Indeed, \citet{2022A&A...659A..26B} and \citet{2024MNRAS.tmp..356L} argue on the existence of two different types of diffuse emission based on the \EWha bimodal distributions centered at $\sim 1$ and $\sim 10$ \AA, which according to the classification by \citet{2018MNRAS.474.3727L}, corresponds to the so-called hDIG ionisation (component dominated by HOLMES), and the ionisation of SF complexes (component dominated by unresolved \hii region, or photons leaked from them), respectively.

The increase in the \sii/\ha and \nii/\ha ratios with distance from the midplane is commonly used to support HOLMES as an ionisation source in the eDIG \citep{2011MNRAS.415.2182F,2022A&A...659A.153R}, as well as the radial behaviour of these lines in low-inclination spiral galaxies \citep{2022A&A...659A..26B}. Nevertheless, the behaviour of the line ratios shown in Fig. \ref{fig:IC1553_lines} (and related figures in the appendices) reveals that the ionisation of the eDIG is highly dependent on the distribution of \hii regions in the disc (Sec. \ref{sec:sec_3}). The \nii/\ha, \sii/\ha, and \oiii/\hb ratios attain their highest values at the greatest distances from the midplane and in regions with lower \SBha. This observation is incompatible with a smooth distribution of HOLMES throughout the thick disc and lower halo. Under the HOLMES scenario, higher \oiii/\hb ratios would be expected in the vicinities of the thick disc, along with a smooth distribution of characteristic high ionisation line ratios within the halos. Furthermore, the few high \oiii/\hb ratio bins near the galactic planes of the galaxies are clearly associated with SF regions within the discs, as indicated by the emission-line maps of eBETIS (see Fig. \ref{fig:IC1553_maps} and related figures in the appendices). Then, the behaviour of the \nii/\ha, \sii/\ha, and \oiii/\hb ratios alone is not sufficient evidence to conclusively support HOLMES as an ionisation source.

In addition, as explained in \citetalias{2024A&A...687A..20G}, the conventional proxy employed to distinguish between star-forming regions, HOLMES, and mixed regimes—the \ha\ equivalent width (\EWha)—may be unreliable. The \EWha\ is a parameter that scales with the SFR per unit mass, namely the specific SFR (sSFR), and serves as a proxy for the stellar birthrate parameter  $b-$Scalo, which is the ratio of the present to past-average SFR \citep{1994ApJ...435...22K,1998ARA&A..36..189K}. Observationally, the \EWha\ is defined as the \ha\ emission-line luminosity normalised to the adjacent continuum flux, effectively quantifying the SFR per unit luminosity \citep{1998ARA&A..36..189K}. Given the relationship between \EWha\ and surface mass density—a projection of the local mass-metallicity-EW relation (\mze) \citep{2012A&A...539A..73R}—low \EWha\ values $\sim 1$ correspond to very high surface mass densities $\Sigma_{\rm Lum} > 10^3$ M$_{\odot}$ pc$^{-2}$, characteristic of early-type galaxies and the bulges of late-type galaxies. This corresponds precisely to the locus of the hDIG distribution at \EWha\ $\sim 1$\r{A} (see Fig. 2 in \citealt{2018MNRAS.474.3727L} and Fig. 9 in \citealt{2024MNRAS.tmp..356L}).

In the low S/N regimes characteristic of the DIG, the low values of \EWha\ employed to distinguish HOLMES from other regimes (\EWha $\sim 3$\r{A}) are highly contingent upon the stellar populations and the techniques utilised for SSP synthesis modelling (\citetalias{2024A&A...687A..20G}). Moreover, as noted by \citep{2022MNRAS.513.2904T}, \EWha\ is an inadequate tracer for discerning whether an \ha photon has originated in dense gas (\hii region) or diffuse gas (DIG region). Crucial considerations must be acknowledged, as the EW – a non-additive, ratio quantity – is calculated over binned sections along the line-of-sight of the halos of galaxies, which exhibit significantly different depths and properties.

\begin{figure}[t!]
\centering
\includegraphics[width=\columnwidth]{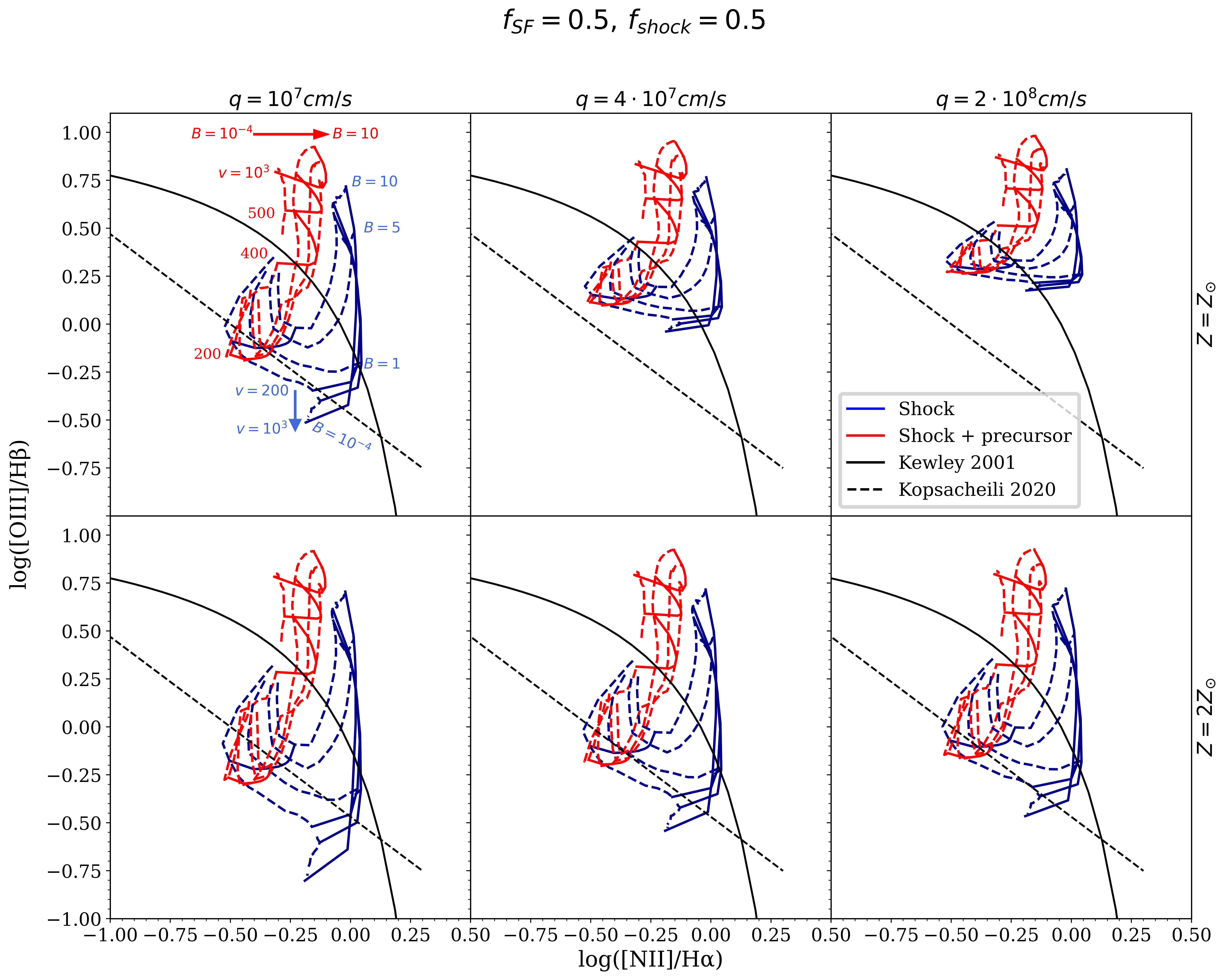}
\includegraphics[width=\columnwidth]{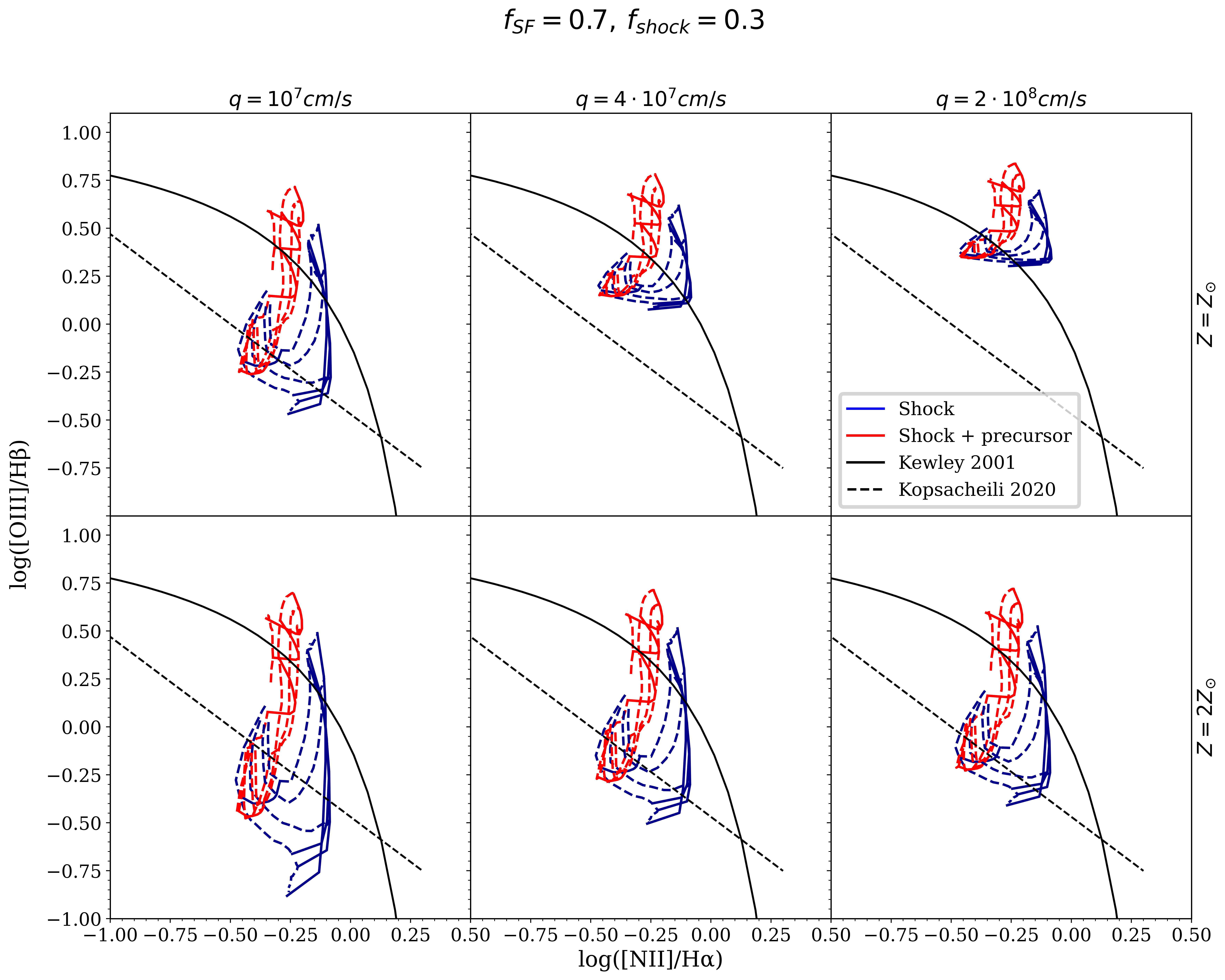}
\caption{Showcase of the star formation + fast shocks hybrid models for the \nii BPT. The upper panels correspond to the hybrid models with a contribution of fast shocks of 50\% and 50\% star formation. The lower panels correspond to 30\% shocks and 70\% star formation. The fast shocks models from \citet{2008ApJS..178...20A} used correspond to shock velocities from 200 to 1000 km/s and magnetic fields from 10$^{-4}$ to 10 in units of $\mu$G$\cdot$cm$^{3/2}$, assuming shocks + a precursor (red curves) and only shocks (blue lines). The red arrow indicates the direction of increasing magnetic field in the shocks + precursor models, and blue arrow indicates the direction of increasing shock velocity in the shock only models. The dashed oblique line in the \nii/\ha diagnosis represent the separation of shock excited (e.g. supernova remnants) from photoionised regions (e.g. \hii regions) from \citet{2020MNRAS.491..889K}. The classic demarcation between \hii regions photoionisation and AGNs from \citet{2001ApJ...556..121K} is also plotted in all diagnoses. Every panel represent the hybrid models for a fixed metallicity and ionisation parameter: first and second row correspond to Z = Z$_\odot$ and Z = 2Z$_\odot$, first, second and third column correspond to q = 10$^7$, 4$\cdot$10$^7$ and 2$\cdot$10$^8$ cm/s respectively.}
\label{fig:models_descrip}     
\end{figure}

\begin{figure*}[ht!]
\centering
\includegraphics[width=\textwidth]{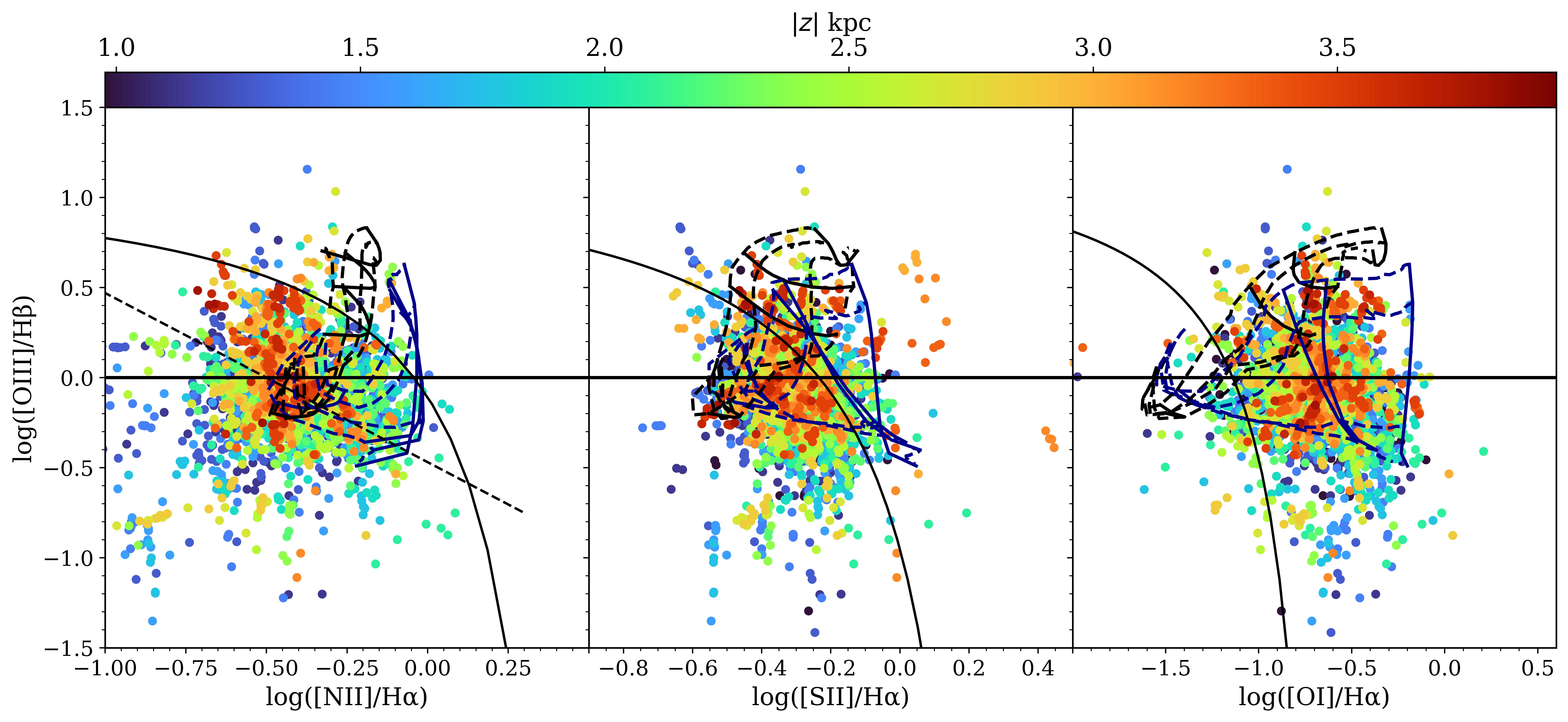} 
\caption{BPT for all eDIG bins of IC1553, with the colours coding the vertical distance (both above and below) from the midplane. The plotted hybrid models correspond to a 40\% of the flux due to fast shocks with pre-ionisation (curves lines) and with front shock only (blue curves), and with the 60\% of the flux due to star formation with Z = Z$_\odot$ and q = $10^7$ cm/s. Solid curves correspond to the shocks winds from 200 to 1000 km/s, and dashed curves represents magnetic field intensities from 0.0001 to 10 $\mu$G$\cdot$cm$^{3/2}$. The \citet{2020MNRAS.491..889K} and \citet{2001ApJ...556..121K} lines are also plotted, as in Figure \ref{fig:models_descrip}.}
\label{fig:IC1553_BPT_total}     
\end{figure*}

\subsection{Hybrid SF and shocks ionisation mechanism of the eDIG}

The detection of high ionisation states in the galactic halo of the MW and several other galaxies, either in emission (e.g. O$^{5+}$; \citealt{1992AAS...180.2505S,2000ApJ...528..310S}) or absorption (e.g. O VI, N V, C IV; \citealt{1974ApJ...193L.121J,1982ApJ...260..561P,2014ApJ...789..131H,2016ApJ...833...58H}), implies that interfaces between hot (10$^6$ K) and cooler ($\leq 10^4$ K) gas are prevalent throughout the ISM at significant $z$ distances from the galaxies' midplanes. The gas at intermediate temperatures of $\sim 10^5$ K at these interfaces generates extreme ultraviolet (EUV) ionising radiation. The radiation of these hot-cool gas interfaces in the halo may constitute a significant source of ionising radiation for distant clouds and regions where LyC ionising radiation from OB stars in the midplane is either shielded or absent.

Different types of interfaces have been proposed, each exhibiting distinct characteristics contingent on the physical processes and dynamical state of the boundary region between the hot and cooler gas. These include evaporative layers (e.g. \citealt{1977ApJ...211..135C}), cooling-condensation fronts \citep{1991PASP..103..923S}, and turbulent mixing layers (TML, e.g. \citealt{1990MNRAS.244P..26B,1993ApJ...407...83S,1998ApJ...501..137R,2001ApJ...551...57C}; \citealt{2009ApJ...695..552B}). \citet{1993ApJ...407...83S} proposed that self-photoionisation through EUV line emission of collisionally excited gas in mixing layers (with a temperature $\sim 10^{5.0-5.5}$ K) between hot ($\sim 10^6$ K) and cold ($\sim 100$ K) gas can reproduce many optical/UV/EUV emission line ratios as observed in the DIG of several galaxies. However, theoretical models for TML are intricate and may lead to over-ionisation of the hot gas and under-ionisation of the cool gas, contingent on the specific details of the models \citep{2009RvMP...81..969H}.

Shocks induced in the ISM by feedback mechanisms, such as supersonic winds originating from high-level SF regions, have been proposed as a significant source of heating for the eDIG\footnote{Note that tidal interactions and AGN jets can also produce shock ionisation in the halos of galaxies, with similar effects as discussed in the text (e.g., \citealt{1995ApJ...455..468D,2007ApJ...670.1115S,2011ApJ...734...87R,2016MNRAS.457.1257H,2018ApJ...864...90M}).} \citep{1985Natur.317...44C,1998ApJ...501..137R,2001ApJ...551...57C,2007ApJ...670L.109B}. In this scenario, galactic winds are driven by the stellar winds and subsequent SNe of massive stars. These stellar winds create kpc-sized cavities in the ambient ISM on timescales of several million years \citep[e.g.][]{2023MNRAS.521.5492M}. Subsequently, these cavities are overrun by SN remnants (SNRs) from the most massive progenitor stars. Both processes occur at velocities exceeding the local sound speed, $v_s$, in the ISM, thereby generating strong shocks.

In an analogous way to TML, fast shocks induce comparable ionised conditions, characterised by an ionisation equilibrium between the hot and cold gas phases and slow mixing within the turbulent layer \citep{1993ApJ...407...83S, 2003A&A...406..505R,2019ApJ...885..160B,2023A&A...678A..84D}. Given that shock heating is significant only if the kinetic energy is efficiently thermalised, both photoionisation and shocks can be regarded as "thermal" heating sources of the ISM. Crucially, the observed rising trend in the \oiii/\hb emission line ratio with height can be expected as a mixing sequence between the predominant shock ionisation in the halo and photoionisation in the disk. Shocks, therefore, emerge as the most promising secondary ionisation mechanism for the eDIG.

In star-forming galaxies, large-scale winds driven by bursts of star formation in the galactic plane create extended wind-driven filaments present in the halo \citep{2009ApJ...696..214S,2016MNRAS.457.1257H}. The ionisation of these wind-blown filamentary structures is primarily dominated by fast shocks. This is because the final stages of the wind evolution, driven by star formation, are marked by a flare-up that overshadows the main ionising source, such as the star formation activity of the \hii regions observed in the midplane \citep{2010ApJ...711..818S}. For the same reason, the contribution of fast shocks to the ionisation of the eDIG is also expected to be less significant near the galactic plane \citep{2010ApJ...711..818S,2016MNRAS.457.1257H}.

\begin{figure}[t!]
    \centering
    \includegraphics[width=\columnwidth]{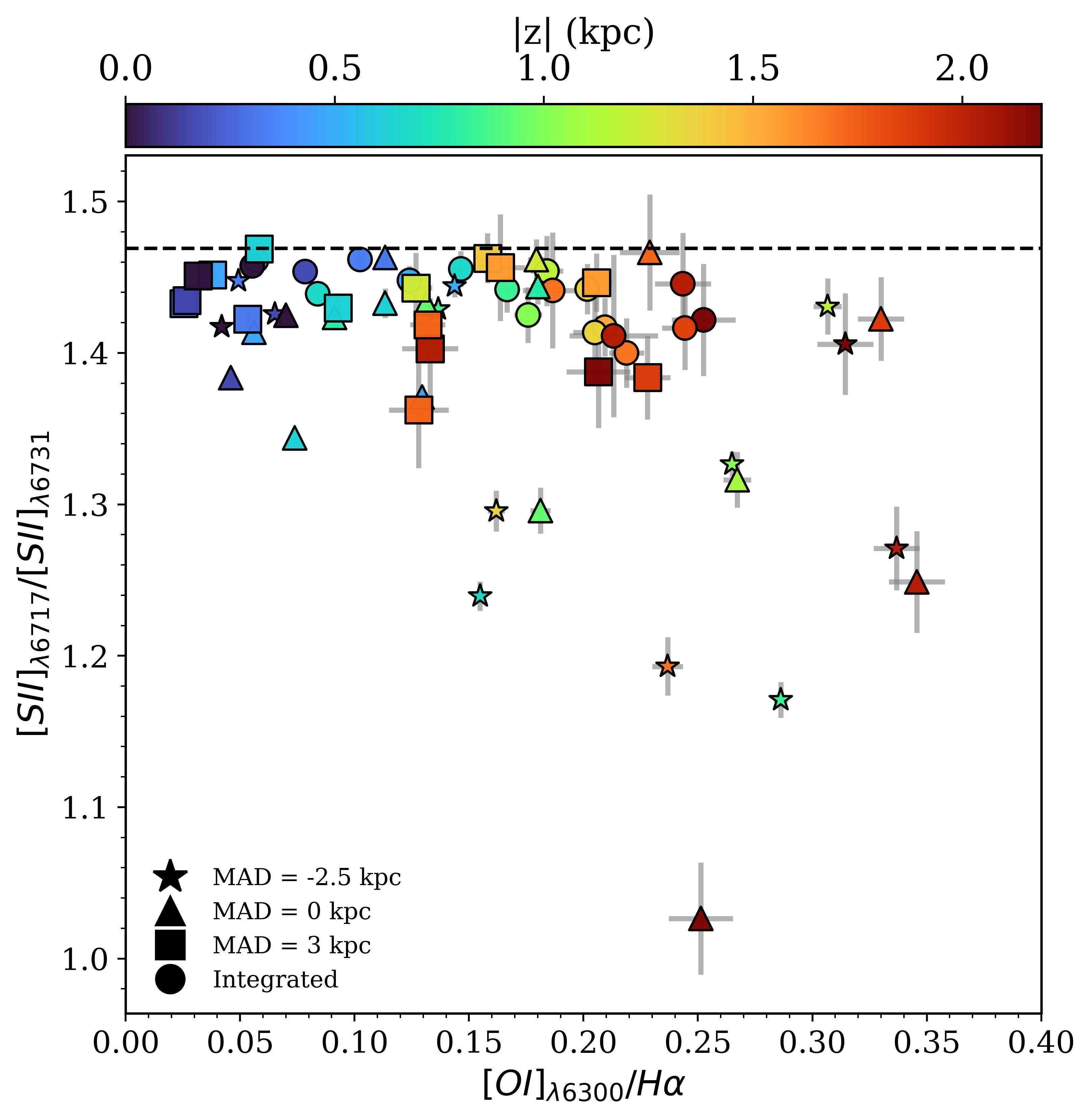} 
    \includegraphics[width=\columnwidth]{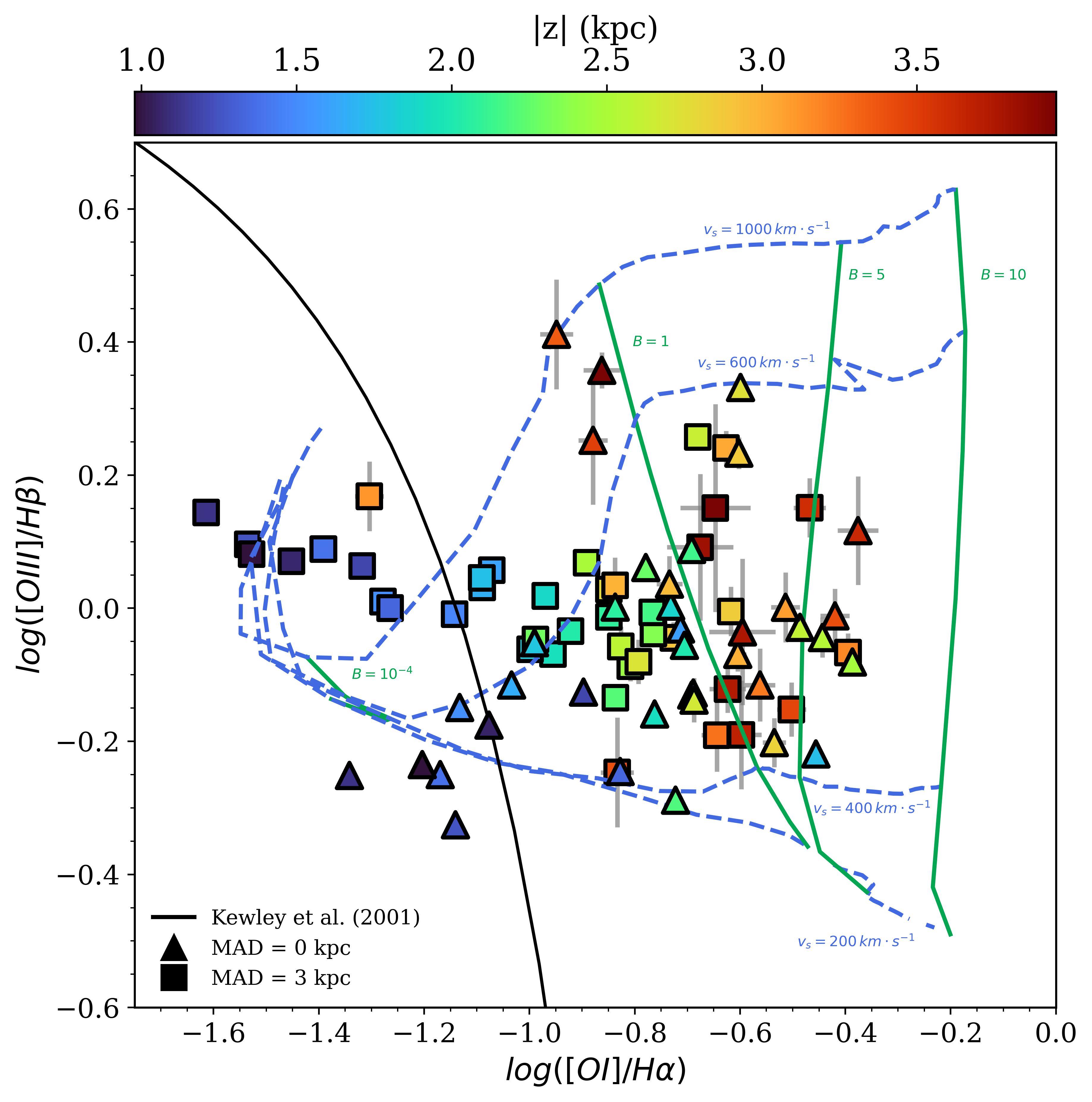}    
    \caption{Top: Electron density sensitive \sii$_{\lambda\,6717}$/\sii$_{\lambda\,6731}$ line ratio vs. shocks-proxy \oi$_{\lambda\,6300}$/\ha line ratio for IC1553. Dashed horizontal line represents the theoretical low-density limit for the \sii doublet, corresponding to 1.469 \citep{2006agna.book.....O}. Colours and symbols are similar to Figure \ref{fig:IC1553_T_S+S}.
    Bottom: \oiii/\hb vs. \oi/\ha diagnostic diagram for the bins of IC1553 at the minor-axis, centred in the biconical structure seen in the \nii/\ha map (triangles) and in the region between 2.5 < MAD < 4.5 kpc (squares), coincident with the high-level SF region visible in the disk of the galaxy. Curves represent the hybrid model for 40\% shocks (without precursor) and 60\% star formation with Z = Z$_\odot$ and q = 10$^7$ cm/s.
    \label{fig:IC1553_SII_OI}.}
\end{figure}

The role of fast shocks as a potential alternative ionisation mechanism for the eDIG, as well as the impact of these filaments in the halo, has been studied by several authors  \citep{1998ApJ...501..137R,2001ApJ...551...57C, 2019ApJ...882...84L, 2023A&A...678A..84D}. For instance, \citet{Martin1997} studied the excitation of the DIG in 14 star-forming dwarf galaxies and concluded that photoionisation due to star formation is the dominant mechanism. Shocks were only considered as a secondary ionisation source for gas with very low surface brightness. In this scenario, shocks could contribute up to 30-50\% of the total ionisation budget.

As stated in the previous sections, the eBETIS sample presents diverse ionised structures in the halo, including filaments that extend more than 2 kpc from the midplane. Thus, we explore the combination of photoionisation from \hii regions and ionisation due to fast shocks as simultaneous contributors to the ionisation mechanism of the eDIG. This is achieved by constructing a set of hybrid models that incorporate both star formation and fast shock mechanisms.

We firstly consider the photoionisation models for low-metallicity star-forming galaxies of \citet{2010AJ....139..712L}. The grids of this models predict the line ratios involved in the typical BPT diagram \citep{1981PASP...93....5B} for a pure photoionisation regime due to the star formation with ionisation parameters (q) ranging from 10$^7$ to 2$\cdot$10$^8$ cm/s and metallicities (Z) from 0.001 to 0.04. Each model grid is computed for electron densities $n_e$ of 10, 10$^2$, 10$^3$ and 10$^4$ cm$^{-3}$, assuming either continuous star formation or an instantaneous burst of star formation at 0 Myr. 

On the other hand, we consider the fast shocks models of \citet{2008ApJS..178...20A}. This models predict the flux of the ionising radiation produced by a shock. The flux is dependent on the shock velocity ($f\propto v_s^3$). Therefore, if the shock velocity surpasses the velocity of the photoionisation front (in the case of a low ionisation parameter), the ionising photons are absorbed by the surrounding gas, altering its ionisation state. In another scenario, if $v_s \approx 170$ km/s, the ionisation front velocity now exceed the velocity of the shock, pre-ionising the surrounding gas and changing the optical emission lines observed \citep{1980ARA&A..18..219M}. The fast shocks models of \citet{2008ApJS..178...20A} consider this two scenarios, with preshock densities ranging from 0.01 to 1000 cm$^{-3}$, shock velocities from 100 to 1000 km/s and magnetic field (B/n$^{1/2}$) from 10$^{-4}$ to 100 $\mu$G$\cdot$cm$^{3/2}$.

Our hybrid models are constructed as follows: we consider both shock models, with and without assuming pre-ionisation. Then, for each shock model, we assume that the predicted flux corresponding to shock wind velocity of 200, 400, 500 and 1000 km/s and magnetic field of 10$^{-4}$, 1.0, 5.0 and 10 $\mu$G$\cdot$cm$^{3/2}$ contributes only a fraction $f_{shock}\in [0,1]$ of the total observed ionising flux. The remaining fraction 1 - $f_{shock}$ corresponds to the flux predicted by a grid model with a certain ionisation parameter and metallicity from \citet{2010AJ....139..712L}.

Figure \ref{fig:models_descrip} shows an example of hybrid models for the \nii/\ha BPT, with Z = Z$_\odot$ and 2Z$_\odot$, q = 10$^7$, 4$\cdot$10$^7$ and 2$\cdot$10$^8$ cm/s, and $f_{shock}$ = 0.5 and 0.3. It shows that decreasing $f_{shock}$ shifts the models towards the OB-stars regime, as defined by the \citet{2001ApJ...556..121K} curve, indicating that photoionisation by star formation becomes more significant in the hybrid models. Additionally, increasing the ionisation parameter in the models tends to flatten them vertically, resulting in higher predicted values for the high excitation \oiii/\hb ratio. Besides, increasing the metallicity tends to expand the models, predicting a wider range of both \oiii/\hb and \nii/\ha ratios.

For instance, Figure \ref{fig:IC1553_BPT_total} shows the BPT diagram for all the eDIG bins for IC1553\footnote{Note that these results do not concur with the diagnosis of \citet{2022A&A...659A.153R} for this galaxy, whose BPT diagrams are predominantly populated within the star formation regime. However, they align more closely with the subsequent study by \citet{2023A&A...678A..84D}. The locus in all three panels of the BPT diagram corresponds with the findings of this latter study, being proximate to the \citet{2001ApJ...556..121K} line for \sii/\ha, and achieving values up to 0 (in log scale) for \oi/\ha. Nevertheless, in comparison to both studies, we observe a greater number of data points in the high excitation regimes for \sii/\ha and \oi/\ha, corresponding to bins at larger distances. We attribute this to our enhanced adaptive binning method, which enables sampling data with S/N > 2 at greater distances, and to our exclusion of bins situated within the galactic plane, as discussed in section \ref{sec:sec_3}.}. The models plotted represent the predicted fluxes for a combination of fast shocks with pre-ionisation (black curve lines) and with front shock only (blue curve lines) weighted with $f_{shock}$ = 0.4, plus a photoionisation model correspondent to Z = Z$_\odot$ and q = 10$^7$ cm/s, for instantaneous burst at 0 Myr and $n_e$ = 100 cm$^{-3}$. Thus, the fluxes predicted for this particular hybrid models has the form: $F_{hyb} = 0.4F_{shock} + 0.6F_{SF}(n_e,q,Z)$. Solid curves correspond to the shocks winds from 200 to 1000 km/s, and dashed curves represents magnetic field intensities from 0.0001 to 10 $\mu$G$\cdot$cm$^{3/2}$. The parameters were selected to best fit the points in the BPT diagram, particularly the \sii/\ha and \oi/\ha diagnostics, as they are more sensitive to changes in ionisation regimes. The points in the \sii/\ha and \oi/\ha diagnostics tend to deviate more from photoionisation models than the \nii/\ha diagnostic if an additional source of ionisation, such as AGN or fast shocks, is present \citep{2023A&A...678A..84D}. The $f_{shock}$ for fitting each BPT was determined by performing several trials with different $f_{shock}$, selecting the one that best fits the data points in the \sii/\ha and \oi/\ha diagnostics.

The presence of shocks in the eDIG is further indicated by the asymmetric increase of \oi/\ha with respect to height, as illustrated in Fig. \ref{fig:IC1553_lines} for IC1553 (and in the corresponding figures in the appendices for the entire eBETIS sample). Given that neutral O and H are coupled via charge-exchange reactions, the observed increase in \oi/\ha suggests a multiphase medium, characteristic of shock-compressed regions \citep{2001ApJ...551...57C, 2019ApJ...885..160B}.

The top-panel of Figure \ref{fig:IC1553_SII_OI} shows the observed \sii$_{\lambda\,6717}$/\sii$_{\lambda\,6731}$ doublet ratio versus the \oi$_{\lambda\,6300}$/\ha line ratio for various binned sections in the halo of IC1553 at different galactocentric distances. The \sii doublet is employed to infer the electron gas density; this ratio diminishes as the electron gas density increases, with a low-density limit of approximately 1.5, corresponding to an electron density of $n_e \leq 10$ cm$^{-3}$ \citep{2006agna.book.....O}. Most of the eDIG regions near the midplane of IC1553 exhibit \sii line ratios close to this low-density limit, accompanied by low \oi/\ha values, indicative of \hii region emission. However, for bins corresponding to a higher altitude |z| from the midplane, the \sii$_{\lambda\,6717}$/\sii$_{\lambda\,6731}$ ratio decreases to values between 1.0 and 1.3, implying high electron densities in the range of $n_e \sim 200-800$ cm$^{-3}$. 

The most striking feature of this figure is that the regions with low \sii doublet ratios coincide with regions exhibiting high \oi/\ha values (> 0.25), indicative of shock-compressed ionised gas. Therefore, the low doublet \sii ratios, their implied high gas density, and the fact that these regions exhibit higher \oi/\ha values are consistent with a scenario wherein gas emission originates from shocks, likely induced by feedback from high-level SF regions within the galactic disk. 
In this scenario, fast shocks can account for the increase in the \oi/\ha ratio with distance from the midplane, with the highest \oi/\ha ratios corresponding to the shocked interface at the greatest distances from the midplane (this is also seen in the equivalent figures for the rest of the galaxies in the appendices).
This picture can be further validated by comparing the high-density, high \oi/\ha bins values of IC1553 to the hybrid SF-shocks models in the \oiii/\hb vs. \oi/\ha diagnostic diagram shown in the bottom panel of Figure \ref{fig:IC1553_SII_OI}. The triangles correspond to integrated bins of IC1553 at the minor-axis (MAD = 0 kpc, centred in the biconical structure seen in the \nii/\ha map) and the squares to bins at 2.5 < MAD < 4.5 kpc, centred at the high-level SF region visible in the disk of the galaxy in \ha. The curves represent a hybrid model with a 60\% star-formation and 40\% shock contribution (without precursor), with Z = Z$_\odot$ and q = 10$^7$ cm/s. The parameter space of this specific hybrid model aligns precisely with the locus of the eDIG bins, corresponding to fast shocks between 400 and 600 km\,s$^{-1}$ and magnetic field values in the range $B = 10^{-4}$– 5, regardless of the MAD distance and the diverse ionisation conditions observed in the halo in those regions.

These results align with recent numerical simulations of the DIG: \citet{Weber2019} calculated the ionisation structure within the DIG, providing quantitative predictions for diagnostic optical emission lines using advanced 3D non-LTE radiative transfer simulations. These simulations assume ionisation by a matter-bounded stellar population SED, based on a plausible parameter space for the ionising sources ($T_{\rm eff}, Z$), while considering varying escape fractions $f_{\rm esc}$ and clumpiness of the DIG. The findings support the scenario where the DIG is primarily ionised by (filtered) radiation from hot stars within the galactic plane. Consequently, photoionisation heating results in higher temperatures in the DIG compared to \hii regions, leading to stronger emissions of collisionally excited optical lines such as \nii, \oi, \oiii, or \sii. However, these models do not account for the increasing \oiii/\hb ratio at greater distances from the galactic plane, indicating the need for an additional ionising source. In our proposed SF-shock hybrid model, SF photoionisation is identified as the principal ionisation mechanism for the eDIG, while incorporating shock heating and collisional excitation as supplementary heating processes responsible for the observed increases in the \oiii/\hb and \oi/\ha ratios.

In summary, the ionisation of the eDIG is evidently not a phenomenon that can be elucidated by a single mechanism; rather, a multitude of mechanisms is necessary to fully comprehend the nature of the ionisation of extraplanar gas. Additionally, the ionisation mechanisms vary in significance across different galaxies. The prevailing consensus is that photoionisation predominantly governs the excitation of the eDIG, with shocks being considered a secondary source of ionisation. The spatial variation of line ratios, along with their absolute values, suggests the presence of shock excitation. Consequently, a model incorporating star-forming ionisation with shock ionisation as an additional heating source offers the most satisfactory explanation for the relatively high \oi/\ha and \oiii/\hb ratios observed in the lowest surface brightness regions as a function of distance from the midplane in edge-on star-forming galaxies.

\section{Results}
\label{sec:results}

\subsection{The radial variation of shocks contribution and the biconical structure of IC1553} 

\begin{figure}[t!]
\raggedleft
\includegraphics[width=\columnwidth]{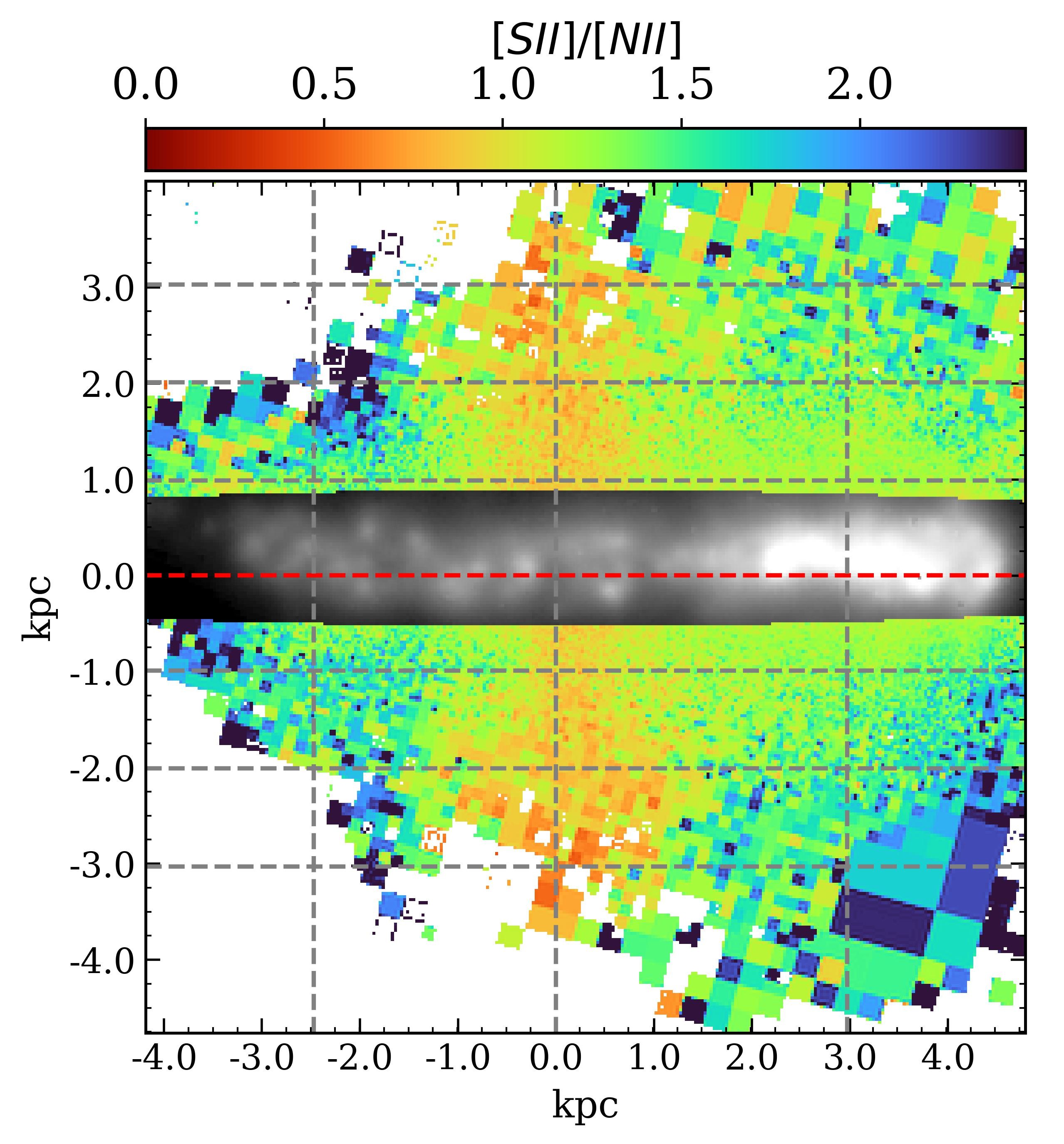} 
\caption{\sii$_{\lambda\,6717}$/\nii$_{\lambda\,6584}$ emission line map for IC1553. Vertical and horizontal lines represented as in Figure \ref{fig:IC1553_maps}.}
\label{fig:IC1553_SIINII}   
\end{figure}

The integration of emission-line ratio maps with hybrid SF+shock models for the eDIG facilitates a comprehensive analysis of the ionisation conditions within targeted regions of specific galaxies. In the case of IC1553, illustrated in the upper-right panel of Figure \ref{fig:IC1553_maps}, there exists a distinct axisymmetric region with elevated \nii/\ha ratios between -0.5 kpc < MAD < 0.5 kpc, centred on the galaxy's minor axis (MAD = 0). This enhanced region forms an apparent biconical structure that appears to emanate from the disk, extending to significant heights above and below the midplane |z| $\sim 3.5$ kpc in both the upper and lower halo of IC1553. This phenomenon is also visible in the \sii/\nii map, as depicted in Figure \ref{fig:IC1553_SIINII}.

This structure was also identified by previous authors \citep{2022A&A...659A.153R,2023A&A...678A..84D}, who suggested that its origin could be attributed to a large off-centred galactic outflow or superbubble. \citep{2022A&A...659A.153R} described this feature as a biconical region of OB–shock ionisation caused by significant bidirectional outflows associated with superbubbles that have breached the disk. However, since there are no corresponding \ha emission enhancements in either the disk or the halo at the same spatial position as the biconical structure, they argue that the absence of a visible starburst responsible for the outflow is due to the starburst having quenched 20 Myr ago, or more. \citet{2023A&A...678A..84D} support the biconical outflow scenario through indirect evidence based on \ha FWHM, which is sensitive to the velocity of gas flows in the presence of shocks. Nevertheless, the enhancement in \ha FWHM identified by \citet{2023A&A...678A..84D} (see their figure 10) does not morphologically coincide with the \nii/\ha or \sii/\nii structure observed in IC1553.

To elucidate the origin of this structure, we can examine the morphological information provided by the emission-line maps of IC1553, as presented in Figures \ref{fig:eBETIS_sample} and \ref{fig:IC1553_maps}. Firstly, as previously discussed, the distribution of \hii regions in the galactic plane directly influences the morphology and ionisation conditions of the eDIG. The \ha\ map of IC1553 shows no evidence of a structure, outflow, or filaments extending from the galactic disk centred on or near the minor axis of the galaxy. Indeed, the strongest emission is observed at the boundaries of the high-level SF regions in the southern part of the galaxy, extending radially within a MAD range of 1.5 to 4.5 kpc, and vertically up to $\geq 3$ kpc.

\begin{figure*}[t!]
\centering
\includegraphics[width=\textwidth]{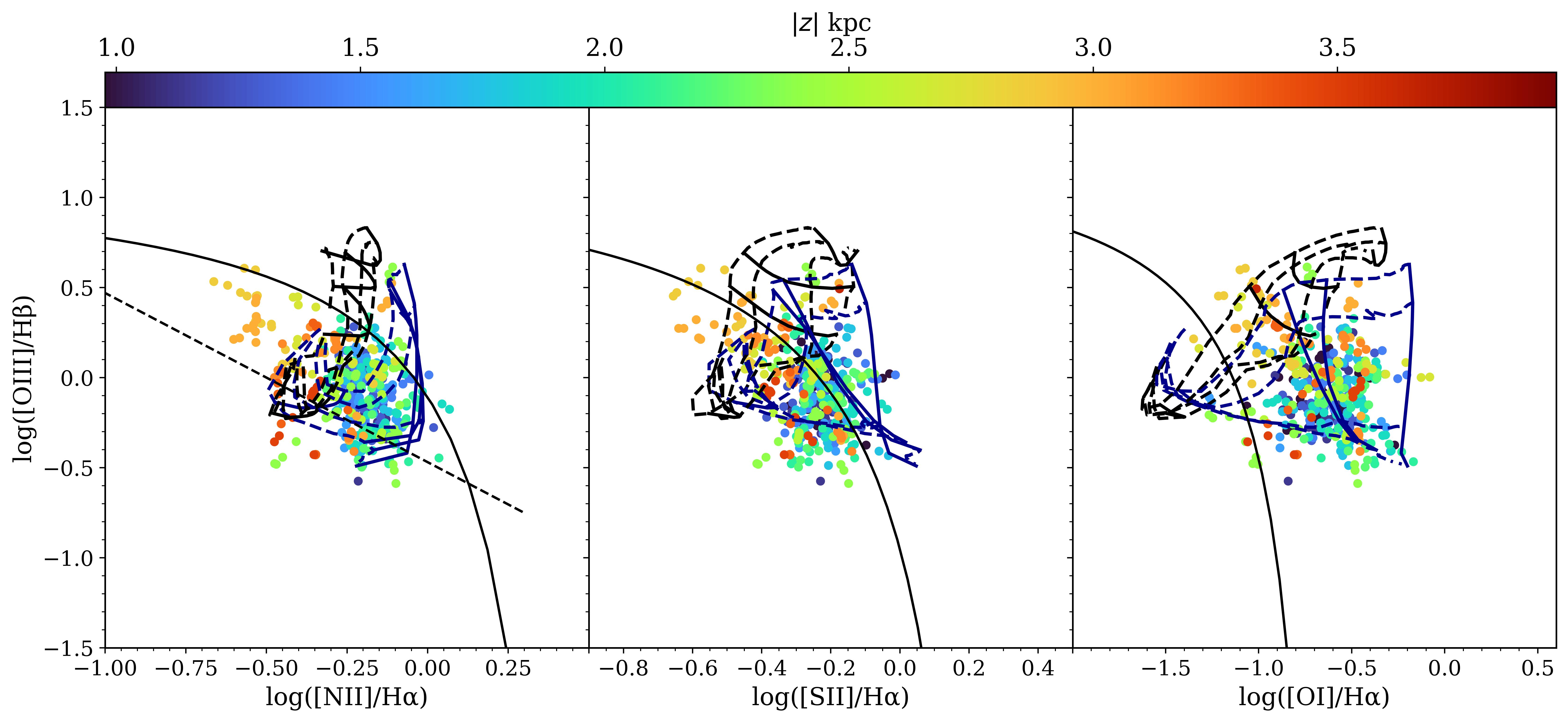} 
\includegraphics[width=\textwidth]{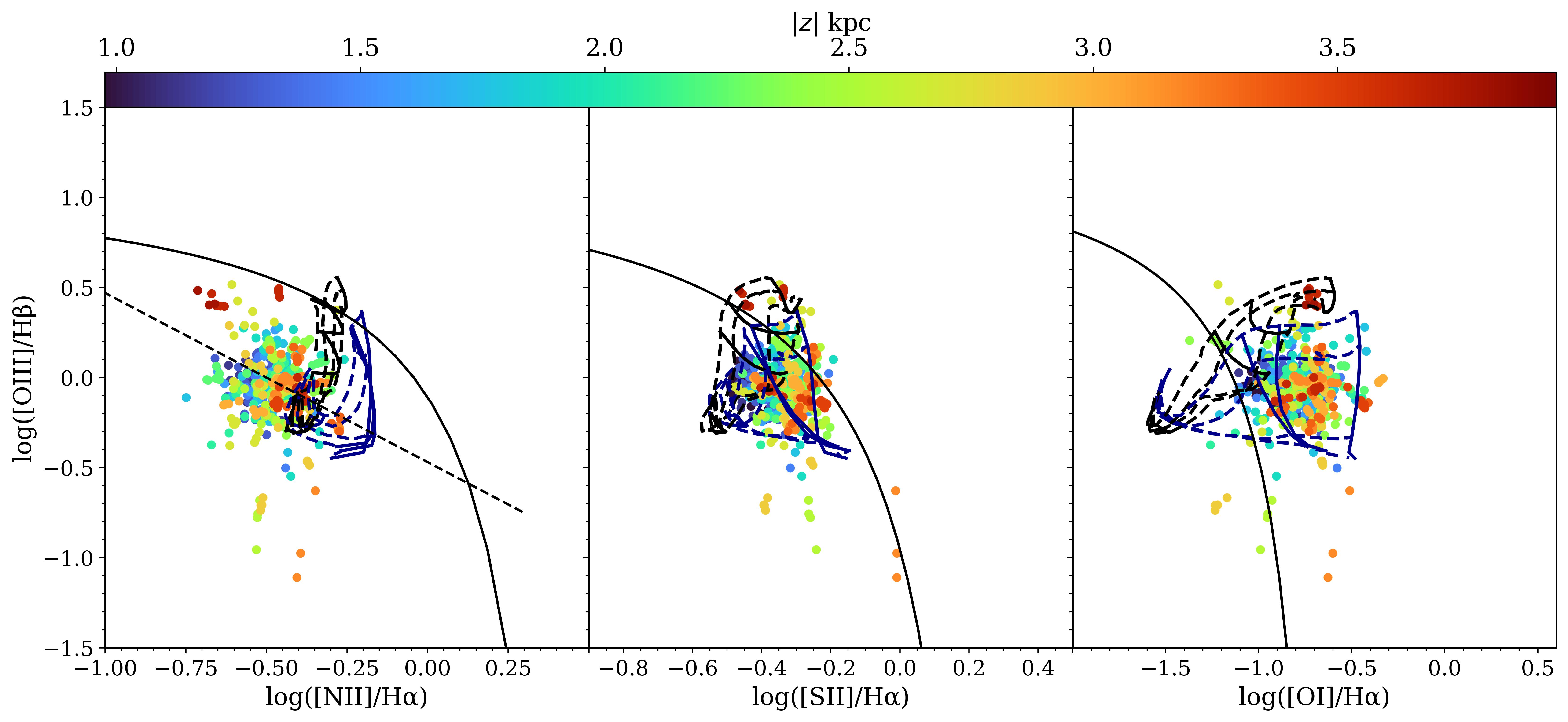} 
\caption{IC1553 BPT with hybrid models, similar to Figure \ref{fig:IC1553_BPT_total}. Top: eDIG bins corresponding to the biconical structure seen in the \nii/\ha map, between -0.5 kpc < MAD < 0.5 kpc. The hybrid models that best fit the data correspond to 40\% shocks and 60\% SF with Z = Z$_\odot$ and q = $10^7$ cm/s. Bottom: eDIG bins corresponding to the high-SF feature on southern edge of the disk, between 2.5 kpc < MAD < 3.5 kpc. The hybrid models that best fit the data correspond to 20\% shocks and 80\% SF with Z = Z$_\odot$ and q = $10^7$ cm/s.
}
\label{fig:IC1553_BPT_MAD}     
\end{figure*}

Secondly, while the structure in the \nii/\ha emission-line map appears to symmetrically originate from the disk at the galaxy's minor axis (MAD $\approx 0$ kpc), the \sii/\ha map reveals a less defined biconical structure with a discernible offset at MAD $\sim -1.2$ kpc. This same structure is observed in the \sii/\nii map, but with an opposite offset at MAD $\sim 0.2$ kpc. Additionally, the upper part of the \oi/\ha map hints at this structure, displaying an offset at MAD $\sim -1.4$ kpc relative to \nii/\ha. Most strikingly, the \oiii/\hb line-ratio map reveals no such structure on either side of the midplane. The high \oiii emission relative to H is a key observational feature of galactic outflows, regardless of whether its origin is due to SF \citep[e.g.][]{1988ARA&A..26..145T, 1989ApJ...345..372N} or AGNs \citep[e.g.][]{2020AJ....159..167L, 2024A&A...687A..20G}. The \oiii/\hb map of IC1553 exhibits the typical increase of the ratio with distance from the midplane in a homogeneous manner, independent of radial distance, but without any characteristic signature of an outflow at MAD $\sim 0$. Therefore, based on the information provided by the 2D distribution of the eDIG, there is no morphological congruence among the different emission-line maps to definitively confirm the existence of an outflow in IC1553.

The \sii/\nii map can also elucidate the origin of this structure. Given the similar ionisation potentials of \sii and \nii (10.4 eV and 14.5 eV, respectively), \sii/\nii is largely independent of electron temperature. Consequently, variations in \sii/\nii imply changes in local ionisation conditions or metallicity (since nitrogen is a secondary nucleosynthetic product while sulphur is a primary product) but primarily reflect variations in S$^+$/S \citep{2009RvMP...81..969H}. In a photoionisation model, \sii/\nii is expected to increase with distance from the ionising source, aligning qualitatively with a smooth transition from predominantly doubly ionised to singly ionised states (S$^{++}$ $\rightarrow$ S$^{+}$), as expected in dilute photoionisation models (e.g. low ionisation parameters, $q$, \citealt{1994ApJ...428..647D}. The consistency of \sii/\nii suggests an additional heating source \citep{1992ApJ...400L..33R,1998ApJ...501..137R,1999ApJ...525L..21R,2001ApJ...551...57C}, which in our model would be provided by shocks. Indeed, in the \citet{2008ApJS..178...20A} models, shocks naturally produce a lower \nii flux relative to \sii compared to photoionisation. In IC1553, we observe a symmetrical increase in the \sii/\nii ratio across all galactocentric distances, from values of approximately 1.2 near the galactic disk, up to values $\geq 2$, except in the region associated with the biconical structure at a MAD of approximately 0.2 where \sii/\nii remains nearly constant at around 0.9 but decreases with height to values around 0.5. An increase in the \sii/\nii ratio with height has also been observed at two slit positions perpendicular to the edge-on galaxy NGC\,5775 \citep{2019ApJ...885..160B}. The overall scenario depicted by the \sii/\nii ratio map suggests that sulphur and nitrogen are ionised by a photoionisation field that diminishes with height, yet with localised deficits or enhancements of either \sii or \nii, indicating local changes in ionisation conditions within the complex eDIG halo, likely due to shocks.

To further explain the structures observed in IC1553, we can apply hybrid SF-shock models as previously described. Figure \ref{fig:IC1553_BPT_MAD} depicts the same BPT diagram as in Figure \ref{fig:IC1553_BPT_total}, but limited to the bins of the biconical structure identified in the \nii/\ha map. Specifically, it shows bins between -0.5 kpc < MAD < 0.5 kpc (top) and bins between 2.5 kpc < MAD < 3.5 kpc (bottom), corresponding radially to the regions above and below the bright SF regions near the southern edge of the disk. In these latter regions, outside the biconical structure, the hybrid models that best fit the data indicate 80\% of the ionisation is due to star formation and 20\% to shocks. For regions within the biconical structures, at MAD $\sim$ 0 kpc, the contribution from star formation decreases, with shocks becoming relatively more significant (increasing from 20\% to 40\%). Hence, while the contribution of shocks within the biconical structure is augmented, the increase is modest; SF remains the predominant source of ionisation, and the emission-line ratios do not exhibit the high \oiii/\hb values characteristic of a galactic outflow.

\begin{table*}[t!]
\centering
\caption{Summary of characteristics of the eDIG.}
\resizebox{\textwidth}{!}{%
\begin{tabular}{lll}
\hline
\textbf{galaxy} & \textbf{DIG morphology} & \textbf{MAD, z} \\
 &  & (kpc) \\ \hline
IC217 & \begin{tabular}[c]{@{}l@{}}Disc: Homogeneous distribution of bright \hii regions\\ Halo: Thin eDIG layer, low S/N\end{tabular} & 6, 2 \\
PGC28308 & \begin{tabular}[c]{@{}l@{}}Disc: Homogeneous distribution of bright \hii regions\\ Halo: Extended and homogeneous eDIG\end{tabular} & 8, 2 \\
PGC30591 & \begin{tabular}[c]{@{}l@{}}Disc: Few bright \hii regions \\ Halo: Inhomogeneous extended eDIG\end{tabular} & 4, 2 \\
ESO544-27 & \begin{tabular}[c]{@{}l@{}}Disc: Homogeneous distribution of bright \hii regions\\ Halo: Homogeneous eDIG and extended filaments\end{tabular} & 4, 2 \\
IC1553 & \begin{tabular}[c]{@{}l@{}}Disc: Inhomogeneous distribution of \hii regions with respect the line of sight, hidden \hii regions\\ Halo: Highly asymmetrical eDIG structure with a bicone and extended filaments\end{tabular} & 4, 3.5 \\
ESO443-21 & \begin{tabular}[c]{@{}l@{}}Disc: Homogeneous distribution of bright \hii regions\\ Halo: Extended eDIG with a large filamentary knot\end{tabular} & 8, 5 \\
ESO469-15 & \begin{tabular}[c]{@{}l@{}}Disc: Bright emission around the centre \\ Halo: Several extraplanar \hii regions, mixed eDIG regimes\end{tabular} & 3, 2 \\
ESO157-49 & \begin{tabular}[c]{@{}l@{}}Disc: Homogeneous distribution of bright \hii regions\\ Halo: Tilted and mixed eDIG regimes with extended filaments\end{tabular} & 4, 2.5 \\ \hline
\end{tabular}%
}
\tablefoot{The second column represents the DIG morphology observed in the disc and halo from the \ha images. The third column shows the horizontal and vertical distances covered by the eDIG.}
\label{tab:eDIG_char}
\end{table*}

The absence of distinct filamentary, outflow-like structures in both the \ha flux and the \oiii/\hb emission-line ratio maps, the observed offsets, and the conflicting morphology between the \nii/\ha, \oi/\ha, and \sii/\nii ratios, combined with the lack of direct spatial correlation between the biconical structure and any potential ionisation sources within the disk, and the evidence indicating that the primary ionisation within the structure is attributed to SF photoionisation, suggest an alternative interpretation for the biconical structure observed in the \nii/\ha ratio map. As discussed throughout this paper, the nature and morphology of the eDIG are directly influenced by the local level of SF within the disk, which drives the feedback mechanisms responsible for ionising the gas in the surrounding vicinity.

From the \ha\ flux map of IC1553, the high-level star-forming (SF) regions at a MAD between 2 and 4 kpc appear to be very near the edge of the galaxy's disk. Consequently, the halo line-of-sight line ratios in the vicinity of these regions resemble those of \hii regions (e.g., \nii/\ha\ $\leq 0.4$). This effect is more pronounced in the \oiii/\hb\ and \oi/\ha\ maps, where high \oiii/\hb\ values ($\geq 1.5$) and low \oi/\ha\ values ($\leq 0.1$) are observed immediately above and below the bright \hii regions on the plane at 2 < MAD < 4 kpc. A common observational challenge in edge-on galaxies is the invisibility of individual SF regions in the optical spectrum due to the high optical depth for both blue continuum and \ha\ line emission. This likely occurs at MAD $\geq -2$ kpc, where no bright \hii regions are visible on the galaxy disk in the \ha\ flux map. Nevertheless, the \nii/\ha, \sii/\ha, \oi/\ha, and \oiii/\hb\ line ratios above and below the galaxy plane in this galactocentric range resemble those near the bright \hii regions at MAD > 2 kpc. At the minor axis (MAD = 0 kpc), the galaxy disk is expected to show a lower average star formation rate (SFR) along the line-of-sight due to the presence of the galaxy's bulge, compared to the rest of the galaxy disk. Consequently, the line emission strength near the galaxy's centre would be dominated by the strong emission from the bright visible \hii regions to the south of the galaxy (2 < MAD < 4 kpc) and by obscured \hii regions on the galaxy disks at projected distances $-1 <$ MAD $< -4$ kpc.

In this scenario, the biconical structure observed in \nii/\ha could be an optical artefact, arising from high typical \nii/\ha eDIG background values ($\geq 0.5$) due to the overall feedback processes in the galaxy disk (SF+shocks), and low \nii/\ha values ($\leq 0.4$) from the bright and obscured \hii regions near the minor axis of the galaxy. This results in \hii region-like emission in the vicinity of the galactic plane. This scenario is more evident in the \sii/\ha and \oi/\ha maps, where the background emission clearly permeates the halo with typical eDIG values, and the strong emission from the adjacent \hii regions is seen in projection, exhibiting typical \hii region values. This interpretation accounts for the offset of the biconical structure between the \nii/\ha and \sii/\ha maps, as S and N present similar but distinct ionisation potentials, shedding light on the structure of the \sii/\nii line-ratio map. It also explains the absence of filaments or outflow features in both the \ha flux and \oiii/\hb line-ratio maps.

In the case of the biconical structure of IC1553, it becomes evident that results obtained from the global analysis of a galaxy should not be generalised. Limiting the analysis to specific regions reveals that the ionisation mechanisms present in the eDIG can vary. This finding aligns with our observations in \citetalias{2024A&A...687A..20G}: a group of galaxies exhibiting diverse physical processes can lead to misleading conclusions about the ionisation mechanisms of the DIG. Furthermore, even within a single galaxy, when the resolution is sufficiently high, a global (BPT) analysis of the eDIG can still be misleading, as it encompasses regions of the galaxy that display distinct physical processes and ionisation conditions.

\subsection{General results for the eBETIS sample}

In this section, we summarise the results of the previous analysis for all the galaxies in the sample. The rest of emission line maps, plots and further details of each galaxy can be found in the appendices from \ref{sec:Ap_A} to \ref{sec:Ap_F}.

In general, the linkage between the star formation activity in the disk and the ionisation of the eDIG is evident in all galaxies. The entire sample shows the same tendencies, with the \SBha distribution in the plane and the line ratios distribution being anti-correlated. Additionally, the increase in T$_e$ with distance from the midplane and the increase in S$^+$/S at those MADs corresponding to lower \SBha in the midplane are also general behaviours in the sample. However, each galaxy can present differences in the ionisation structure of the eDIG due to the presence of morphological structures in the halo or asymmetries in the disc. This can cause the contribution of fast shocks to the ionisation budget of the eDIG to vary in different zones of the halo.

\begin{itemize}
    
    \item PGC28308: The bidimensioanl structure of the eDIG is shown as a shift between the \nii/\ha and \sii/\ha MAD distributions at z = 1.5 and 2.5 kpc. In general, the ionisation regimes in the halo are mixed, with contributions of approximately 50\% from fast shocks and 50\% from the star formation. 
    \item PGC30591: The bidimensioanl structure of the eDIG is more pronounced in comparison with PGC28308, with a clearer shift between the MAD distributions at z = 0.7 and 1.5 kpc. Fast shocks contribute a 40\% to the ionisation budget. Furthermore, since the ionisation regimes are not that mixed in comparison with other galaxies, the dependence of these regimes on height is more evident in the BPT diagrams. Bins further from the plane tend to be closer to the fast shock regime in the BPT, coinciding with the increment of $T_e$ in height.
    \item ESO544-27: This galaxy presents a homogeneous distribution of \hii regions with respect to the line of sight, and the \SBha distribution does not show strong asymmetries along the major axis. This homogeneity is reflected in the eDIG as well. The increase in T$_e$ with distance from the midplane is linear and uniform, with a constant S$^+$/S$\sim$0.5. Additionally, the contribution of fast shocks to the ionisation budget is 50\%.
    \item ESO443-21: This galaxy presents a broad knot-shape filamentous structure at a distance of 1.5 kpc, extending up to a height of 4.5 kpc within the range of -1 kpc < MAD < 2.5 kpc. The presence of this knot adds an additional 10\% to the star formation contribution for the ionisation budget of the eDIG, resulting in a 20\% contribution of fast shocks to the ionisation of the eDIG for z > 0 and 30\% for z < 0.
    \item ESO469-15: The presence of extraplanar \hii regions in its halo affects the overall ionisation structure of the eDIG. The line ratios reach lower values compared to the rest of the galaxies, as well as the T$_e$, ranging between 6-8$\cdot$10$^3$K. The contribution of fast shocks is only 30\% throughout the eDIG.
    \item ESO157-49: The galaxy shows an asymmetry in the eDIG between z > 0 and z < 0, being the emission significantly uniform at z > 0 and presenting a tilted and shifted ionisation structures at z < 0, probably due to the influence of its nearby neighbours. In general, the contribution of fast shocks in the ionisation of the eDIG is 50\%, however, its drops to 30\% between 2 kpc < MAD < 3 kpc, where the \SBha is higher. In this latter case, is clear that bins further from the plane tend to be closer to the fast shock regime in the BPT, coinciding with the increment of $T_e$ in height.
    \item IC217: The low data sampling due to poor data quality does not allow for a more in-depth analysis of this galaxy. It exhibits the general characteristics of the rest of the sample described above, with an overall contribution of fast shocks of 50\%.
\end{itemize}

The general characteristics of the eDIG found in the eBETIS sample are summarised in table \ref{tab:eDIG_char}.

\section{Summary and conclusions}\label{sec:conclusion}

In this work, we present the second part of the Bidimensional Exploration of the warm-Temperature Ionised gaS (BETIS) project, focusing on the spatially resolved and spectral study of the extraplanar diffuse ionised gas (eDIG) in a selection of eight edge-on (i > 75°) nearby galaxies. We used the galaxies from the \citetalias{2019A&A...623A..89C} dataset as our sample and applied the methodology described in \citetalias{2024A&A...687A..20G} for adaptively binning the observed datacubes and extracting the binned emission line maps from them.

The \ha emission line maps reveal a complex ionisation structure in the galactic halos composed by diverse structures as filaments or knots that can reach more than 3 kpc from the midplane. The real complexity of the ionisation structure of the eDIG is shown when examining the \nii/\ha, \sii/\ha, \oiii/\hb and \oi/\ha maps, revealing the presence of new structures as an apparent biconical structure in IC1553 or a broad ionised knot of ESO443-21. The behaviour of these line ratios reveals that the ionisation structure of the eDIG is shown to be influenced by the distribution of \hii regions in the galactic plane as seen from the line of sight. In all galaxies, the line ratios increase both vertically and radially, showing an anti-correlation with the \SBha distribution of the galactic disc. These ratios reach higher values at distances further from the \hii regions, both in terms of height (z) and major axis distance (MAD). 

This correlation between the morphological distribution of star-forming regions in the disks and the morphology of the halos presents the strongest evidence supporting the interpretation that energy sources from star formation within the galaxy disks drive the observed disk-halo interaction, being the Lyc photons leaking from OB associations the main ionisation source of the eDIG. 

However, OB-star-driven ionisation can not explain some features found in the eDIG, such as the enhanced \nii/\ha and \sii/\ha and the presence of high ionisation species such as \oiii at higher distances from the midplane. From the \nii/\ha and \sii/\ha ratios we explored the radial and vertical variations of T$_e$ and S$^+$/S. The general behaviour indicates a linear, constant increment of T$_e$ with increasing distance from the galactic plane for all the galaxies. Besides, there is significant variations in T$_e$ and S$^+$/S at different MADs, also linked to the \hii regions distribution with respect the line of sight. The variations in temperature and ionisation structure of the eDIG as a function of spatial position within a single galaxy provide compelling evidence of the intricate dynamical heating structure of the ISM in the halo. These variations show that the eDIG tends towards a lower ionisation state (higher S$^+$/S) compared to classical \hii regions. The anomalous \nii/\ha and \sii/\ha
line ratios in the eDIG can be attributed to an increase in the T$_e$ temperature
in the halo, resulting from photoionisation with a lower ionisation parameter.

To account for the \oiii/\hb and \oi/\ha ratios, various ionisation mechanisms have been considered. One such mechanism, recently highlighted in the literature, involves photoionisation by a hard spectrum emanating from the high-temperature end of the white dwarf distribution and the central stars of planetary nebulae; the low-mass, hot, and evolved (post-AGB) stars known as HOLMES. Nevertheless, the correlation between the ionisation structure of the eDIG and the \hii regions distribution observed in this study does not support a homogeneous distribution of HOLMES across the thick disc and lower halo. This inconsistency, combined with the unreliability of \EWha as a proxy commonly used in the literature to differentiate star-forming regions from HOLMES, effectively negates HOLMES as a plausible ionisation mechanism for the eDIG. Consequently, the behaviour of the \nii/\ha, \sii/\ha, and \oiii/\hb ratios alone is not sufficient evidence to conclusively support HOLMES as an ionisation source. These behaviours should be carefully used in the analysis of the ionisation mechanisms of the (e)DIG.

We propose that shocks induced in the ISM by feedback mechanisms serve as a secondary ionisation source for the eDIG. Given that shock heating becomes substantial only when the kinetic energy is efficiently thermalised, both photoionisation and shocks may be regarded as "thermal" heating sources for the ISM. The observed enhancement in the \oiii/\hb ratio with increasing height can be interpreted as a mixing sequence between the predominant shock ionisation in the halo and the photoionisation in the disk.

To ascertain the impact of shocks on the ionisation budget of the eDIG, we constructed a suite of hybrid models that integrate both star formation and fast shock regimes. We conducted a BPT analysis by computing the hybrid models for each galaxy individually, given that each galaxy may exhibit unique characteristics necessitating a tailored ionisation diagnosis for the eDIG (\citetalias{2024A&A...687A..20G}). Our results indicate that fast shocks significantly contribute to the ionisation budget of the eDIG, with contributions ranging from 20\% to 50\% across the sample.

The presence of shocks is further corroborated when examining the density-sensitive \sii doublet ratio as a function of the \oi/\ha ratio. For all galaxies, most of the eDIG regions near the midplane exhibit \sii line ratios close to the low-density limit, accompanied by low \oi/\ha values, which are indicative of \hii region emission. However, regions with low \sii doublet ratios –typically at high midplane distances– coincide with regions exhibiting high \oi/\ha values, indicative of shock-compressed ionised gas. Therefore, the low \sii doublet ratios, along with their implied high gas densities and the fact that these regions exhibit higher \oi/\ha\ values, support a scenario in which wherein gas emission originates from shocks, likely induced by feedback from high-level star-forming regions within the galactic disk. In this scenario, fast shocks can account for the increase in the \oi/\ha\ ratio with distance from the midplane, with the highest \oi/\ha\ ratios corresponding to the shocked interface at the greatest distances from the midplane.

The contribution of fast shocks can vary significantly across different regions of the halo, with variations reaching up to 20\% within the same galaxy. This variability arises from the distinct properties, physical processes, and structures inherent to each galaxy that affect the ionisation of the extraplanar gas. These structures include filaments, knots, extraplanar \hii regions, and nearby neighbouring galaxies. Nonetheless, the distribution of \hii regions within the disc consistently exerts a substantial influence on the ionisation of the halo gas.

A prime example of this phenomenon is IC1553. In the \nii/\ha emission line map, it reveals a biconical central structure within the halo, previously interpreted as indicative of outflows or superbubbles. Nonetheless, at the bicone's location (MAD = 0), the galaxy's disk is expected to exhibit a lower specific star formation rate (sSFR) along the line-of-sight due to the presence of the galaxy’s bulge. This contrasts with the intense emission from the bright adjacent star-forming regions at the edge of the disk, observed in projection, which exhibit typical \hii region (low) emission values. This scenario is more pronounced in the \sii/\ha and \oi/\ha maps, where the background emission conspicuously permeates the halo with typical extended eDIG values. These observations, coupled with the absence of any structure in \ha, \oiii/\hb, the consistent \sii/\nii ratio, and the lack of spatial correlation with potential ionisation sources within the disk, indicate that this bicone is an optical artefact resulting from the projection effect relative to the line-of-sight of the high-level SF regions in the galaxy's disk, in relation to a pervasive background eDIG emission in the halo.

From this perspective, it can be inferred that the extended eDIG emission comprises multiple overlapping components contingent upon the line of sight: eDIG emanating from low-density gas, typified by elevated \nii, \sii line ratios, and gas ionised through star formation (SF) activity proximate to the midplane of edge-on galaxies, witht typical \hii line emission values. The overlaying patterns vary according to the morphology and extent of SF regions within the galactic plane, resulting in a complex morphology of the eDIG concerning line emissions.

To advance the Bidimensional Exploration of the warm-Temperature Ionised gaS (BETIS) project, in forthcoming papers in this series we will examine the kinematic reflection of the eDIG connection with the galactic plane. Specifically, we will analyse radial variations in rotational velocity gradients (referred to as "lag"), which are correlated with the SFR and associated with morphological structures within the halo. Additionally, we will investigate the energy deposition into the ISM through SF-related feedback mechanisms, the primary source of the eDIG ionisation. Moreover, we will explore how the (e)DIG influences the determination of various parameters, including chemical abundances and SFR.

\begin{acknowledgements}
Based on the observations at ESO with program IDs: 096.B-0054(A) and 097.B-0041(A). R.G.D. acknowledges the CONAHCyT scholarship No. 1088965 and INAOE for the PhD program. The authors also acknowledges Manuel Zamora and Raúl Naranjo for allowing us the usage of the Mextli cluster at INAOE. R.G.D and L.G. acknowledge financial support from AGAUR, CSIC, MCIN and AEI 10.13039/501100011033 under projects PID2020-115253GA-I00, PIE 20215AT016, CEX2020-001058-M, and 2021-SGR-01270.

\end{acknowledgements}

\bibliographystyle{aa}
\bibliography{aa}

\appendix 

\section{PGC28308}
\label{sec:Ap_A}

\begin{figure}[ht!]
\centering
\includegraphics[width=\columnwidth]{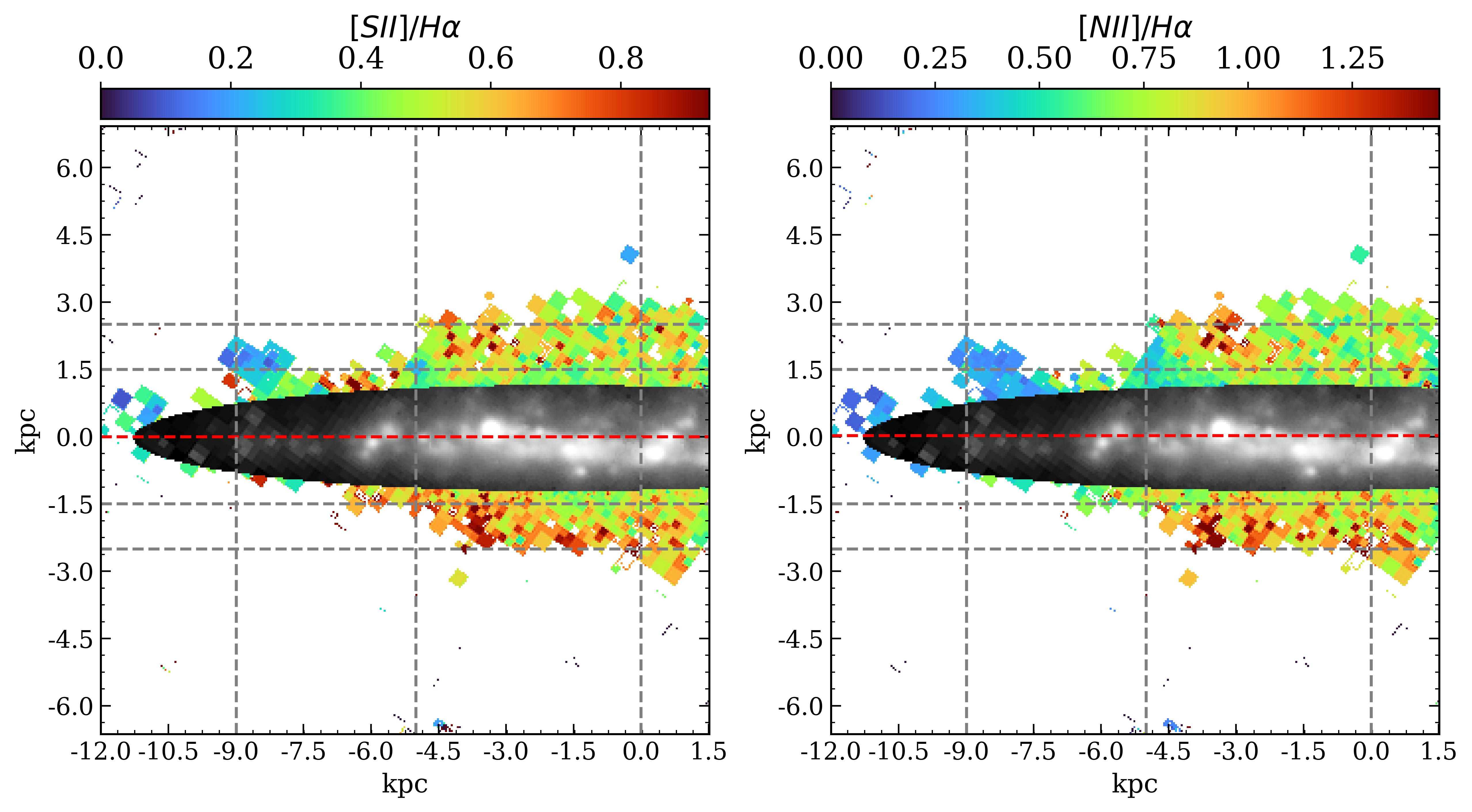}
\includegraphics[width=\columnwidth]{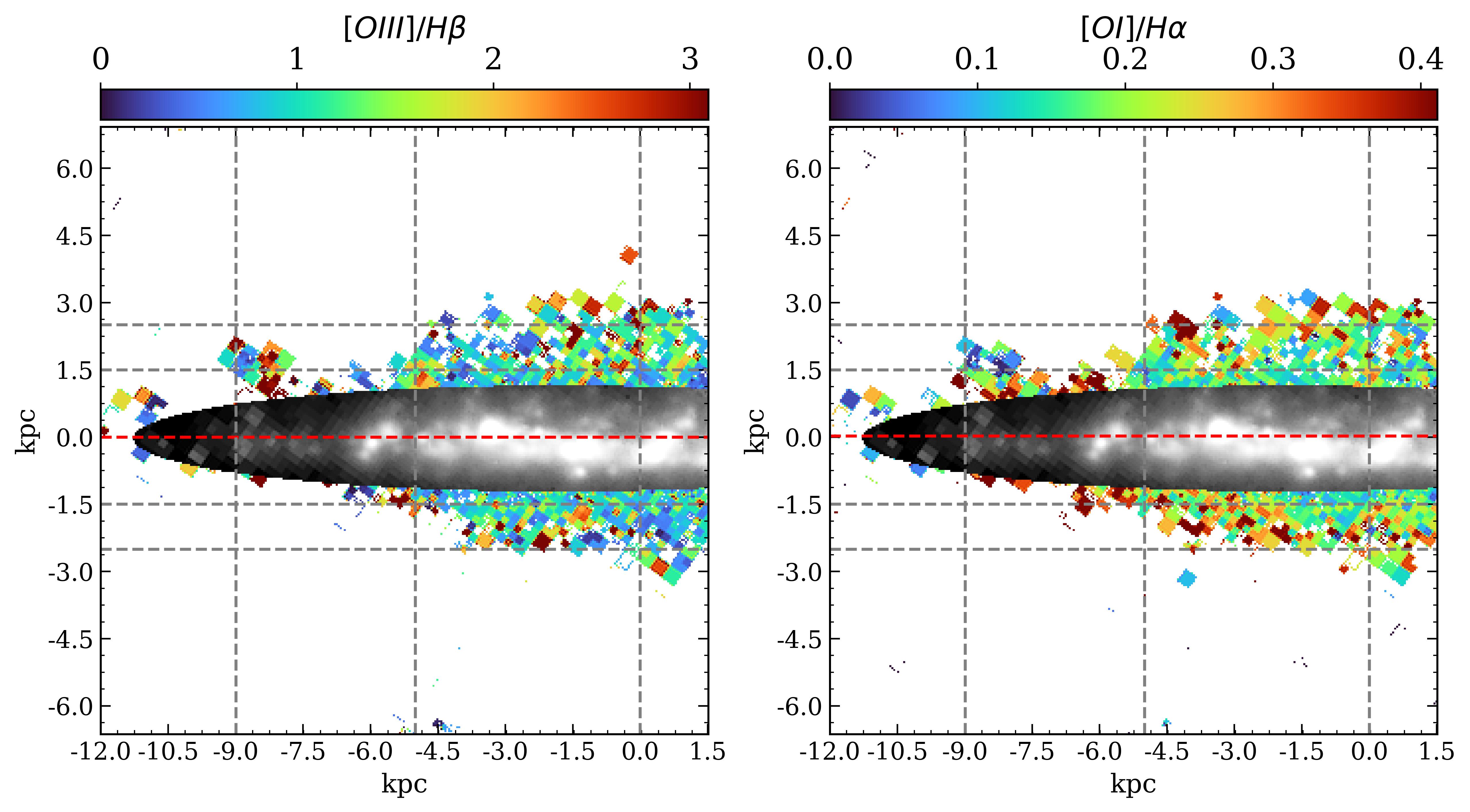} 
\captionof{figure}{PGC28308 line ratio maps, similar to Figure \ref{fig:IC1553_maps}. The grey dashed lines indicates
the heights with respect the midplane z = $\pm$1.5, $\pm$2.5 and major axis distances MAD = -9, -5, 0 kpc.}\label{fig:PGC28308_maps}
\end{figure}

At a distance of 45.22 Mpc, PGC28308 is the most distant galaxy in the sample \citep{2008Ap.....51..336K}. Since all galaxies in the sample have the same total exposure time (except IC217; \citetalias{2019A&A...623A..89C}), this results in lower S/N, a higher average bin size (175.4 pc), and a greater loss of information after performing the S/N and relative error cutout (see section \ref{sec:data}).

These cutouts left us with data up to |z| = 2.5 kpc between -4.5 kpc $\lesssim$ MAD $\lesssim$ 1.5 kpc, and |z| < 1.5 kpc at MAD < -4.5 kpc for the \sii/\ha and \nii/\ha line ratio maps (Figure \ref{fig:PGC28308_maps}).

\begin{figure}[!t]
\centering
    \includegraphics[width=0.9\columnwidth]{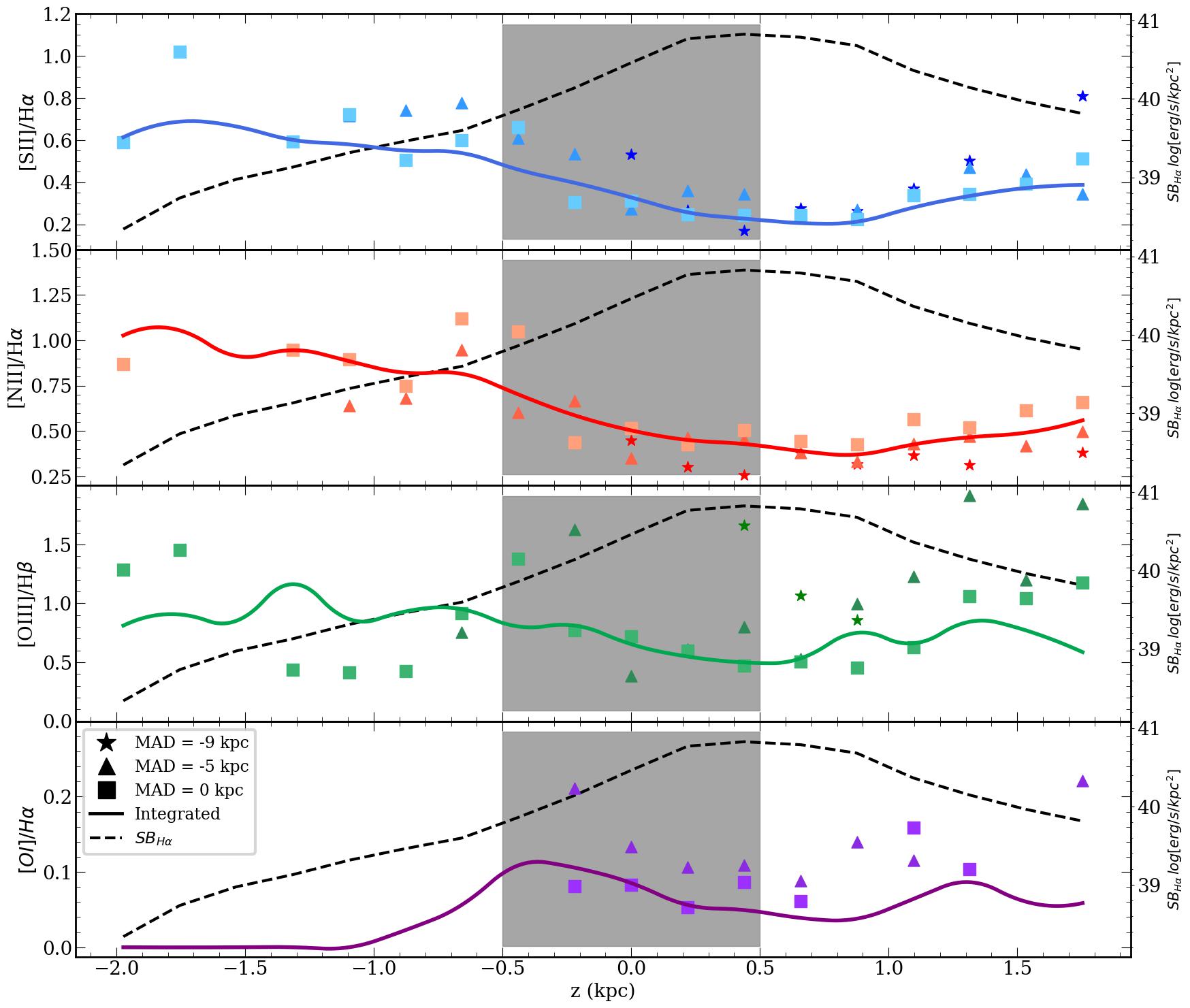}
        \caption{PGC28308 line ratio distributions of the with respect the distance from the midplane. Similarly to Figure \ref{fig:IC1553_lines}, for MAD = -9 (stars), -5 (triangles) and 0 (squares) kpc.}
        \label{fig:PGC28308_lines}
        
\end{figure}

\begin{figure}[!h]
        \centering
        \includegraphics[width=0.9\columnwidth]{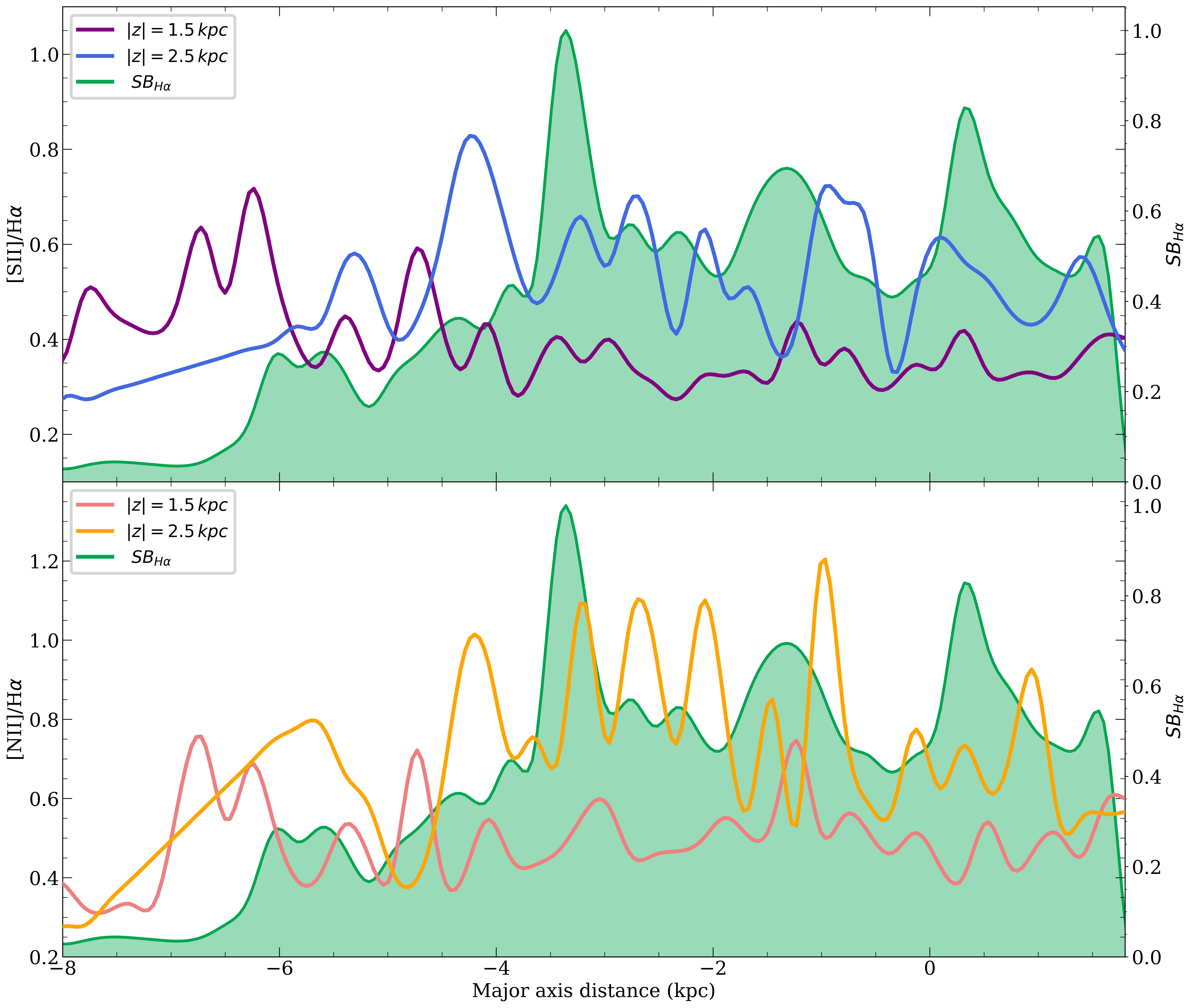}
        \caption{PGC28308 MAD distribution for z = 1.5 and 2.5 kpc.}
        \label{fig:PGC28308_MAD}
\end{figure}

With this data, we can observe that the trend of line ratios is consistent with what is found in the rest of the galaxies, reaching minimum values where the \SBha is maximum, and with the \sii/\ha and \nii/\ha ratios increasing in height, reaching values of 0.7 and 1.1 respectively (see Figure \ref{fig:PGC28308_lines}). The \oiii/\hb ratio remains approximately constant around 1.0 for |z| > 0.7 kpc and, for the \oi/\ha ratio, the distribution remains flat at 0 for z < -1 kpc, probably due to the low S/N of the \oi line, but reaches values of $\sim$ 0.08 at z > 0.7 kpc. In all cases, the squares, triangles, and stars representing the height distributions at MAD = 0, -3, and -9 kpc, respectively, follow a similar trend as the integrated distribution, which is also in line with what we observed in the rest of the galaxies. The radial dependence of the line ratios is not as clear as for the case of IC1553 by only considering Figure \ref{fig:PGC28308_lines}. However, a clearer scenario can be seen in Figure \ref{fig:PGC28308_MAD}. Both \sii/\ha and \nii/\ha distributions consistently show higher ratios at z = 2.5 than at z = 1.5. Moreover, the differences between these distributions becomes more pronounced as the \SBha reaches higher values (at MAD$\sim$-3 kpc and MAD$\sim$0.5 kpc. Examination of the MAD distributions also reveals the bidimensional structure of the eDIG, as the positions of relative maxima shift when the height increases in both \sii/\ha and \nii/\ha distributions.

\begin{figure*}[t!]
\centering
\includegraphics[width=\textwidth]{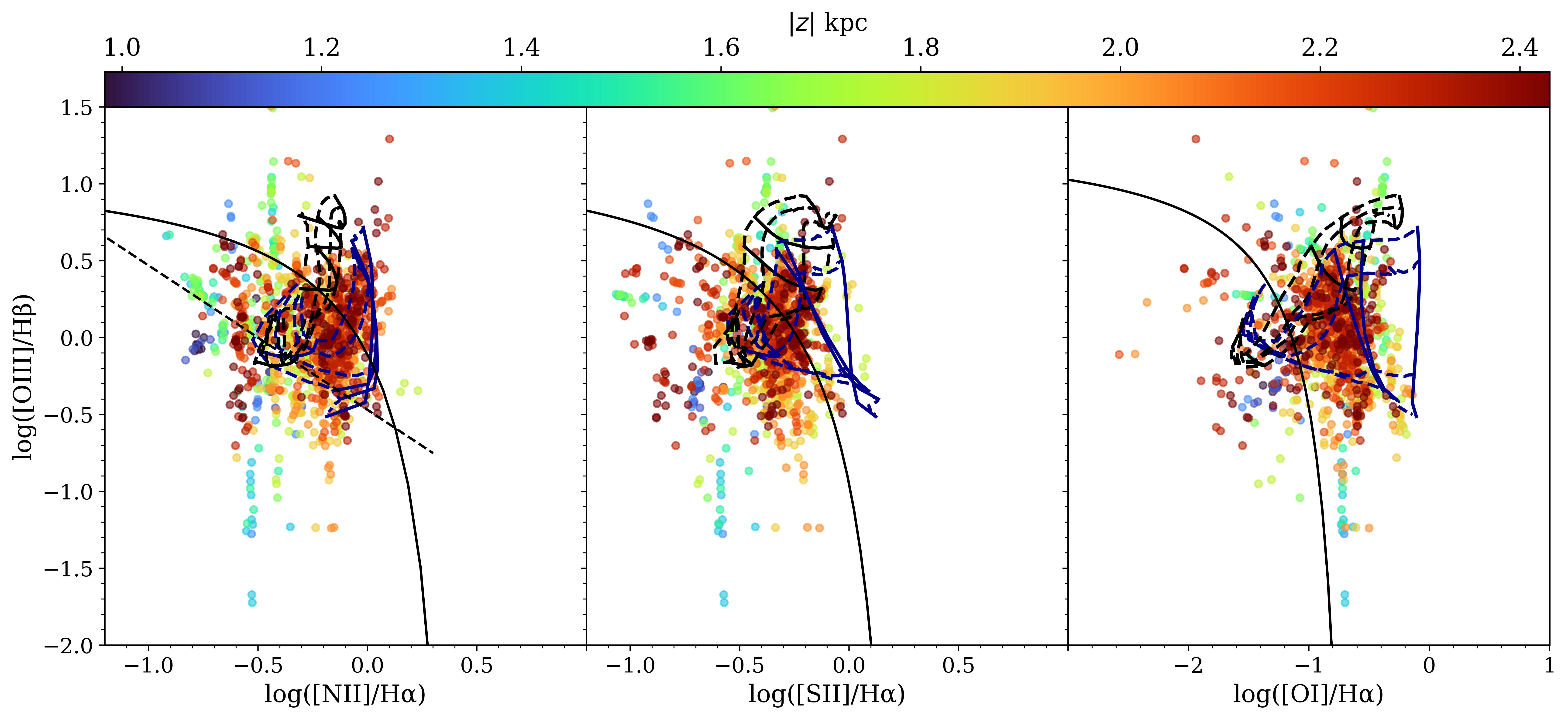} 
\caption{PGC28308 BPT with hybrid models, similar to Figure \ref{fig:IC1553_BPT_total}. 50\% fast shocks and 50\% star formation with Z = Z$_\odot$ and q = $10^7$ cm/s.}
\label{fig:PGC28308_BPT}     
\end{figure*}

\begin{figure}
\includegraphics[width=0.9\columnwidth]{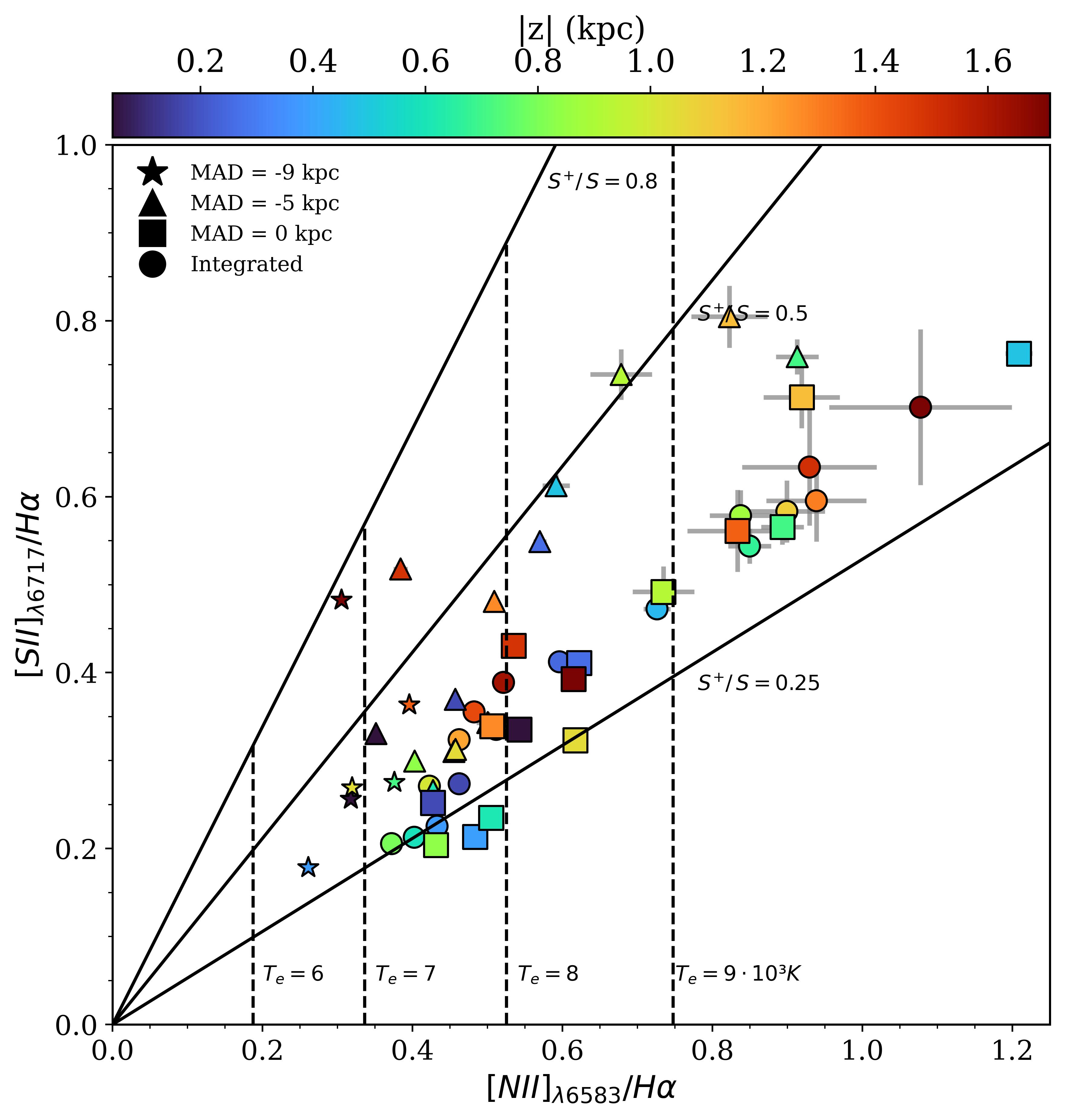}
        \caption{\nii$_{\lambda\,6583}$/\ha vs. \sii$_{\lambda\,6717}$/\ha for PGC28308. Similarly to Figure \ref{fig:IC1553_T_S+S}, for MAD = -9 (stars), -5 (triangles) and 0 (squares) kpc.}
        \label{fig:PGC28308_T_S+S}   
\end{figure}

\begin{figure}
\centering
\includegraphics[width=0.9\columnwidth]{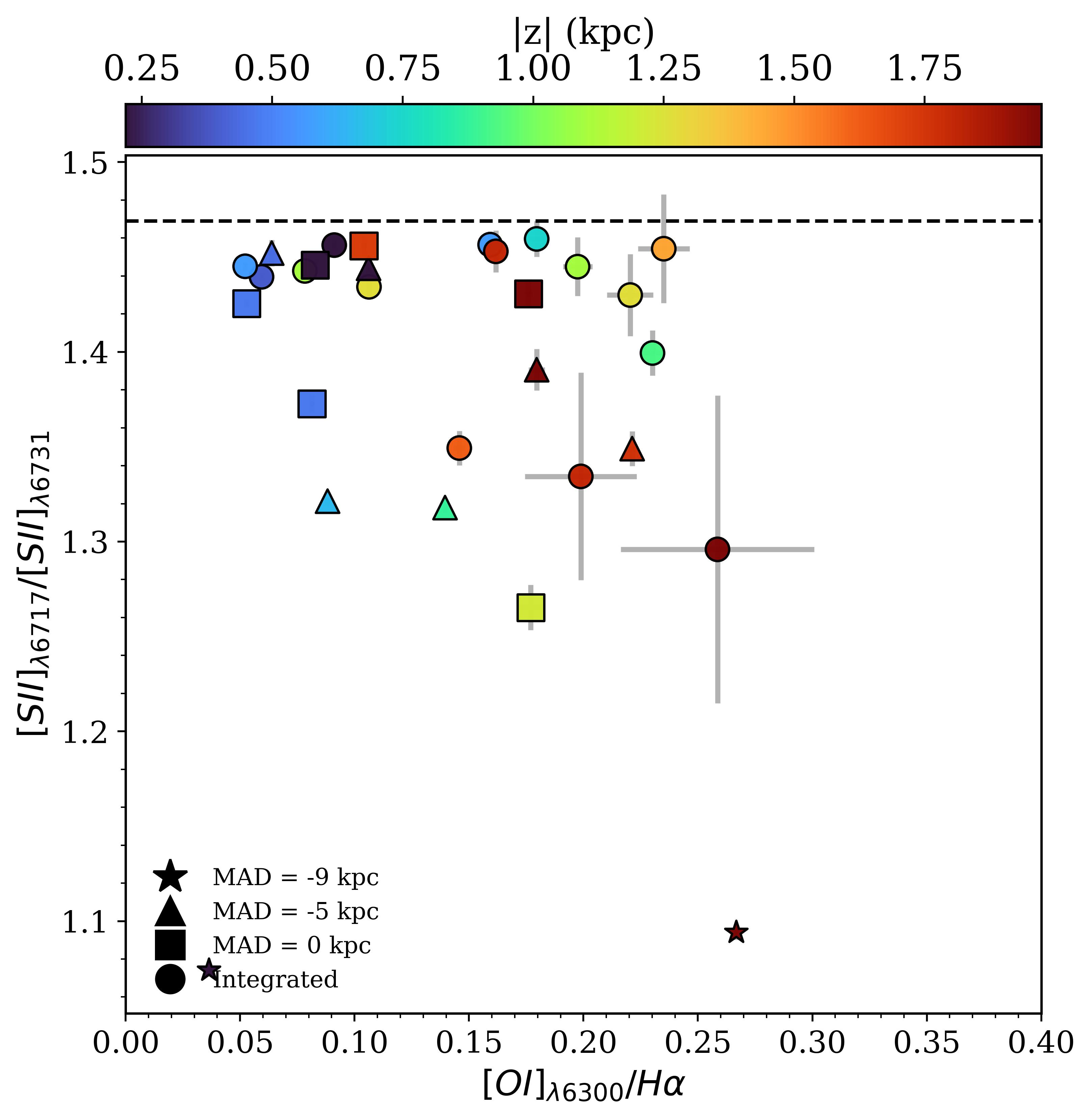} 
\caption{\sii$_{\lambda\,6717}$/\sii$_{\lambda\,6731}$ vs \oi/\ha for PGC28308. Similar to Figure \ref{fig:IC1553_SII_OI}.}
\label{fig:PGC28308_SII_OI}     
\end{figure}

As the sampling of this galaxy is lower due to the resolution and S/N cutouts, Figure \ref{fig:PGC28308_T_S+S} does not show a dependence of the T$_e$ with |z| as clear as for IC1553. However, it is clear the dependence of the MAD with the ionisation ratio S$^+$/S. At MAD = 0 kpc (squares) S$^+$/S$\sim$0.3, in contrast with MAD = -5 kpc (triangles), with S$^+$/S$\sim$0.5. Although both MAD are within zones of the galactic plane exhibiting high \SBha, Figure \ref{fig:PGC28308_maps} shows that the emission from \hii regions is displaced relative to the line of sight. Specifically, the \hii regions nearest to the -1.5 line at MAD = 0 kpc, indicating that the eDIG below the galactic plane at that MAD is highly ionised due to ongoing star formation. This ionisation leads to a reduction in the S$^+$/S ratio and consequently enhances the S$^{++}$/S ratio \citep{1980JPhB...13L.543B, 1999ApJ...523..223H, 2006agna.book.....O}. In general, the eDIG bins are located in the BPT diagram between both classical regimes of star formation and AGN (see Figure \ref{fig:PGC28308_BPT}). The hybrid models that best fit the data correspond to 50\% of ionisation due to fast shocks, and 50\% star formation with Z = Z$_\odot$ and q = 10$^7$ cm/s.

\FloatBarrier

\section{PGC30591}
\label{sec:Ap_B}
\raggedbottom

\begin{figure}
\includegraphics[width=\columnwidth]{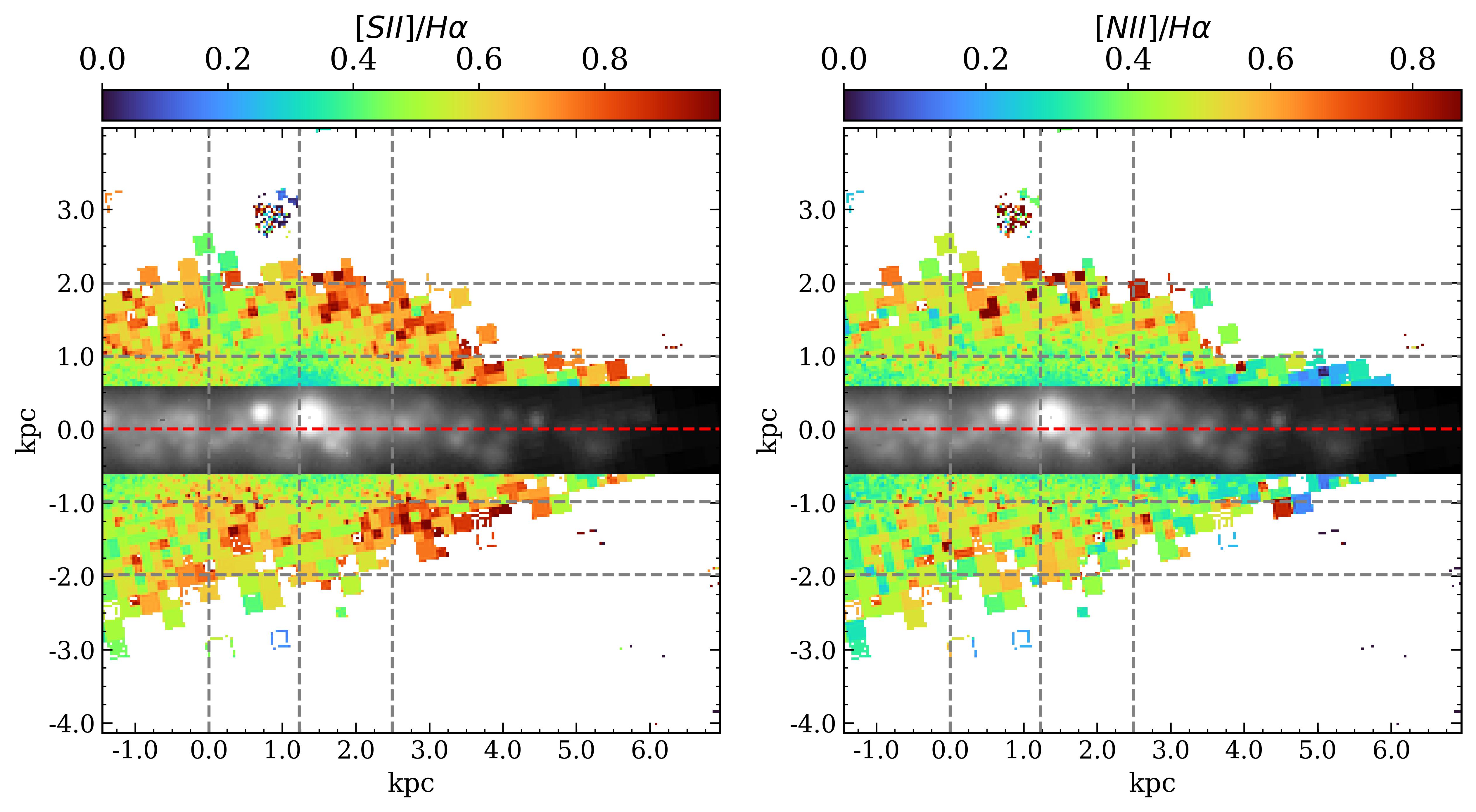}
\includegraphics[width=\columnwidth]{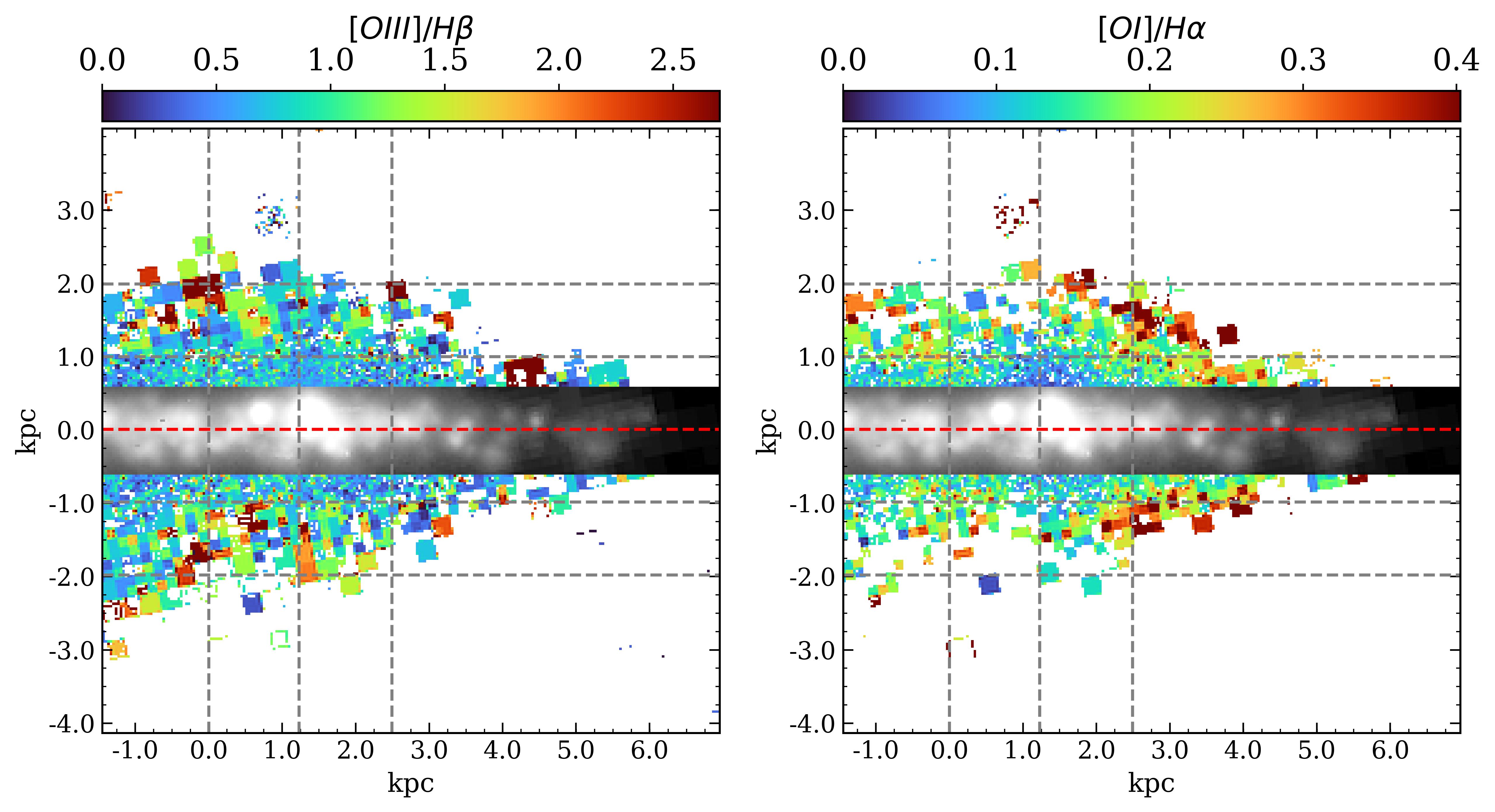} 
\parbox{\columnwidth}{\captionof{figure}{PGC30591 line ratio maps, similar to Figure \ref{fig:IC1553_maps}. The grey dashed lines indicates the heights with respect the midplane z = $\pm$1, $\pm$2 and major axis distances MAD = 0, 1.25, 2.5 kpc.}
\label{fig:PGC30591_maps}}
\end{figure}

As PGC28308, PGC30591 is a late-type spiral galaxy, although it is relatively closer in comparison \citep{1991rc3..book.....D}. This results in a higher S/N and better data sampling after implementing the cutouts of \ref{sec:data}, with bins reaching |z| up to 2 kpc between -1.5 kpc < MAD < 3 kpc. From the line of sight, the distribution of \hii seems homogeneous along the major axis, but with a brighter region between 1 kpc < MAD < 2 kpc, near to the limit of z = 0.5 kpc (see Figure \ref{fig:PGC30591_maps}). The influence of this region causes a clear decrease in the \nii/\ha, \sii/\ha and \oi/\ha line ratios between z = 0.5 kpc and z = 1.5 kpc in comparison with the rest of the eDIG (see Figure \ref{fig:PGC30591_maps}).

\begin{figure}[t!]
\centering
\includegraphics[width=\columnwidth]{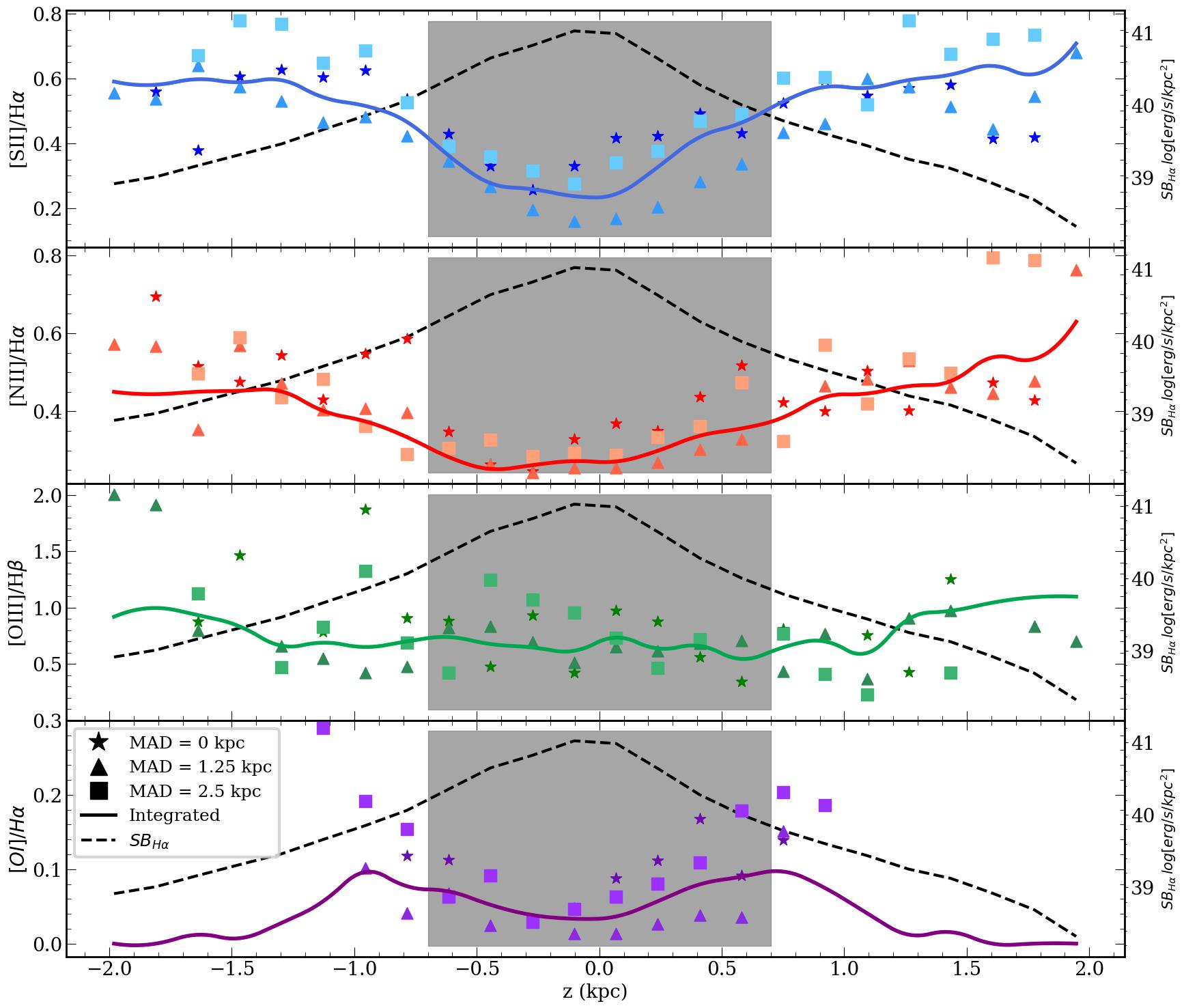}
\caption{PGC30591 line ratio distributions of the with respect the distance from the midplane. Similarly to Figure \ref{fig:IC1553_lines}, for MAD = 0 (stars), 1.25 (triangles) and 2.5 (squares) kpc.}
\label{fig:PGC30591_lines}     
\end{figure}

Figure \ref{fig:PGC30591_lines} shows the same anti-correlation than in the other cases; the line ratios reaching minimum values where the \SBha is maximum, and increasing in height. \nii/\ha and \sii/\ha remains approximately flat at z < - 1 kpc, reaching values of 0.3 and 0.6, respectively, but continuing to increase beyond these values at z > 1 kpc. \oi/\ha also increases up to 0.1 at |z| = 1 kpc and then deceases dramatically. The \oiii/\hb distribution remains flat around 0.6 for |z| < 1 kpc, and reach $\sim$ 1 at |z| > 1 kpc. The distributions at MAD = 0, 1.25 and 2.5 kpc (stars, triangles and squares respectively) follow the same trend that the integrated distribution, as for the rest of the galaxies. 

\begin{figure}[ht!]
        \centering
        \includegraphics[width=0.9\columnwidth]{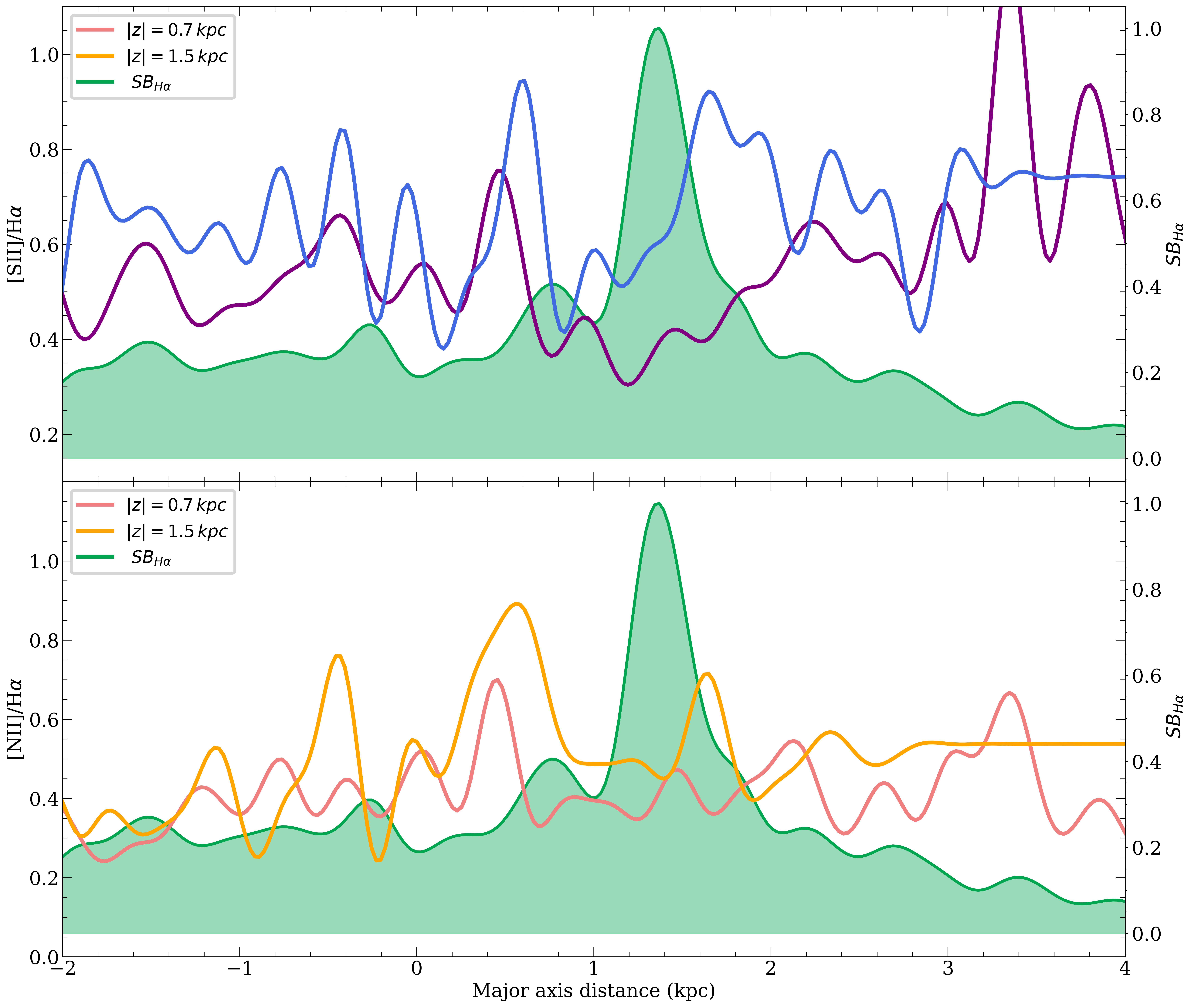}
        \caption{PGC30591 MAD distribution for z = 0.7 and 1.5 kpc.}
        \label{fig:PGC30591_MAD}
\end{figure}

Additionally, the distribution of triangles, where the most luminous star-forming regions are located in the plane, consistently exhibits lower line ratios compared to the other distributions. This aligns with what has been observed for IC1553 (see section \ref{sec:sec_3}). This can also be seen in the MAD distribution (Figure \ref{fig:PGC30591_MAD}). In addition to the typical increase in the value of the line ratios at higher heights observed in all galaxies of the sample, between 1 kpc < MAD < 2 kpc, where the surface brightness of \SBha reaches its maximum values, both line ratios for both heights decrease. 

Furthermore, the difference between the distributions for z = 0.7 kpc and 1.5 kpc becomes maximum for the \sii/\ha ratio (similarly to IC1553, with up to 0.5 dex). We can also observe a shift in the relative maxima of the \nii/\ha MAD distribution (more pronounced than for PGC28308) between 1 kpc < MAD < 2 kpc: where the \SBha distribution reaches its maximum at MAD $\simeq$ 0.4 kpc, the \nii/\ha distribution at z = 0.7 kpc reaches its relative maximum (0.45) in that interval at MAD $\simeq$ 0.55 kpc, and for z = 1.5 kpc, the relative maximum (0.65) is reached at MAD $\simeq$ 0.7 kpc. This shift of $\sim$ 0.15 kpc can also be observed in the absolute maximum at MAD $\simeq$ 0.5 kpc. Thus, this shift in the MAD distribution for different heights may be attributed to the spherical structure of ionisation between z = 0.5 kpc and z = 1.5 kpc, situated above the brighter star-forming region between 1 kpc < MAD < 2 kpc (see Figure \ref{fig:PGC30591_maps}). 

\begin{figure*}[t!]
\centering
\includegraphics[width=\textwidth]{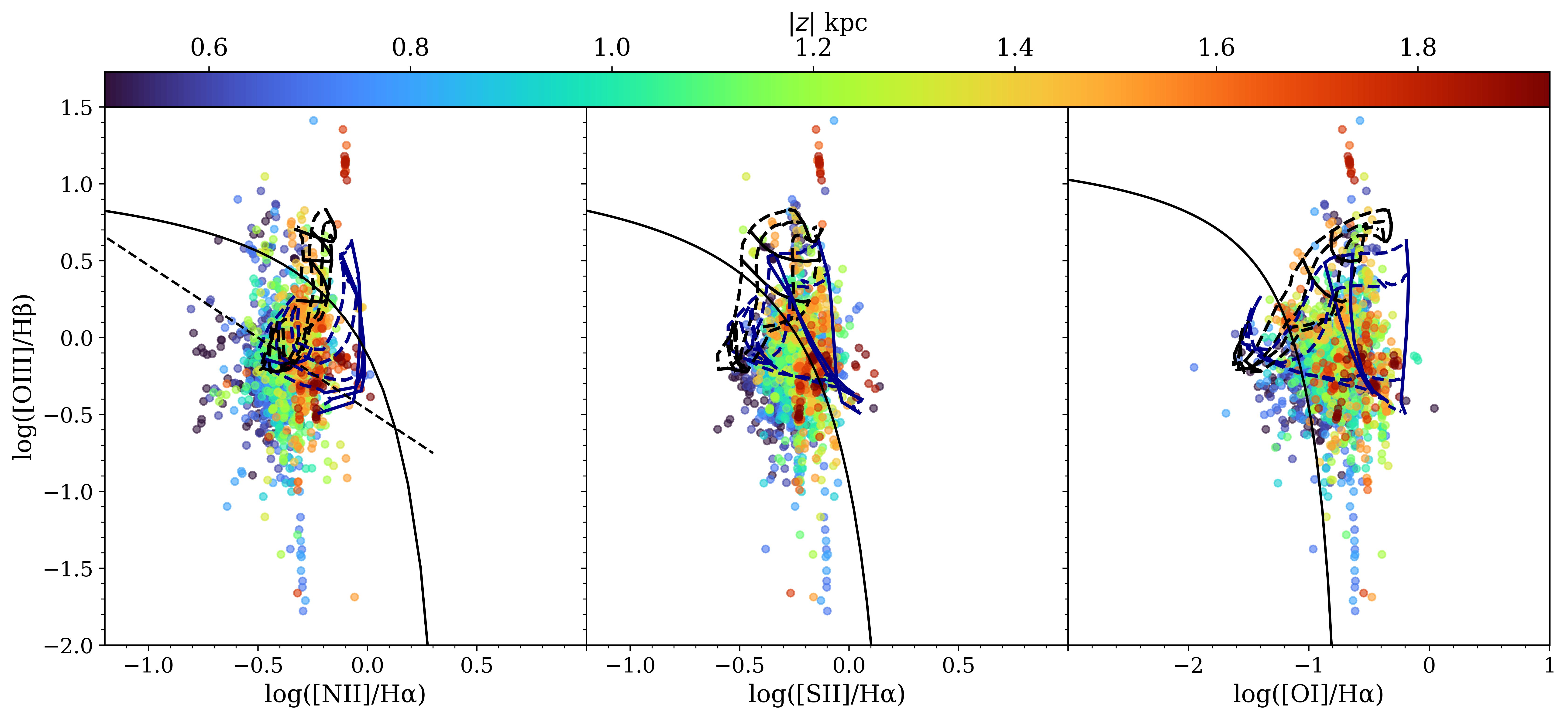} 
\caption{PGC30591 BPT with hybrid models, similar to Figure \ref{fig:IC1553_BPT_total}. 40\% fast shocks and 60\% star formation with Z = Z$_\odot$ and q = $10^7$ cm/s.}
\label{fig:PGC30591_BPT}     
\end{figure*}

\begin{figure}
    \includegraphics[width=0.9\columnwidth]{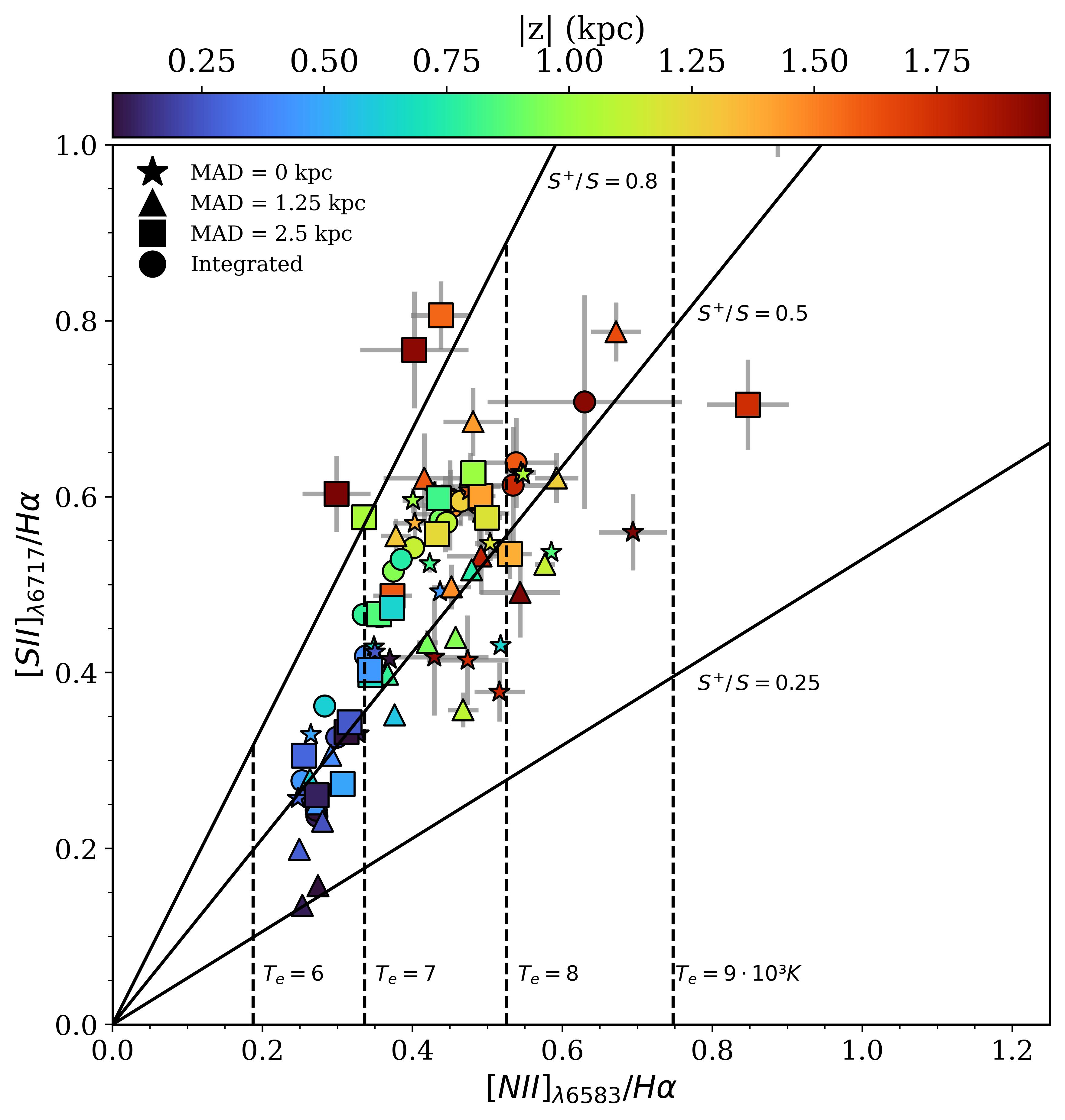}
        \caption{\nii$_{\lambda\,6583}$/\ha vs. \sii$_{\lambda\,6717}$/\ha for PGC30591. Similarly to Figure \ref{fig:IC1553_T_S+S}, for MAD = 0 (stars), 1.25 (triangles) and 2.5 (squares) kpc.}
        \label{fig:PGC305918_T_S+S}  
\end{figure}

\begin{figure}
\centering
\includegraphics[width=0.9\columnwidth]{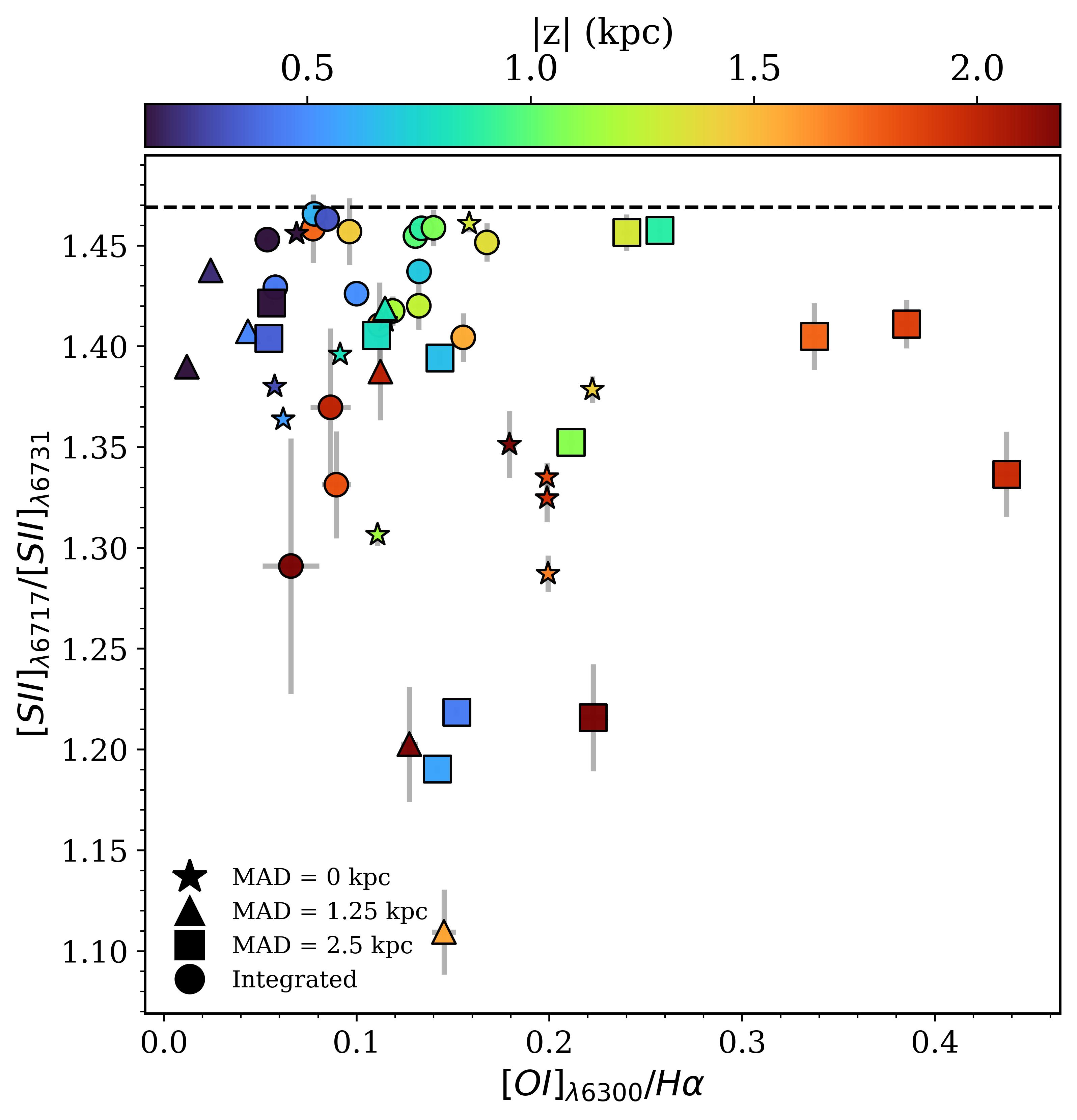} 
\caption{\sii$_{\lambda\,6717}$/\sii$_{\lambda\,6731}$ vs \oi/\ha for PGC30591. Similar to Figure \ref{fig:IC1553_SII_OI}.}
\label{fig:PGC30591_SII_OI}     
\end{figure}

In Figure \ref{fig:PGC305918_T_S+S}, the T$_e$ follows a dependency on |z|, increasing linearly with respect to the distance from the midplane and reaching values up to 9$\cdot10^3$ K. This result is also consistent with the findings in IC1553 and other authors \citep{1985ApJ...294..256R,2005ApJ...630..925M}. However, the ionisation fraction S$^+$/S does not seem to exhibit strong differences for different MADs, as all the points for S$^+$/S fall between approximately 0.35 and 0.8. Since the differences in the \ha emission of the galactic plane at differences MADs is not that pronounced (not as IC1553 at least), we expect Figure \ref{fig:PGC305918_T_S+S} to be more similar to PGC28308 (Figure \ref{fig:PGC28308_T_S+S}) than IC1553 (Figure \ref{fig:IC1553_T_S+S}), and S$^+$/S to be mixed at different distances from the midplane. Similar results are expected for the BPT diagram (Figure \ref{fig:PGC30591_BPT}). The best fit hybrid models correspond to 40\% of ionisation due to fast shocks, and 60\% star formation with Z = Z$_\odot$ and q = 10$^7$ cm/s. In this case, the dependence on height is more evident, as there is not a strong mixture of different ionisation conditions, and the data sampling due to S/N is better than for PGC28308. The figure illustrates that the points nearest to the galactic plane tend to align closer to the classical star-forming regime of the BPT diagram, and vice versa.

\FloatBarrier

\section{ESO544-27}
\label{sec:Ap_C}

\begin{figure}[t!]
\centering
\includegraphics[width=\columnwidth]{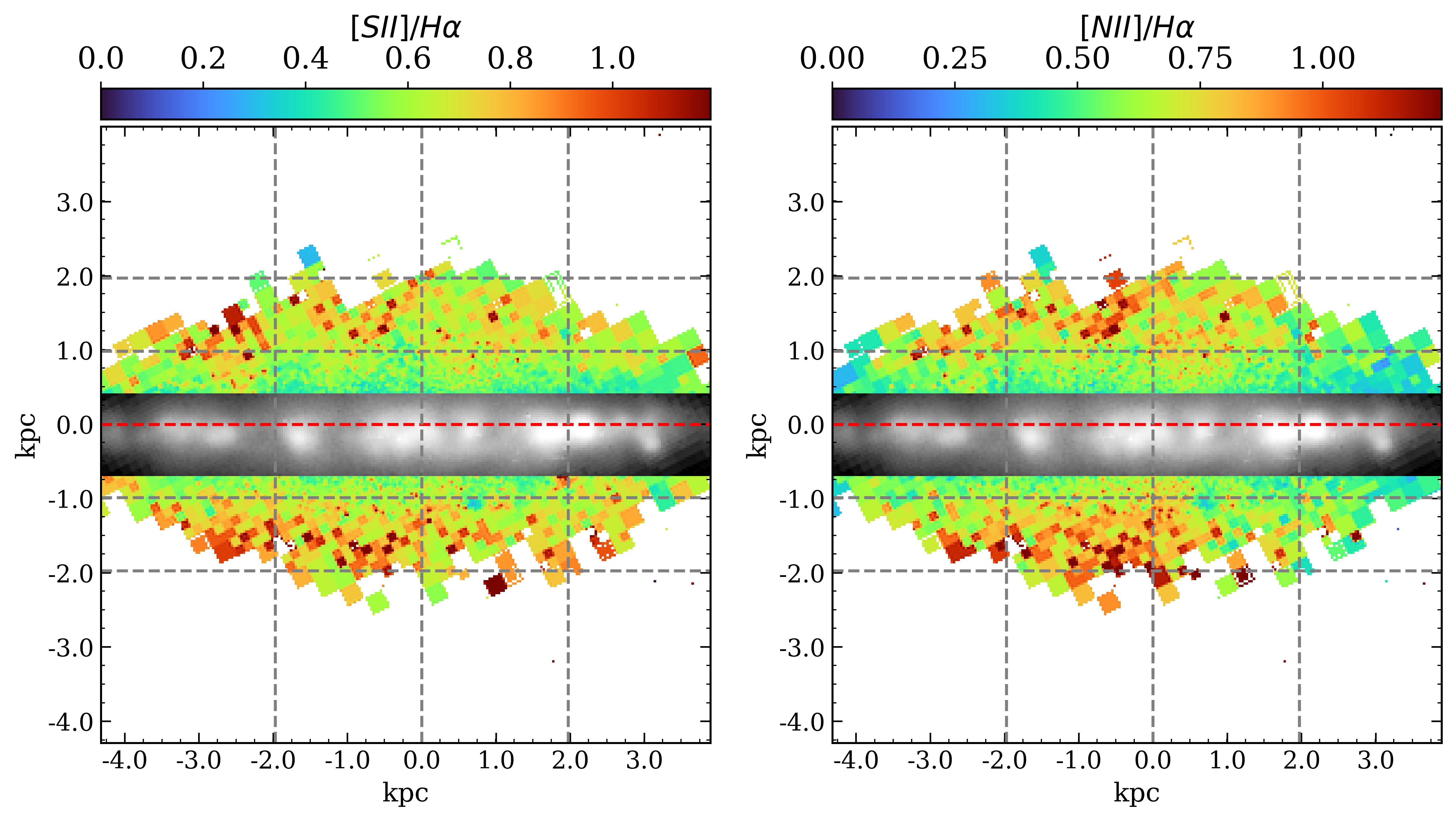}
\includegraphics[width=\columnwidth]{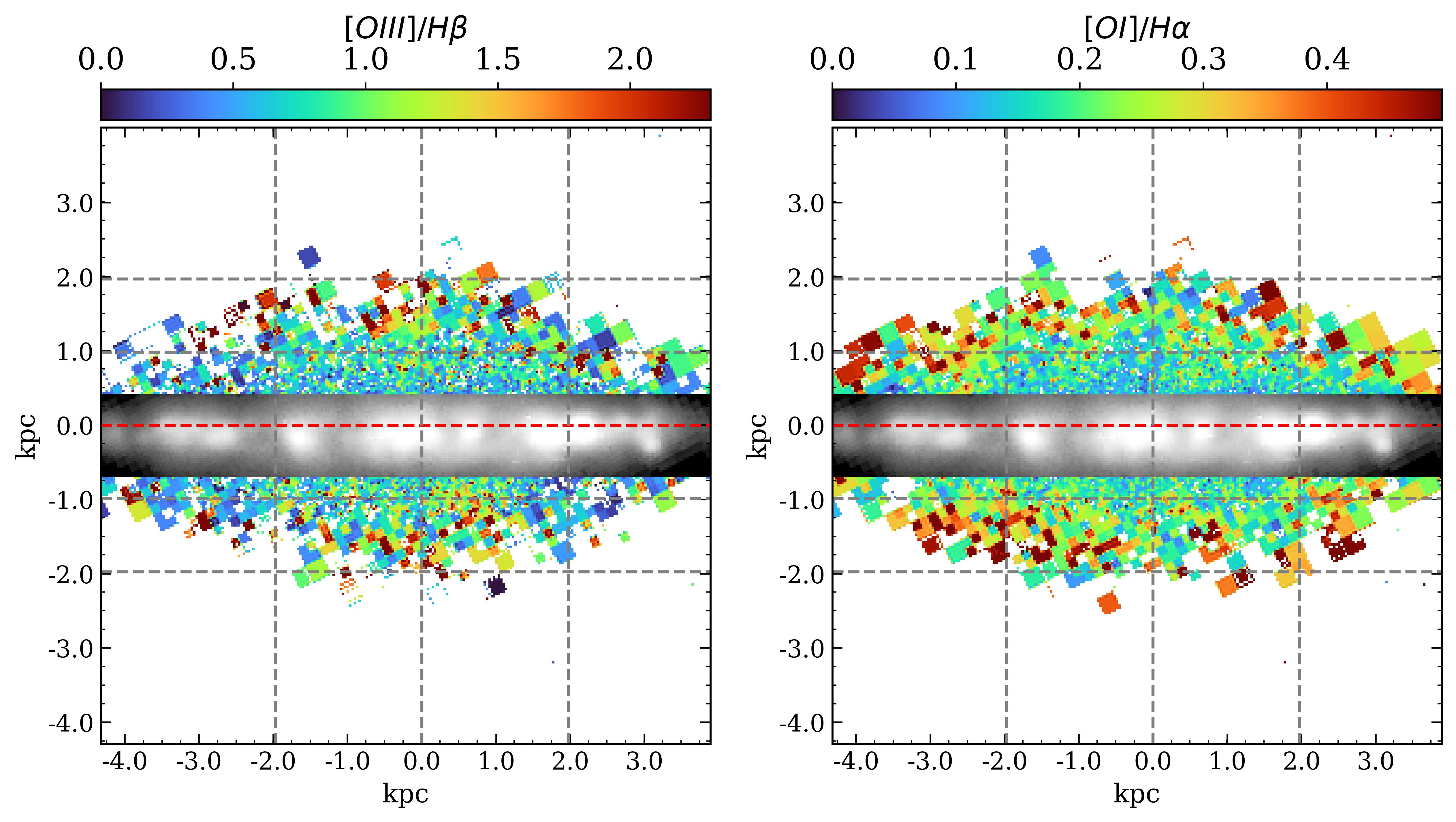} 
\parbox{\columnwidth}{\captionof{figure}{ESO544-27 line ratio maps, similar to Figure \ref{fig:IC1553_maps}. The grey dashed lines indicates
the heights with respect the midplane z = $\pm$1, $\pm$2 and major axis distances MAD = -2, 0, 2 kpc.}\label{fig:ESO544-27_maps}}
\end{figure}

ESO544-27, along with ESO469-15, is the most edge-on galaxy in the sample, with $i = 90$°. classified as an Sb galaxy \citep{2004BSAO...57....5M}, exhibits a more homogeneous emission from the galactic plane, mildly symmetric with respect to the centre of the galaxy, at MAD = 0 kpc. In consequence, the eDIG line maps exhibits the same symmetry (see figure Figure \ref{fig:ESO544-27_maps}), being evident the increase in the line ratios with respect to the distance from the midplane solely by examining the maps.

\begin{figure}[t!]
\centering
\includegraphics[width=\columnwidth]{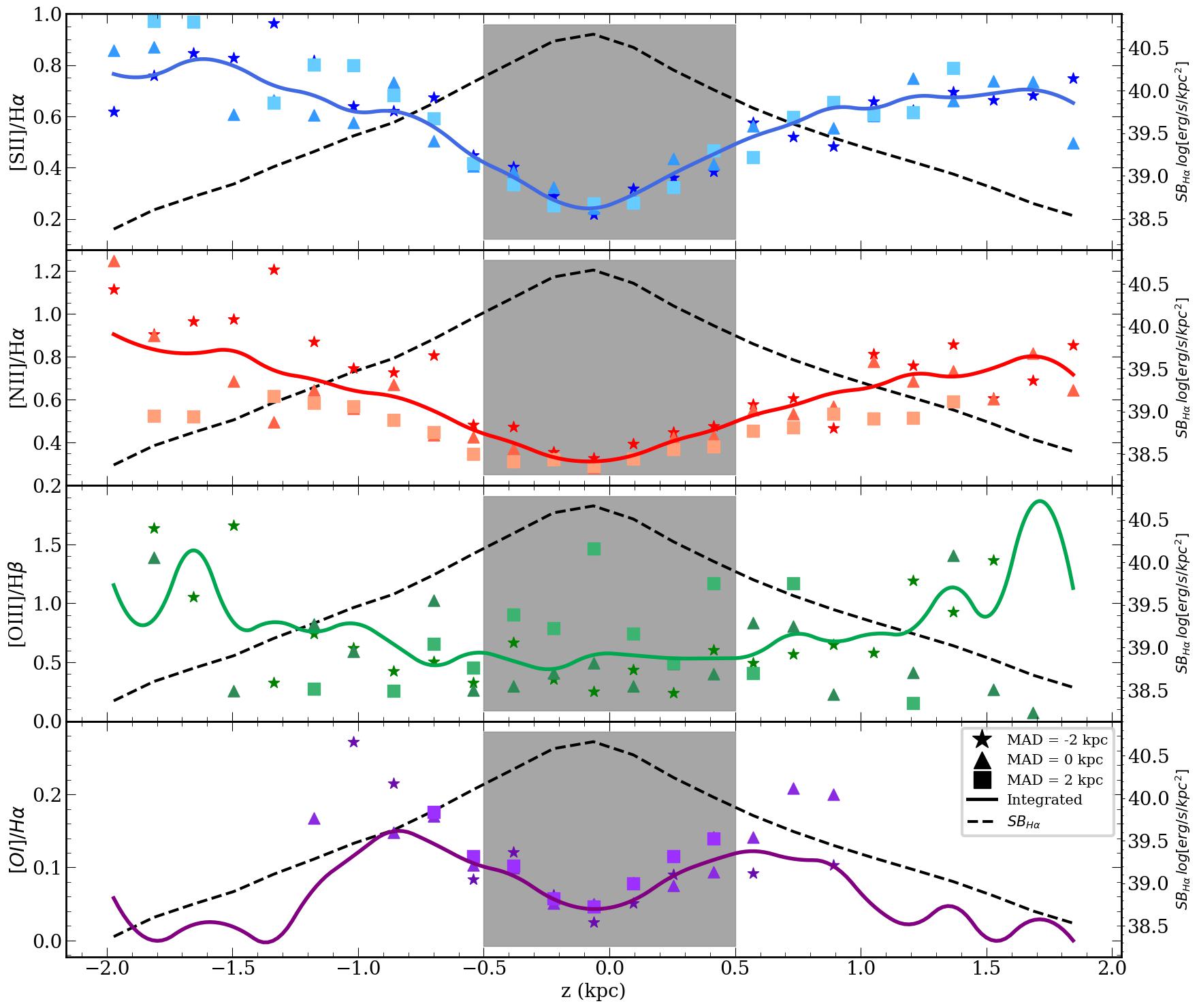}
\caption{ESO544-27 line ratio distributions of the with respect the distance from the midplane. Similarly to Figure \ref{fig:IC1553_lines}, for MAD = -2 (stars), 0 (triangles) and 2 (squares) kpc.}
\label{fig:ESO544-27_lines}     
\end{figure}

\begin{figure}[ht!]
        \centering
        \includegraphics[width=\columnwidth]{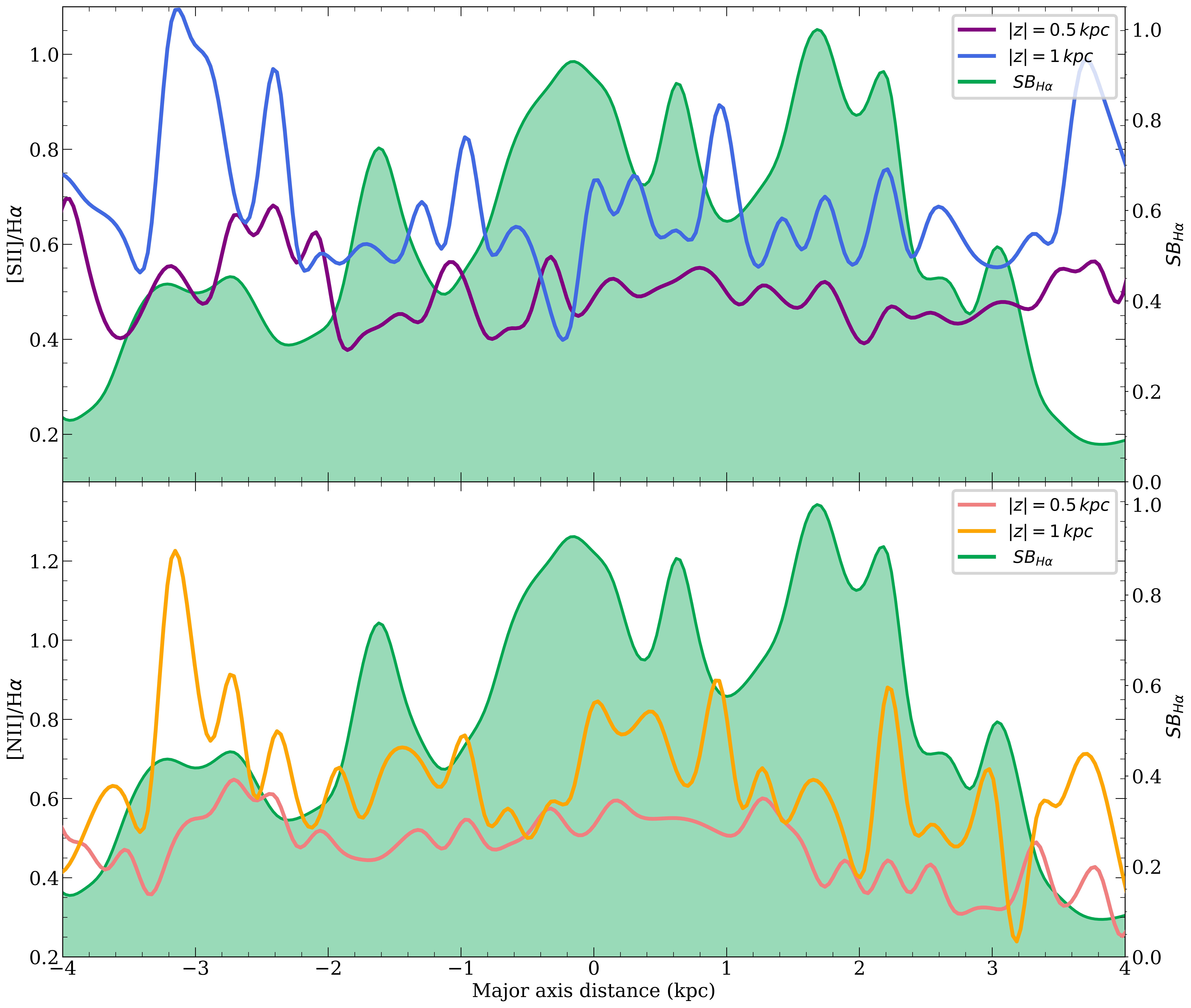}
        \caption{ESO544-27 MAD distribution for z = 0.5 and 1 kpc.}
        \label{fig:ESO544-27_MAD}
\end{figure}

The height distributions of the line ratios (Figure \ref{fig:ESO544-27_lines}) show the same behaviour than in the rest of the galaxies. The line ratios reach minimum values where the \SBha is maximum, and increase in height, up to 0.8 for \sii/\ha, 0.9 for \nii/\ha and 1.5 for \oiii/\hb. \oi/\ha reaches $\sim$0.1 at |z| = 1 kpc and then falls dramatically. The stars, triangles and squares distributions were  strategically selected to be, respectively, in the middle of two star-forming regions (MAD = -2 kpc), the centre of the galaxy (MAD = 0 kpc) and in the middle of a star forming region (MAD = 2 kpc). All distributions are similar, but the stars at MAD = -2 kpc present lower \nii/\ha, specially at higher distances from the midplane. In general, all points are mixed between them, showing no clear differences in the height distributions between different MADs. The MAD distribution (Figure \ref{fig:ESO544-27_MAD}) shows precisely the homogeneity in the eDIG; the \nii/\ha and \sii/\ha distributions remains basically flat along the major axis at both heights, being more consistent the difference between the distributions at both heights. This strengthens the evidence regarding the direct effect of the galactic plane on the ionisation structure of the eDIG. Furthermore, within the ranges of -4 kpc < MAD < -2 kpc and 3 kpc < MAD < 4 kpc, where the \SBha is lower, the disparities between the distributions at both heights are maximal for both \nii/\ha and \sii/\ha distributions. This consistency aligns with findings in other galaxies of the sample.

\begin{figure*}[t!]
\centering
\includegraphics[width=\textwidth]{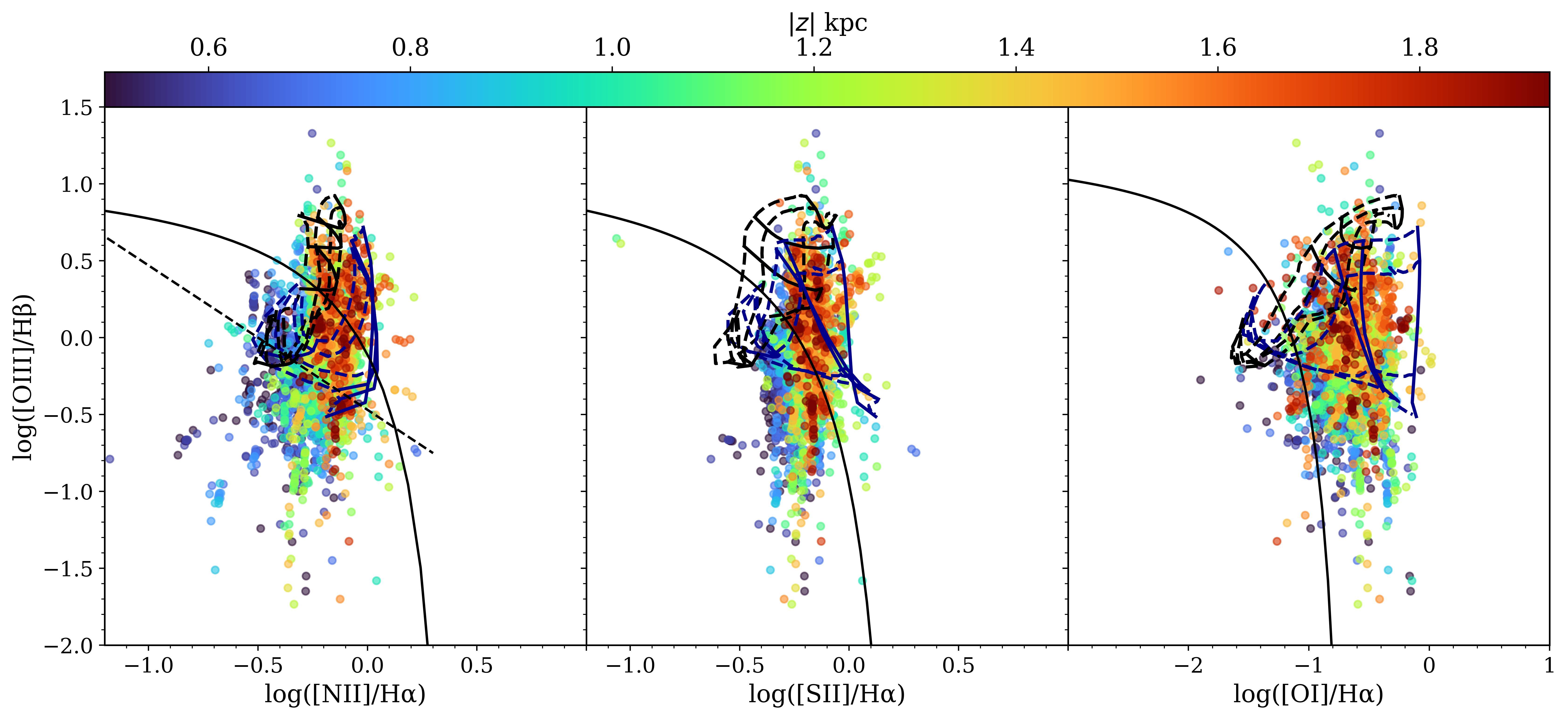} 
\caption{ESO544-27 BPT with hybrid models, similar to Figure \ref{fig:IC1553_BPT_total}. 50\% fast shocks and 50\% star formation with Z = Z$_\odot$ and q = $10^7$ cm/s.}
\label{fig:ESO544-27_BPT}     
\end{figure*}

\begin{figure}
        \includegraphics[width=\columnwidth]{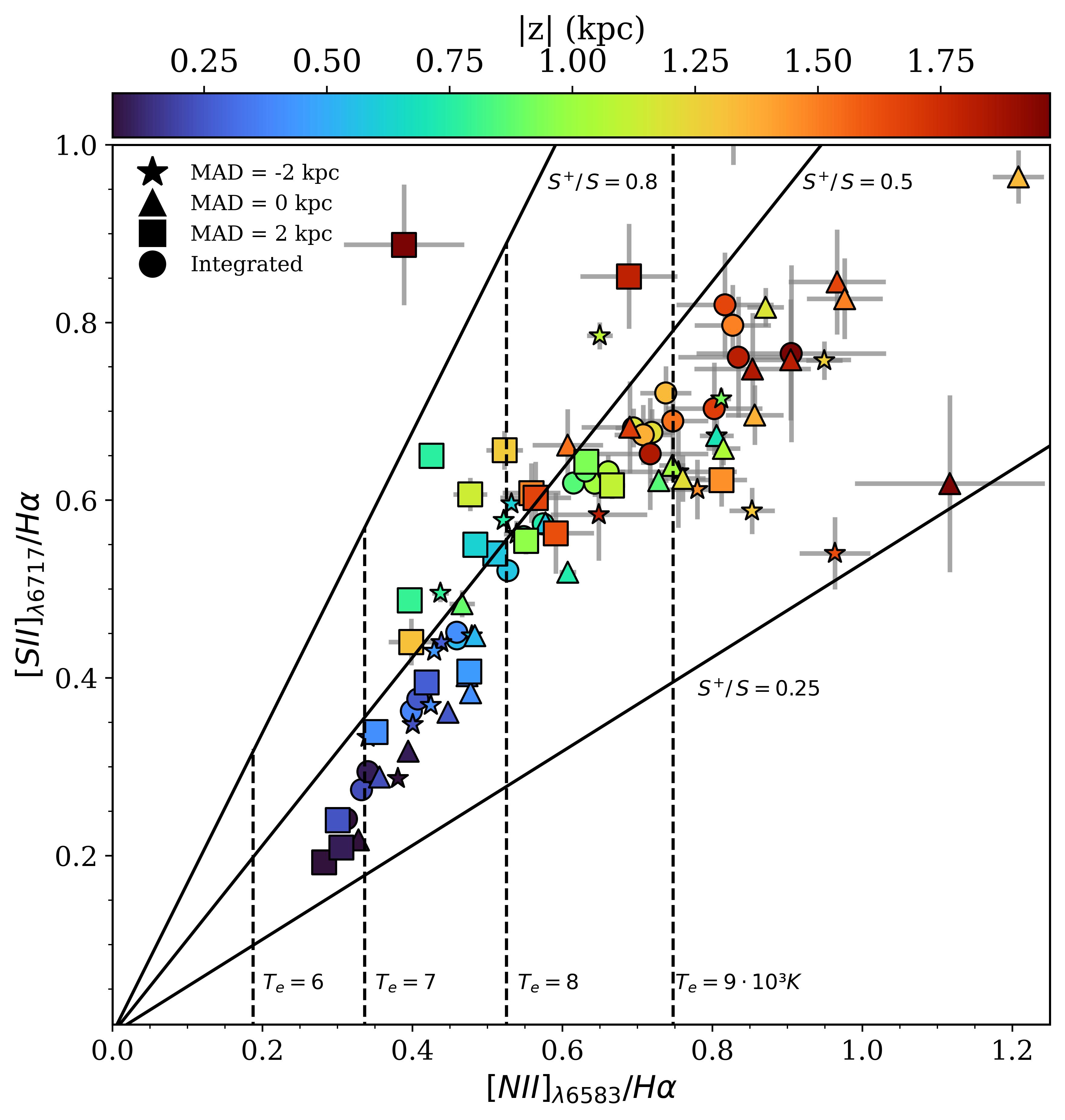}
        \caption{\nii$_{\lambda\,6583}$/\ha vs. \sii$_{\lambda\,6717}$/\ha for ESO544-27. Similarly to Figure \ref{fig:IC1553_T_S+S}, for MAD = -2 (stars), 0 (triangles) and 2 (squares) kpc.}
        \label{fig:ESO544-27_T_S+S}  
\end{figure}

The homogeneity in the eDIG for this galaxy is also evident in figures \ref{fig:ESO544-27_T_S+S} and \ref{fig:ESO544-27_BPT}. The T$_e$ increases lineally with the distance from the midplane, up to $\sim$10$^4$ K, and all the points remain close to the diagonal line corresponding to  S$^+$/S = 0.5. Also, in the BPT, the hybrid models that best fit the data correspond to 50\% of ionisation due to fast shocks, and 50\% star formation with Z = Z$_\odot$ and q = 10$^7$ cm/s. As for PGC30591, the dependence on height is more evident, as there is not a strong mixture of different ionisation conditions, with the points nearest to the galactic plane tending to align closer to the classical star-forming regime of the BPT diagram, and vice versa.

\begin{figure}
\centering
\includegraphics[width=\columnwidth]{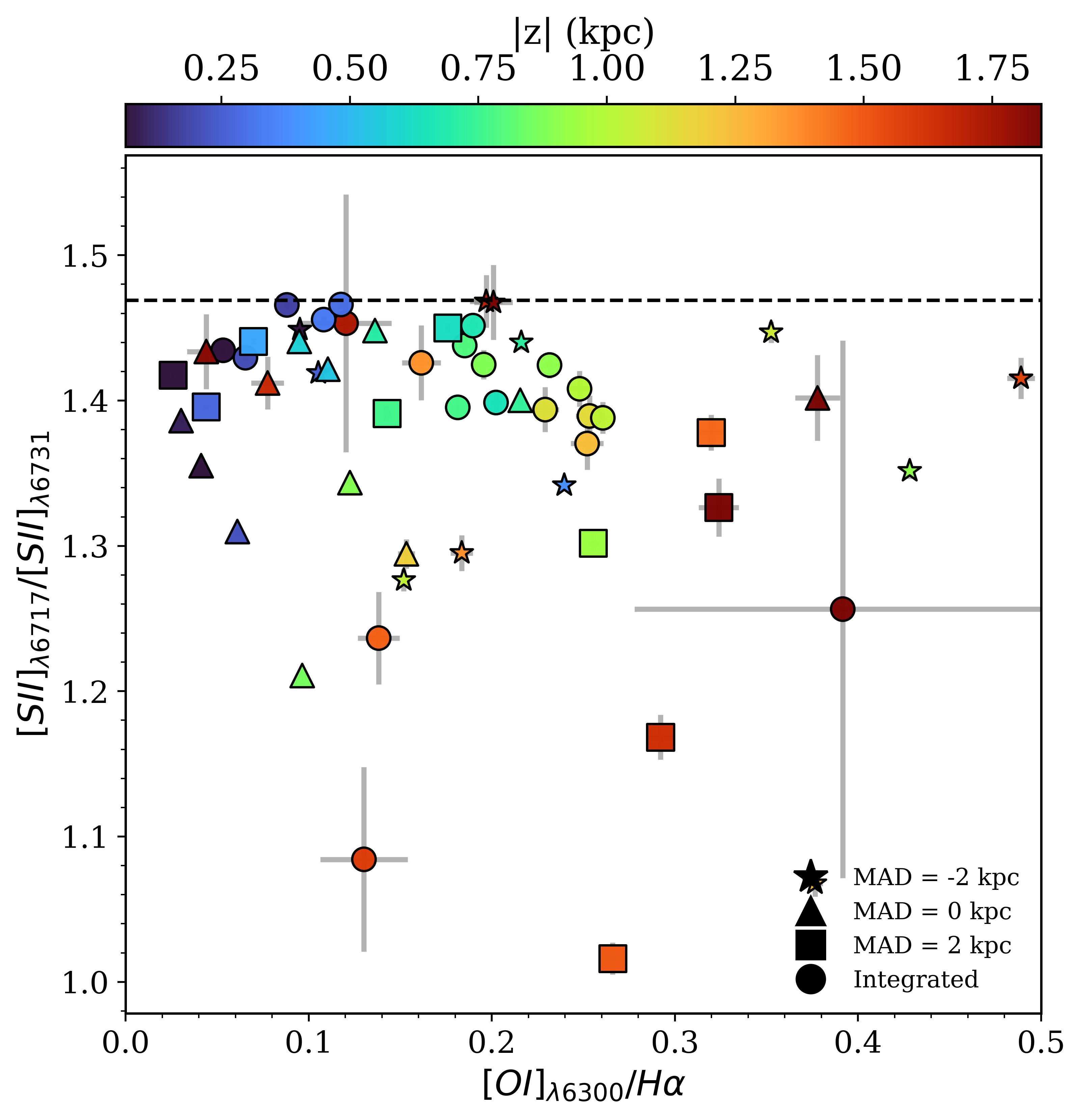} 
\caption{\sii$_{\lambda\,6717}$/\sii$_{\lambda\,6731}$ vs \oi/\ha for ESO544-27. Similar to Figure \ref{fig:IC1553_SII_OI}.}
\label{fig:ESO544-27_SII_OI}     
\end{figure}

\FloatBarrier

\section{ESO443-21}
\label{sec:Ap_D}

\begin{figure}[t!]
\centering
\includegraphics[width=\columnwidth]{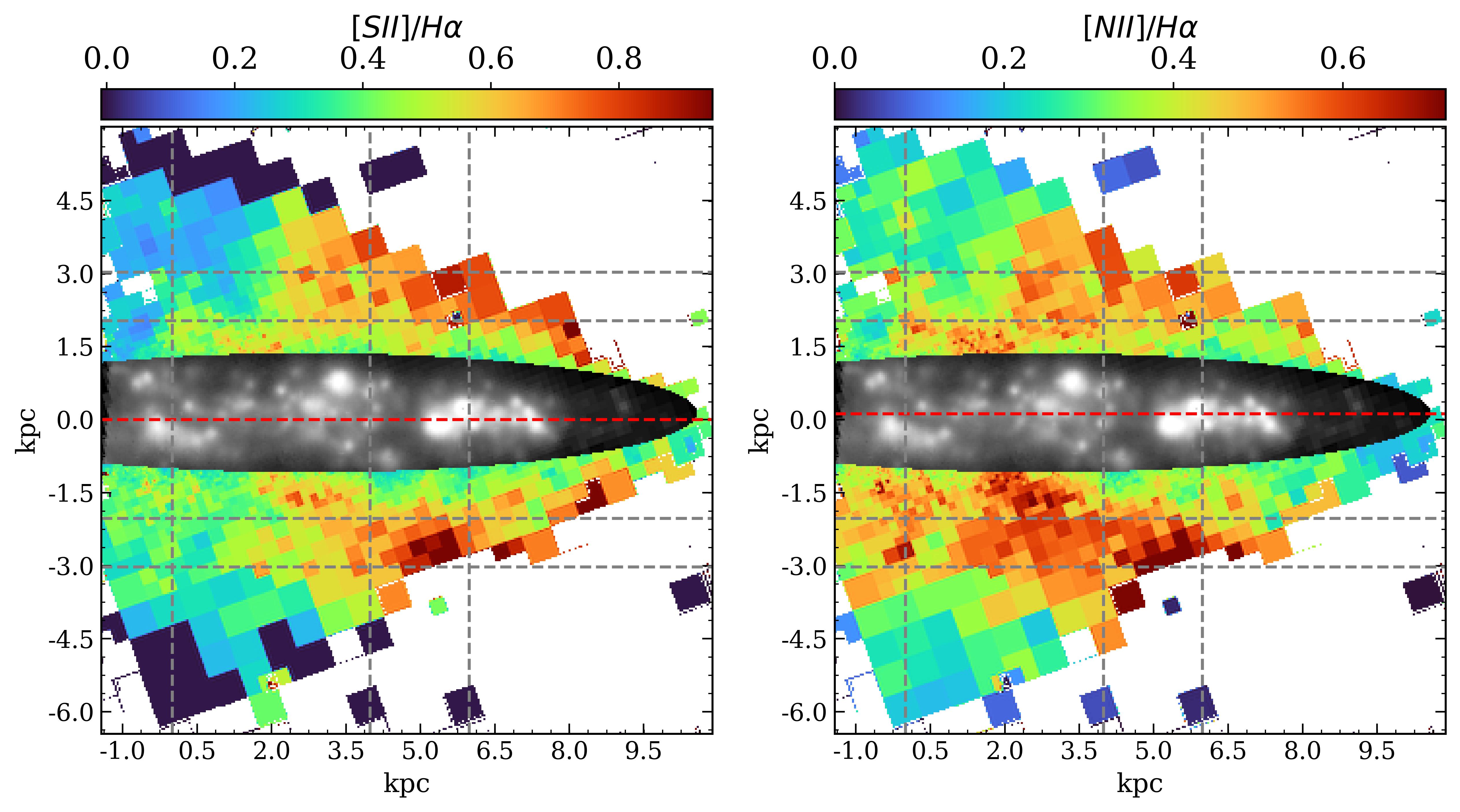}
\includegraphics[width=\columnwidth]{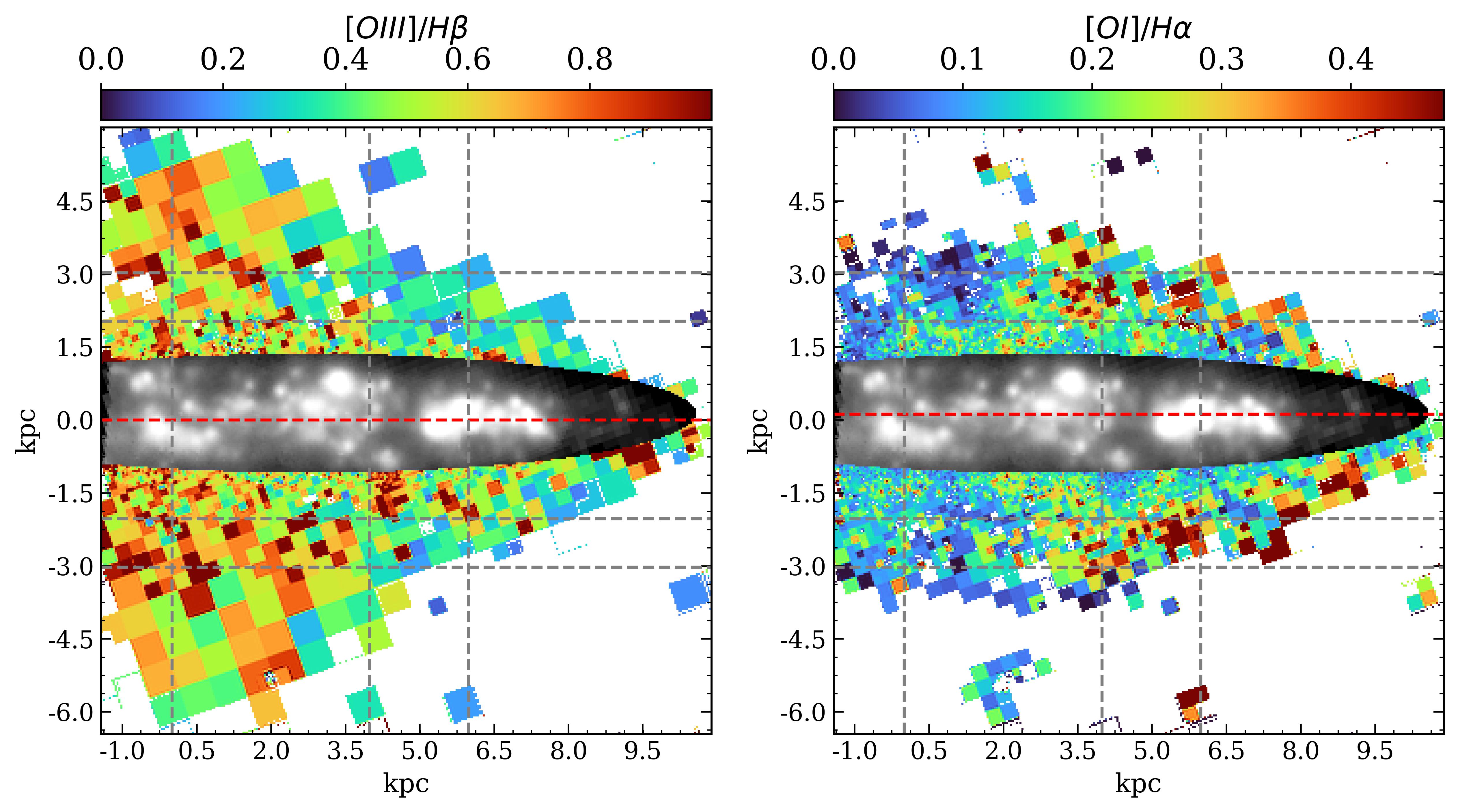} 
\parbox{\textwidth}{\captionof{figure}{ESO443-21 line ratio maps, similar to Figure \ref{fig:IC1553_maps}. The grey dashed lines indicates
the heights with respect the midplane z = $\pm$2, $\pm$3 and major axis distances MAD = 0, 4, 6 kpc.}\label{fig:ESO443-21_maps}}
\end{figure}

The late-type galaxy ESO443-21, along with the irregular galaxy IC1553, stands out as one of the most peculiar galaxies in the eBETIS sample. The excellent S/N of the data for this galaxy enables us to sample the extraplanar gas up to a distance of 4.5 kpc from the midplane. This sampling and high S/N ratio have enabled us to detect a broad knot-shape filamentous structure at a distance of 1.5 kpc, extending up to a height of 4.5 kpc within the range of -1 kpc < MAD < 2.5 kpc. This is evident not only in the \ha map of Figure \ref{fig:eBETIS_sample}, but also when examining the emission line maps in Figure \ref{fig:ESO443-21_maps}, where the structure is clearly visible as a filament characterised by low \sii/\ha and \nii/\ha ratios. 

\begin{figure}[t!]
\centering
\includegraphics[width=\columnwidth]{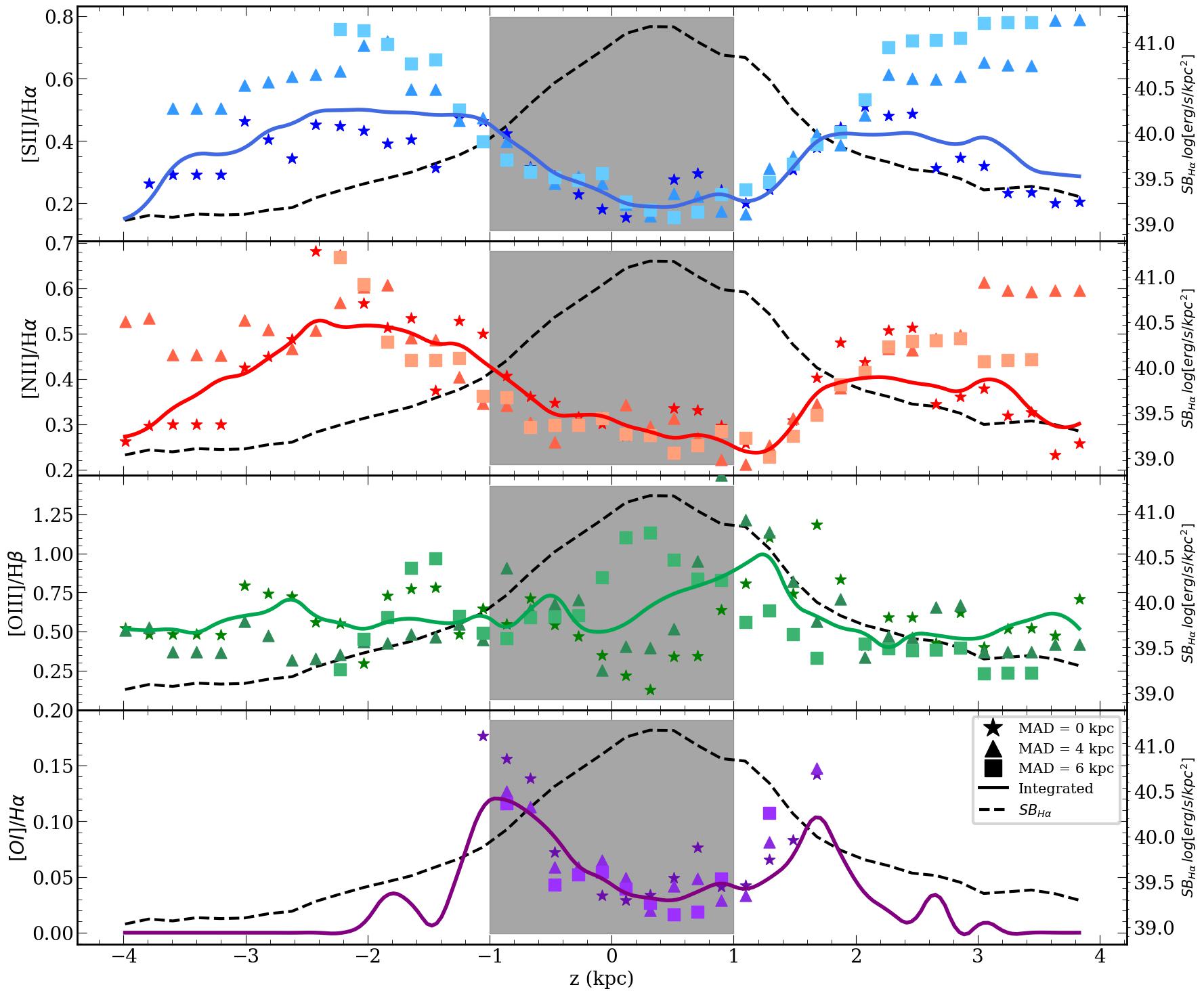}
\caption{ESO443-21 line ratio distributions of the with respect the distance from the midplane. Similarly to Figure \ref{fig:IC1553_lines}, for MAD = 0 (stars), 4 (triangles) and 6 (squares) kpc.}
\label{fig:ESO443-21_lines}     
\end{figure}

\begin{figure}
        \centering
        \includegraphics[width=\columnwidth]{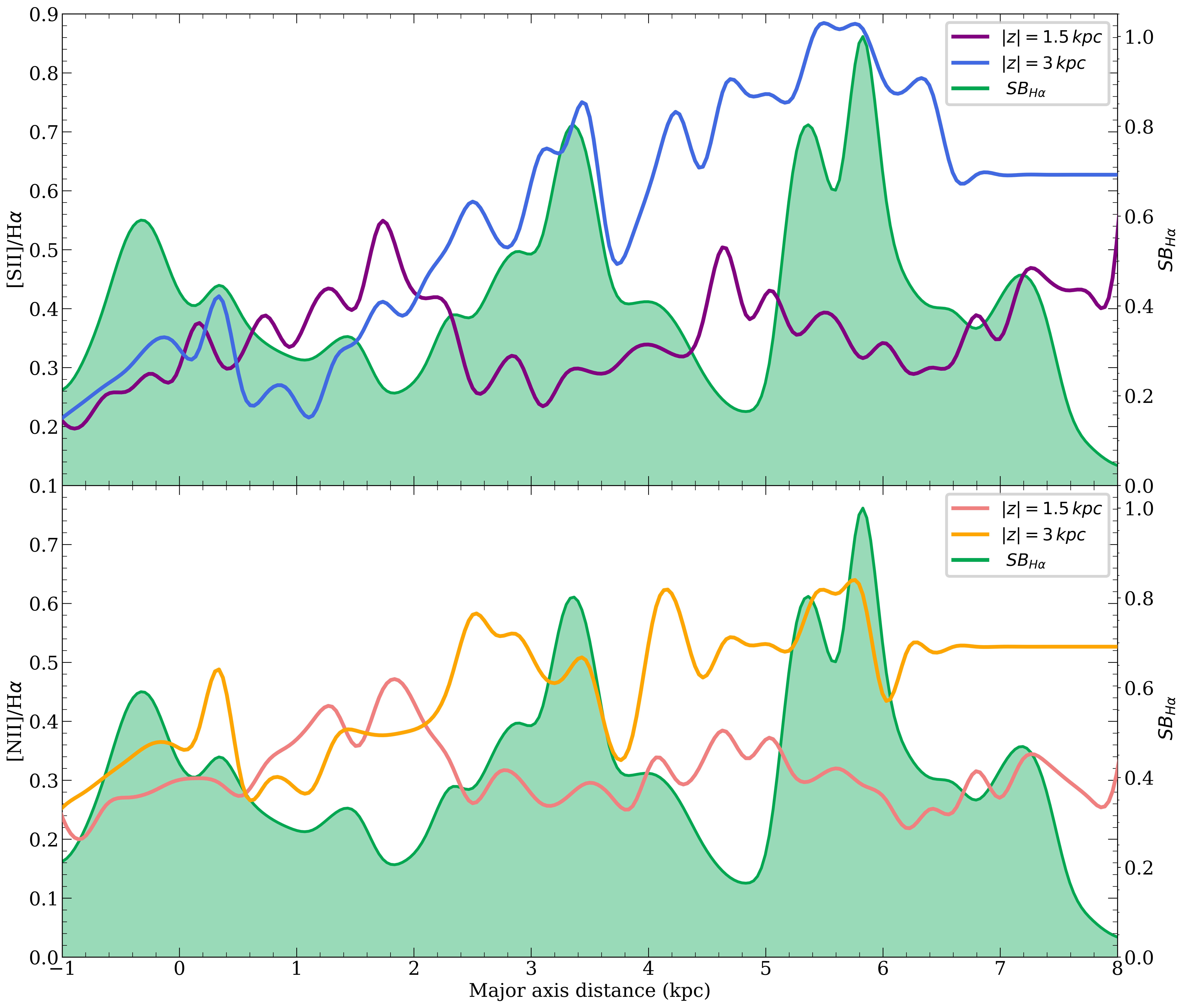}
        \caption{ESO443-21 MAD distribution for z = 1.5 and 3 kpc.}
        \label{fig:ESO443-21_MAD}
\end{figure}

The presence of this knot significantly influences the overall behaviour of the line ratio distributions and the ionisation structure of the eDIG. Figure \ref{fig:ESO443-21_lines} shows that greater distances from the midplane, the \sii/\ha and \nii/\ha  ratios deviate from the observed tendency to increase in other galaxies, and instead begin to decrease at |z| $\lesssim$ 2.5 kpc. Between 2 kpc < z < 4 kpc, where the filament is located, the decrease is more gradual, with a slight increase in the ratios observed at z $\simeq$ 3 kpc. The stars distribution at MAD = 0 kpc consistently falls below the other distributions, particularly between 2 kpc < z < 4 kpc, exhibiting differences of up to 0.6 dex for the \sii/\ha ratio and 0.4 for the \nii/\ha ratio. This emphasises the significant impact of the filament on the ratios. In Figure \ref{fig:ESO443-21_MAD}, the MAD distributions show the impact of the filament as well. The differences in the distributions between z = 1.5 kpc and z = 3 kpc for both line ratios are higher compared to the rest of the galaxies, mainly because of the larger difference in the selected heights. At z = 1.5 kpc, both \sii/\ha and \nii/\ha ratios remain relatively flat. However, at z = 3 kpc, the line ratios experience a significant decline at MAD < 2 kpc, reaching levels typical of the eDIG close to \hii regions found in other galaxies. This effect is particularly notable for \sii/\ha, which drops from 0.9 to 0.2, reversing the trend of increasing line ratios with height. 

\begin{figure}
        \includegraphics[width=\columnwidth]{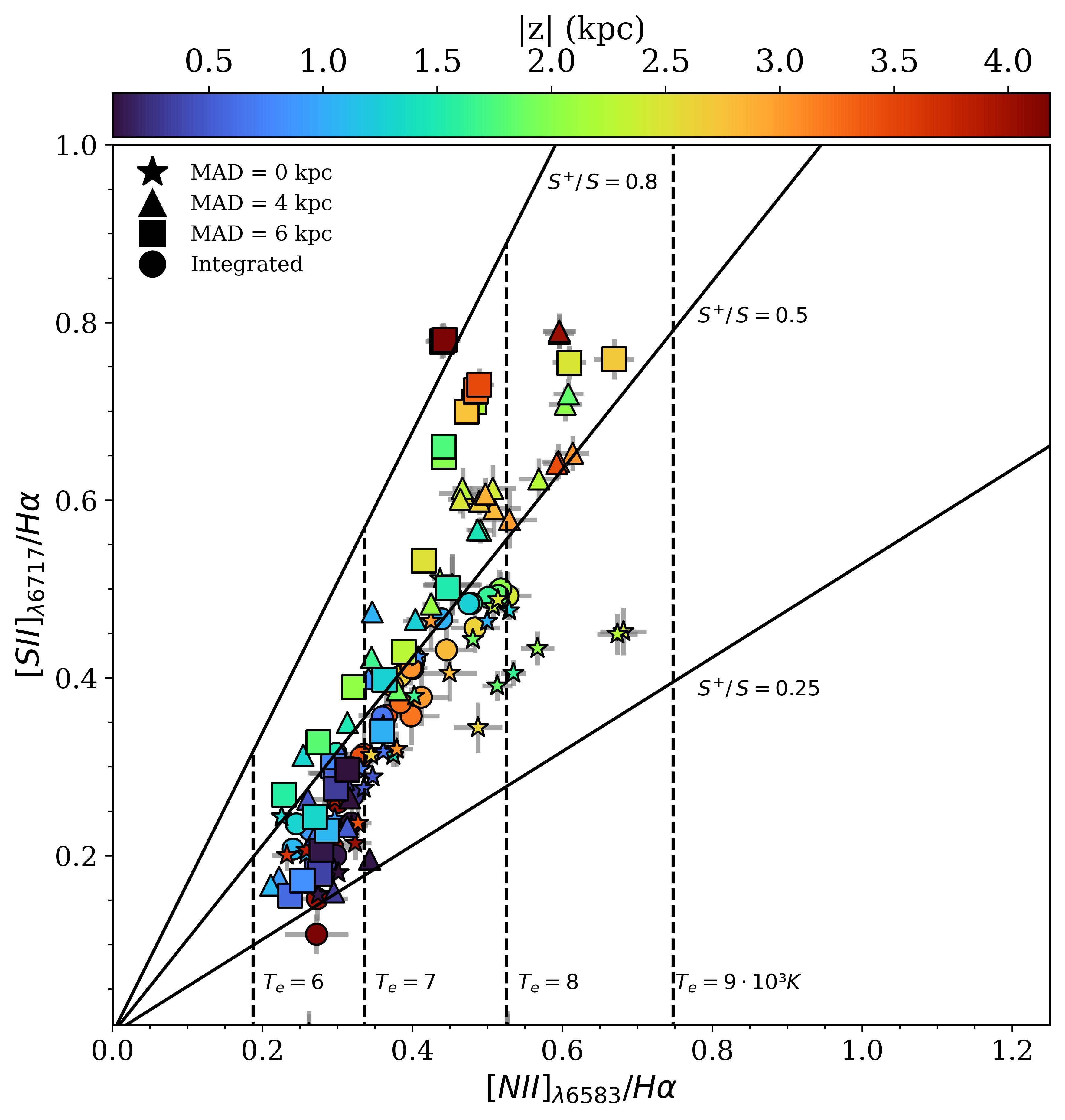}
        \caption{\nii$_{\lambda\,6583}$/\ha vs. \sii$_{\lambda\,6717}$/\ha for ESO443-21. Similarly to Figure \ref{fig:IC1553_T_S+S}, for MAD = 0 (stars), 4 (triangles) and 6 (squares) kpc.}
        \label{fig:ESO443-21_T_S+S}   
\end{figure}

\begin{figure}
\centering
\includegraphics[width=\columnwidth]{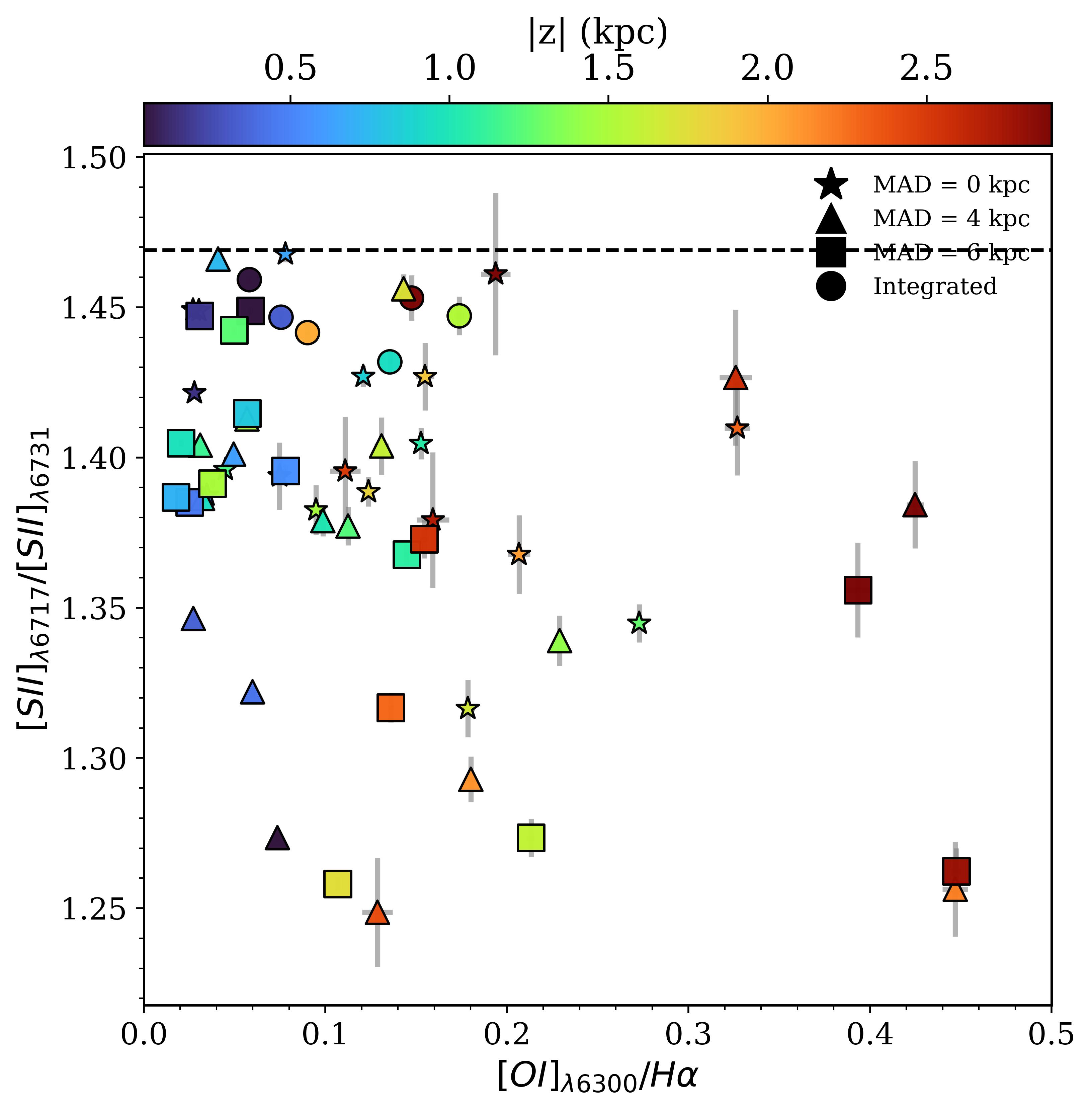} 
\caption{\sii$_{\lambda\,6717}$/\sii$_{\lambda\,6731}$ vs \oi/\ha for ESO443-21. Similar to Figure \ref{fig:IC1553_SII_OI}.}
\label{fig:ESO443-21_SII_OI}     
\end{figure}

In Figure \ref{fig:ESO443-21_T_S+S}, the general trend again indicates an increase in electron temperature with height. However, for the circles representing the integrated fluxes along the major axis, this trend disappears. The S$^+$/S fraction remains approximately constant at 0.5 but increases in height up to 0.8 for greater distances from the midplane. This finding is consistent with what we observed in other galaxies: a higher S$^+$/S ratio when the distance from the \hii regions is greater, indicating an ionisation regime favouring collisions. Additionally, for the stars corresponding to MAD = 0 kpc and between 1.5 kpc < z < 3.5 kpc (where the knot is located), the S$^+$/S ratio drops to $\sim$ 0.3, suggesting a regime more similar to ionisation due to star formation.

\begin{figure*}[ht!]
\centering
\includegraphics[width=\textwidth]{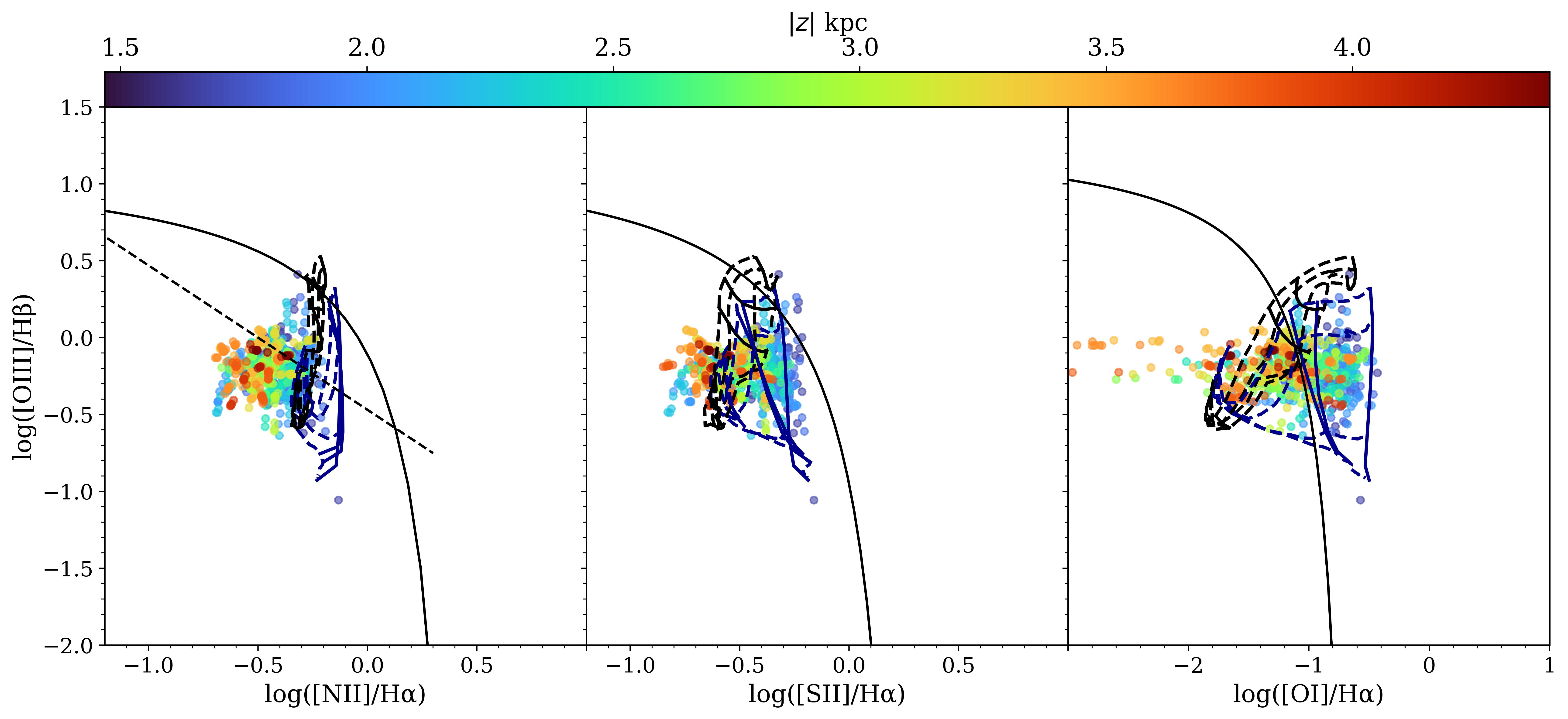} 
\includegraphics[width=\textwidth]{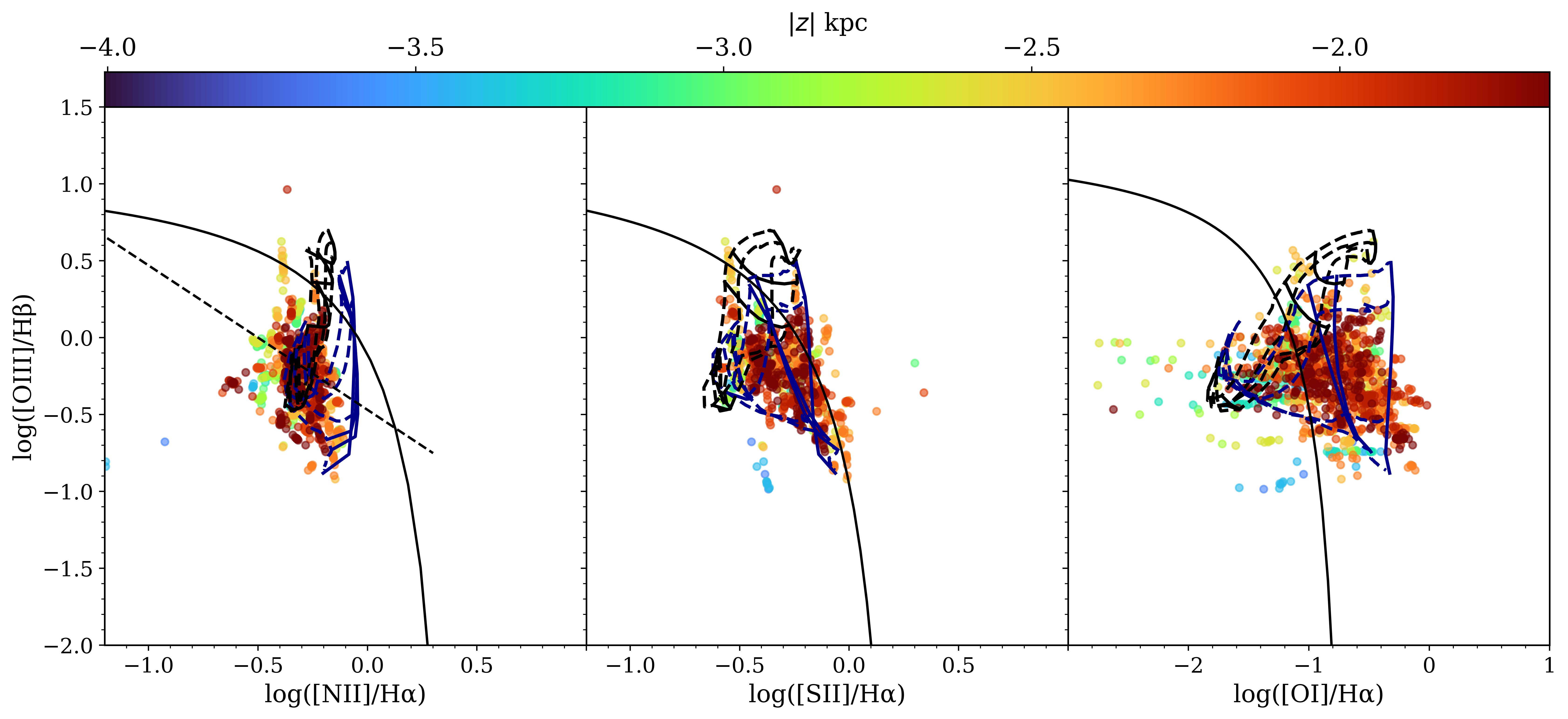} 
\caption{ESO443-21 BPT with hybrid models, similar to Figure \ref{fig:IC1553_BPT_total}. Up: BPT for bins at z > 0. Hybrid models correspond to 20\% fast shocks and 80\% star formation with Z = Z$_\odot$ and q = $10^7$ cm/s. Down: BPT for bins at z < 0. Hybrid models correspond to 30\% fast shocks and 70\% star formation with Z = 2Z$_\odot$ and q = $10^7$ cm/s.}
\label{fig:ESO443-21 BPT}     
\end{figure*}

This last figure suggests an asymmetry in the ionisation of the eDIG due to the filament, indicating different regimes when considering z < 0 or z > 0. To verify this, we present the BPT diagrams with the hybrid models separately for z < 0 and for z > 0.
The best fit hybrid models for the points at z > 0 correspond to 20\% of ionisation due to fast shocks, and 80\% star formation with Z = 2Z$_\odot$ and q = 10$^7$ cm/s. For z < 0 correspond to 30\% of ionisation due to fast shocks, and 70\% star formation with the same parameters. The presence of the knot only affects an additional contribution of star formation to the overall ionisation mechanisms by up to 10\%.

\FloatBarrier

\section{ESO469-15}
\label{sec:Ap_E}
\raggedbottom

\begin{figure}[t!]
\centering
\includegraphics[width=\columnwidth]{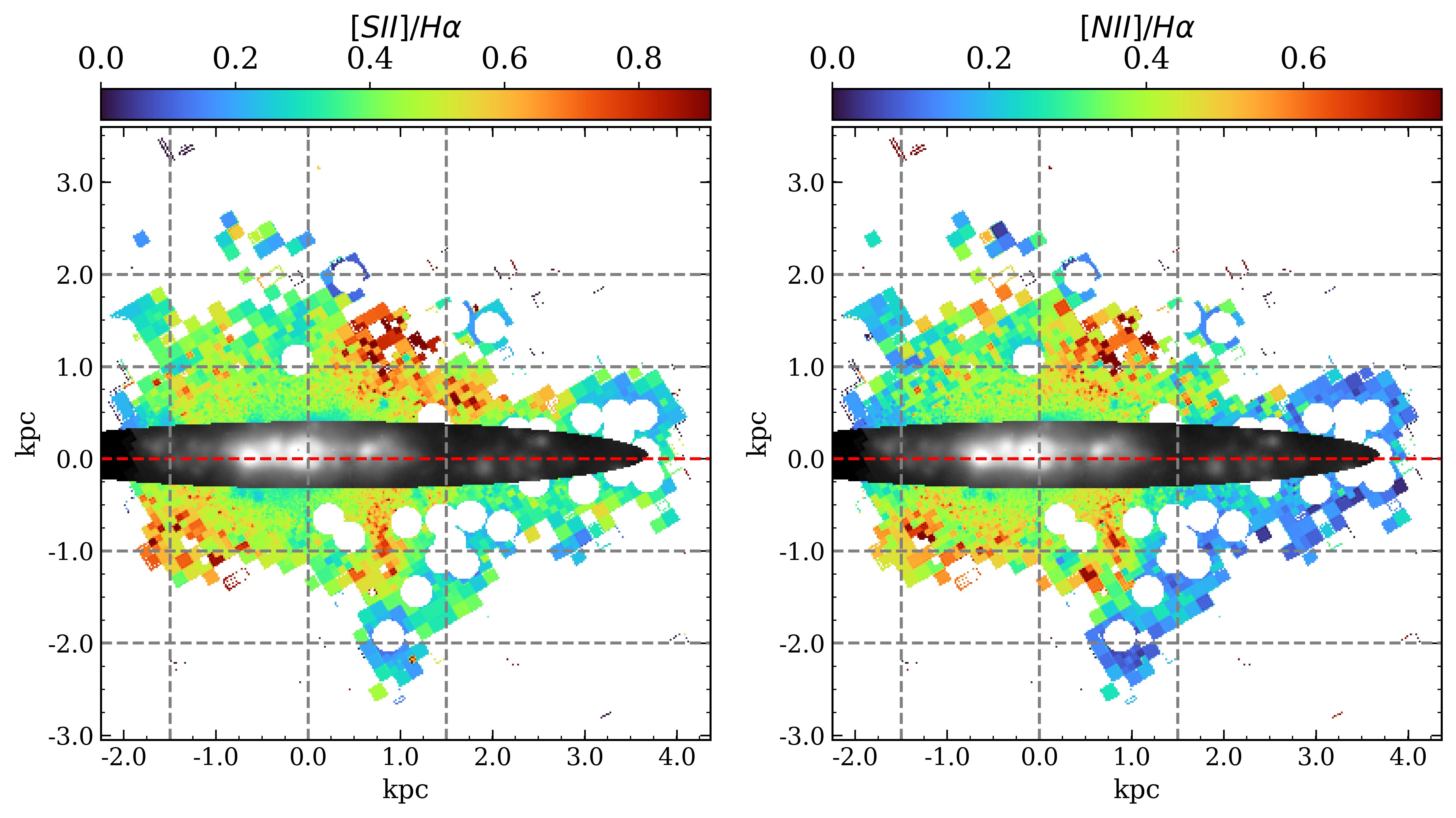}
\includegraphics[width=\columnwidth]{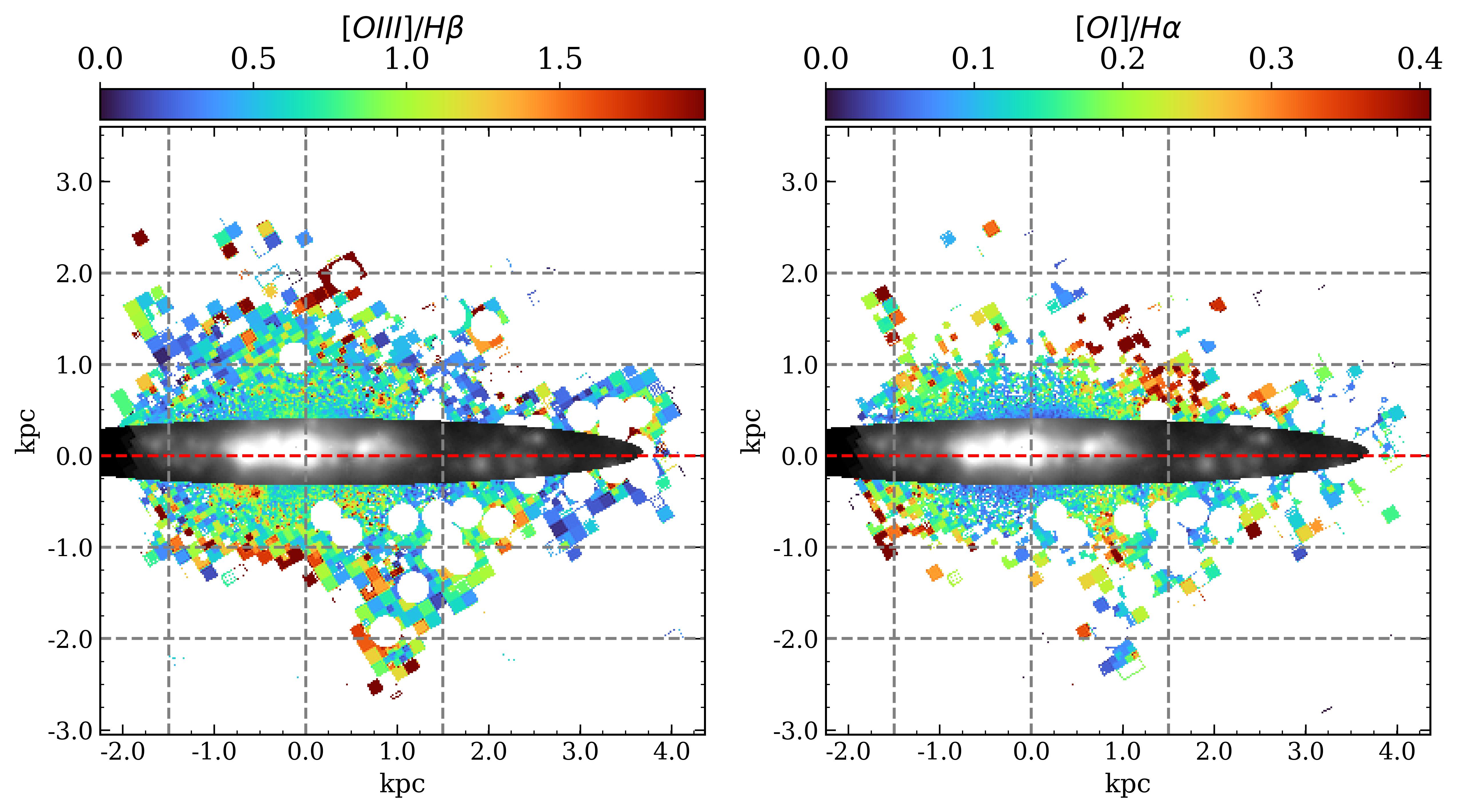} 
\captionof{figure}{ESO469-15 line ratio maps, similar to Figure \ref{fig:IC1553_maps}. The grey dashed lines indicates
the heights with respect the midplane z = $\pm$1, $\pm$2 and major axis distances MAD = -1.5, 0, 1.5 kpc.}\label{fig:ESO469-15_maps}
\end{figure}

\begin{figure}
\centering
\includegraphics[width=\columnwidth]{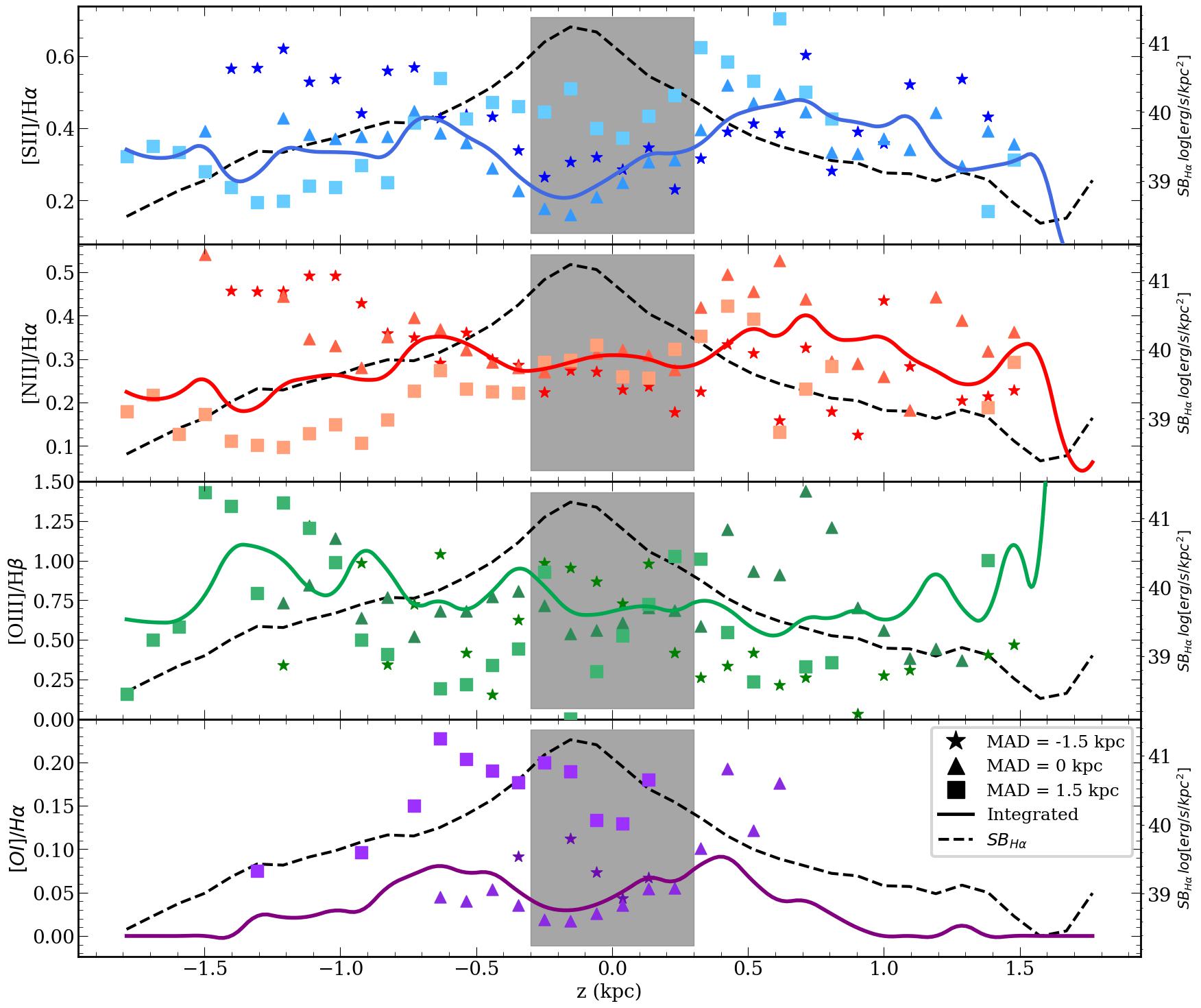}
\caption{ESO469-15 line ratio distributions of the with respect the distance from the midplane. Similarly to Figure \ref{fig:IC1553_lines}, for MAD = -1.5 (stars), 0 (triangles) and 1.5 (squares) kpc.}
\label{fig:ESO469-15_lines}     
\end{figure}

The galaxy ESO469-15 exhibits a noticeable presence of discrete extraplanar \hii regions (also reported by other authors, e.g. \citealt{2022A&A...659A.153R}). Before performing the analysis, we masked all the extraplanar \hii regions using the {\sc python} routine {\sc DAOStarFinder} from the {\sc PhotUtils}\footnote{\href{https://photutils.readthedocs.io/en/stable/api/photutils.detection.DAOStarFinder.html}{{\sc DAOStarFinder}} documentation.} package. Figure \ref{fig:ESO469-15_maps} shows the line maps after the \hii regions removal. The line ratios near to the \hii regions are even lower than those close to the bright galactic centre, being considerably low compared to the rest of the eDIG, specially at z < 0, where most of the extraplanar \hii regions are located. The effect of the extraplanar \hii regions is clear in the line distributions (Figure \ref{fig:ESO469-15_lines}). The trend of the integrated distribution changes continuously along the distance from the midplane, and squares distributions (at MAD = 1.5, where most of the extraplanar \hii regions are located) present considerably lower line ratios in comparison with the rest of the distributions at z < 0, up to 0.6 for the \sii/\ha and \nii/\ha ratios. \oiii/\hb ratio shows a mixed tendency for every distribution, as for the rest of the galaxies. Finally, the \oi/\ha ratio shows lower values at MAD = 0 kpc and z $\simeq$ 0 kpc, since for at higher distances the S/N of the \oi line is not sufficient to sample those regions.

\begin{figure}
        \centering
        \includegraphics[width=\columnwidth]{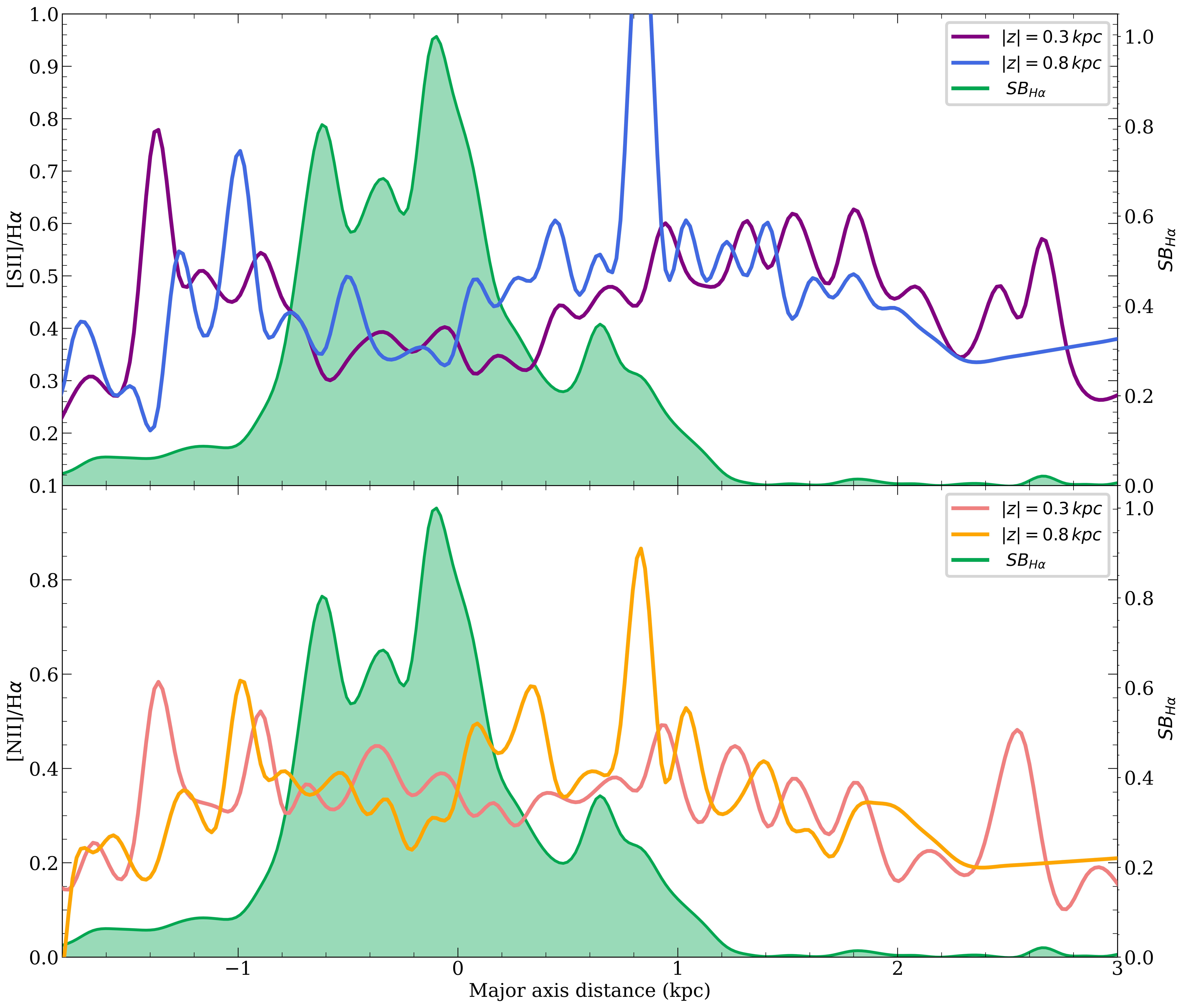}
        \caption{ESO469-15 MAD distribution for z = 0.3 and 0.8 kpc.}
        \label{fig:ESO469-15_MAD}
\end{figure}

The MAD distributions in Figure \ref{fig:ESO469-15_MAD} reinforce what was observed in IC1553, and generally across the entire sample. The \SBha along the major axis of the galaxy is notably more prominent between -1 kpc < MAD < 1 kpc. Subsequently, the \sii/\ha and \nii/\ha ratios experience a sudden decrease at these MADs. 

The T$_e$ in the eDIG (\ref{fig:ESO469-15_T_S+S}) is also influenced by the presence of extraplanar \hii regions, being notably lower for all the eDIG compared to other galaxies, ranging only between 6 and 8 $\cdot10^3$ K. 

\begin{figure*}[t!]
\centering
\includegraphics[width=\textwidth]{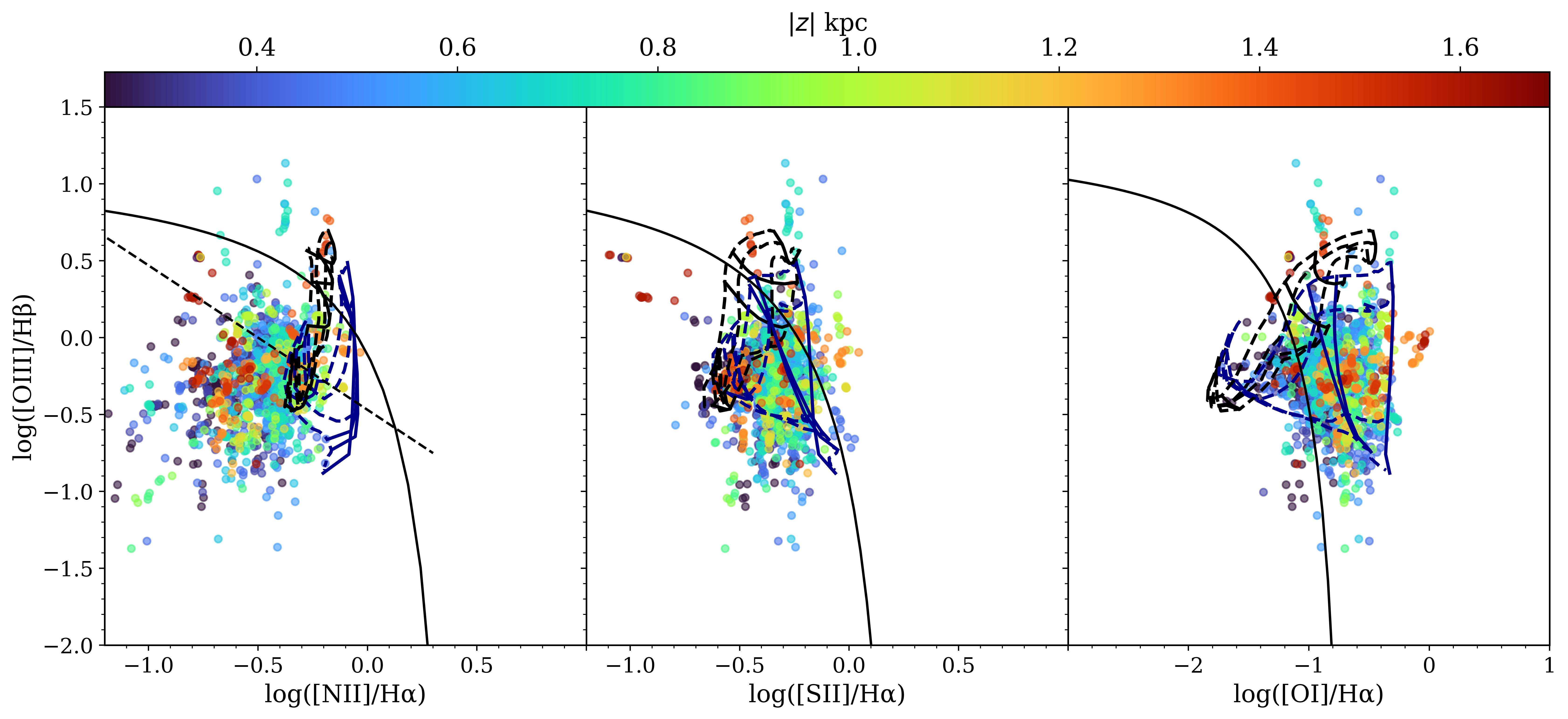} 
\caption{ESO469-15 BPT with hybrid models, similar to Figure \ref{fig:IC1553_BPT_total}. 30\% fast shocks and 70\% star formation with Z = 2Z$_\odot$ and q = $10^7$ cm/s.}
\label{fig:ESO469-15 BPT}     
\end{figure*}

\begin{figure}
        \raggedright
        \includegraphics[width=\columnwidth]{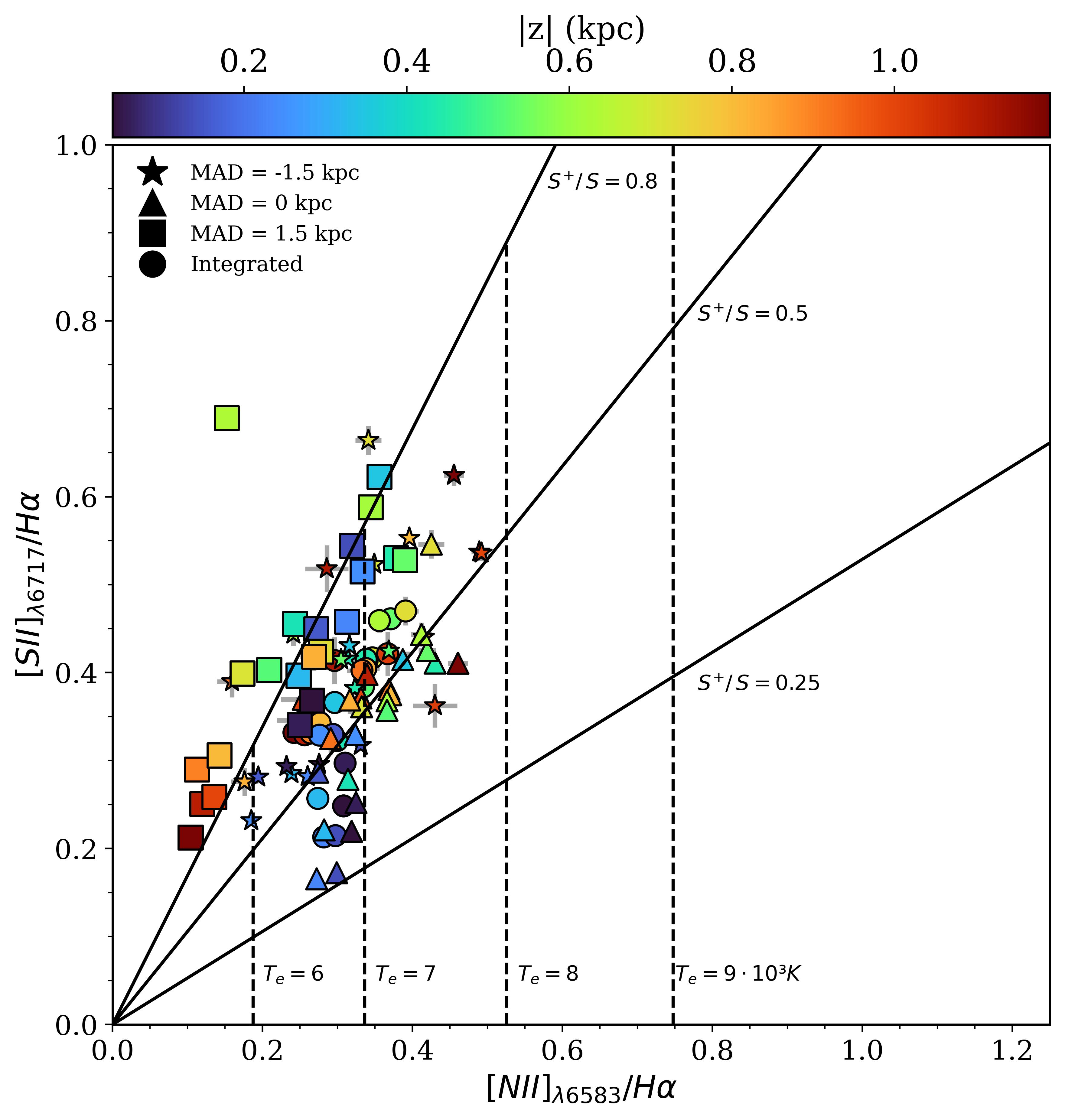}
        \caption{\nii$_{\lambda\,6583}$/\ha vs. \sii$_{\lambda\,6717}$/\ha for ESO469-15. Similarly to Figure \ref{fig:IC1553_T_S+S}, for MAD = -1.5 (stars), 0 (triangles) and 1.5 (squares) kpc.}
        \label{fig:ESO469-15_T_S+S}
\end{figure}

In addition to these results, the hybrid models in the BPT diagram for all the eDIG (Figure \ref{fig:ESO469-15 BPT}) indicate only a 30\% contribution of fast shocks in the ionisation mechanisms of the eDIG. This represents the lowest contribution in the sample when the data in the BPT diagram is not restricted by z or MAD.

\begin{figure}
\centering
\includegraphics[width=\columnwidth]{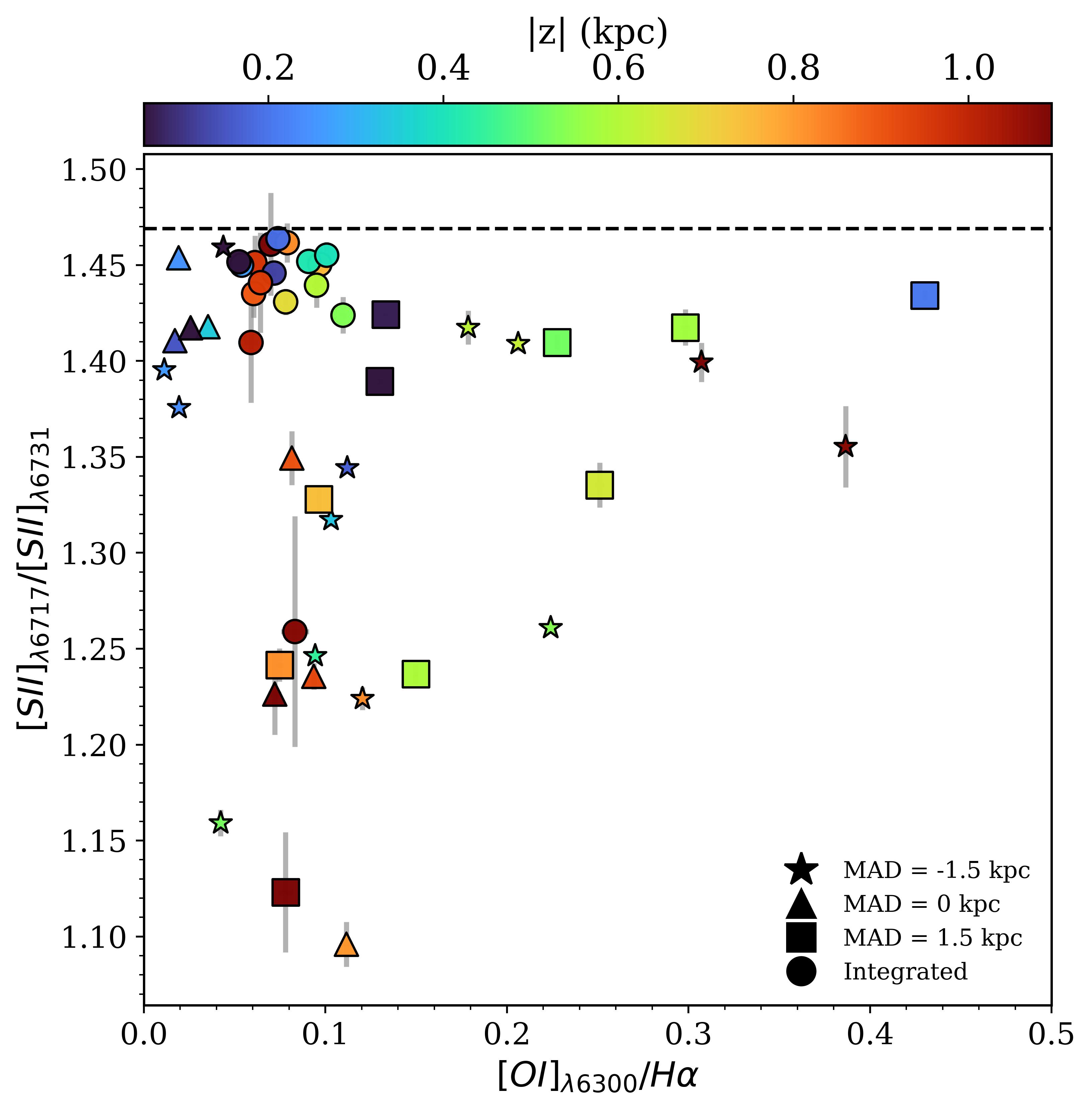} 
\caption{\sii$_{\lambda\,6717}$/\sii$_{\lambda\,6731}$ vs \oi/\ha for ESO469-15. Similar to Figure \ref{fig:IC1553_SII_OI}.}
\label{fig:SO469-15_SII_OI}     
\end{figure}

\FloatBarrier

\section{ESO157-49}
\label{sec:Ap_F}

\begin{figure}[t!]
\centering
\includegraphics[width=\columnwidth]{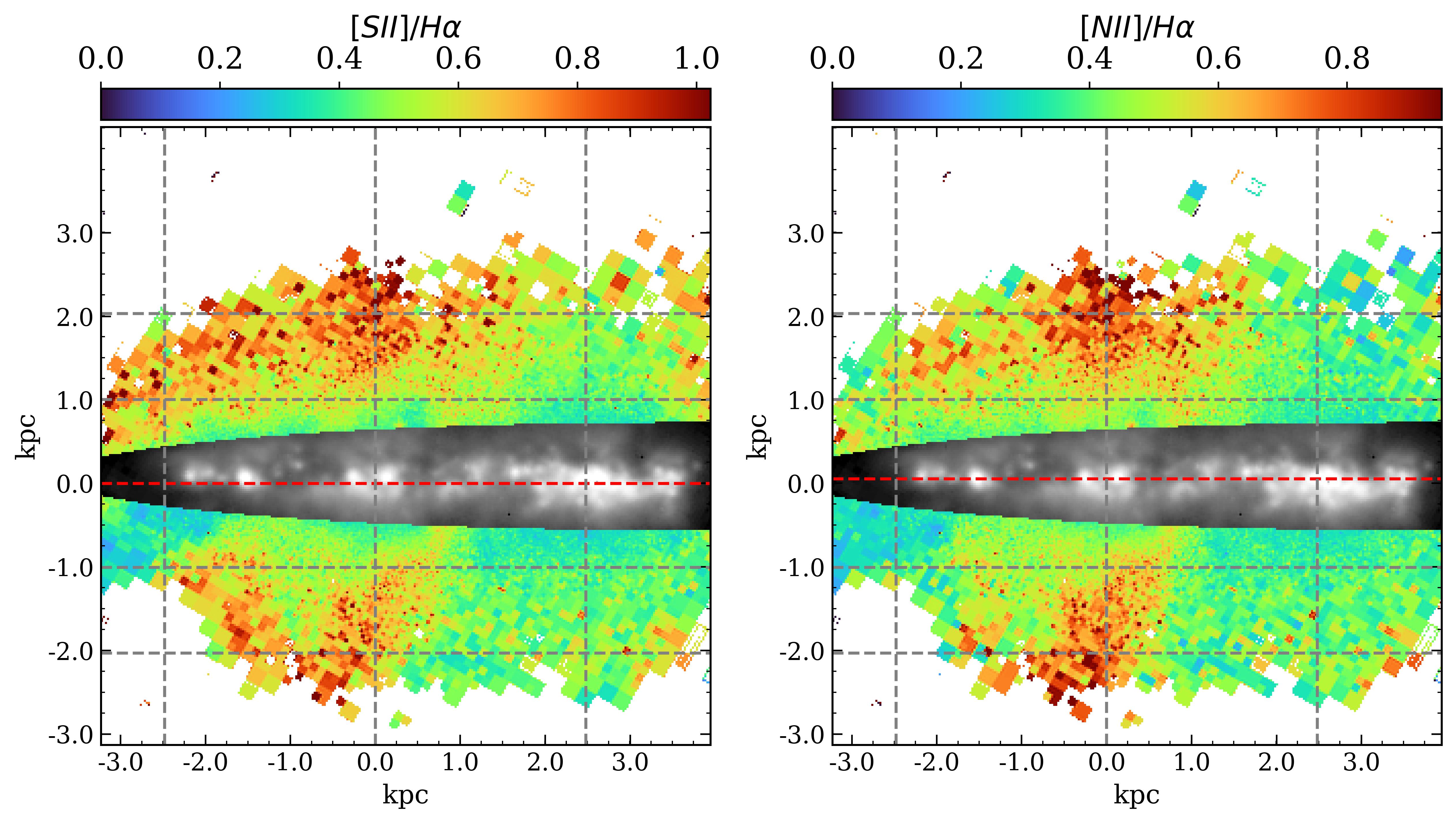}
\includegraphics[width=\columnwidth]{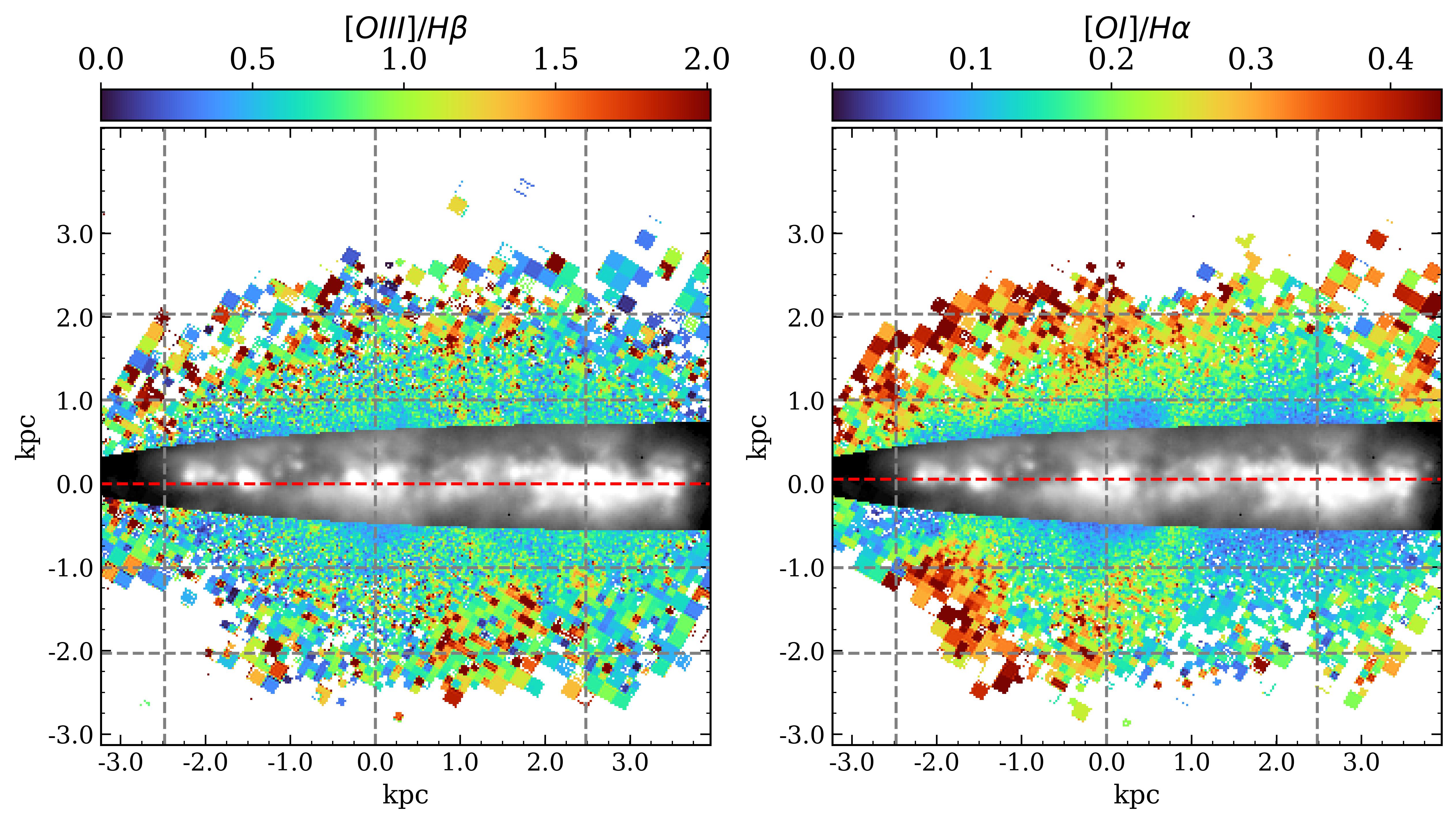} 
\captionof{figure}{ESO157-49 line ratio maps, similar to Figure \ref{fig:IC1553_maps}. The grey dashed lines indicates
the heights with respect the midplane z = $\pm$1, $\pm$2 and major axis distances MAD = -2.5, 0, 2.5 kpc.}\label{fig:ESO157-49_maps}
\end{figure}

\begin{figure}[t!]
\centering
\includegraphics[width=\columnwidth]{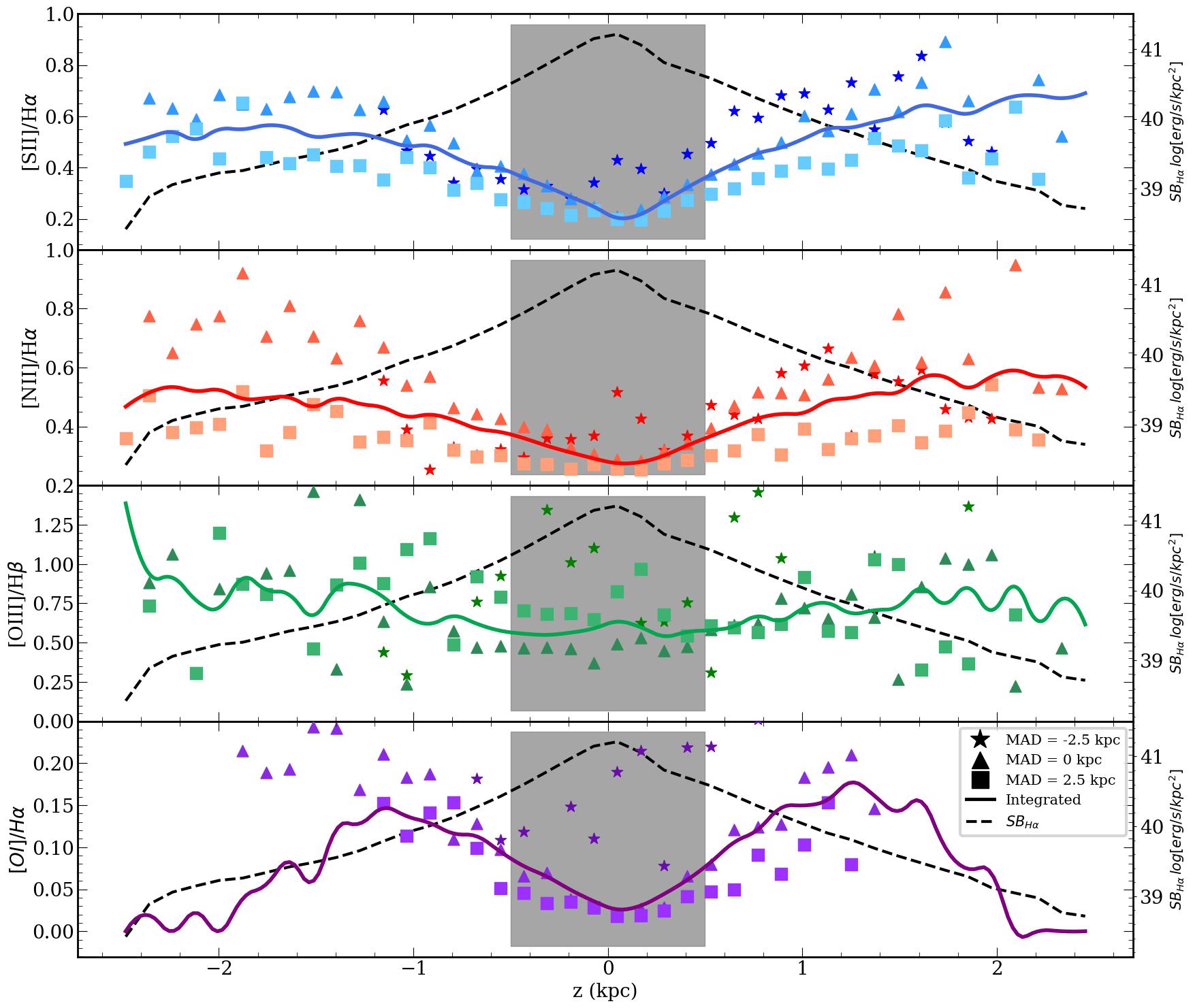}
\caption{ESO157-49 line ratio distributions of the with respect the distance from the midplane. Similarly to Figure \ref{fig:IC1553_lines}, for MAD = -2.5 (stars), 0 (triangles) and 2.5 (squares) kpc.}
\label{fig:ESO157-49_lines}     
\end{figure}

\begin{figure}
        \centering
        \includegraphics[width=\columnwidth]{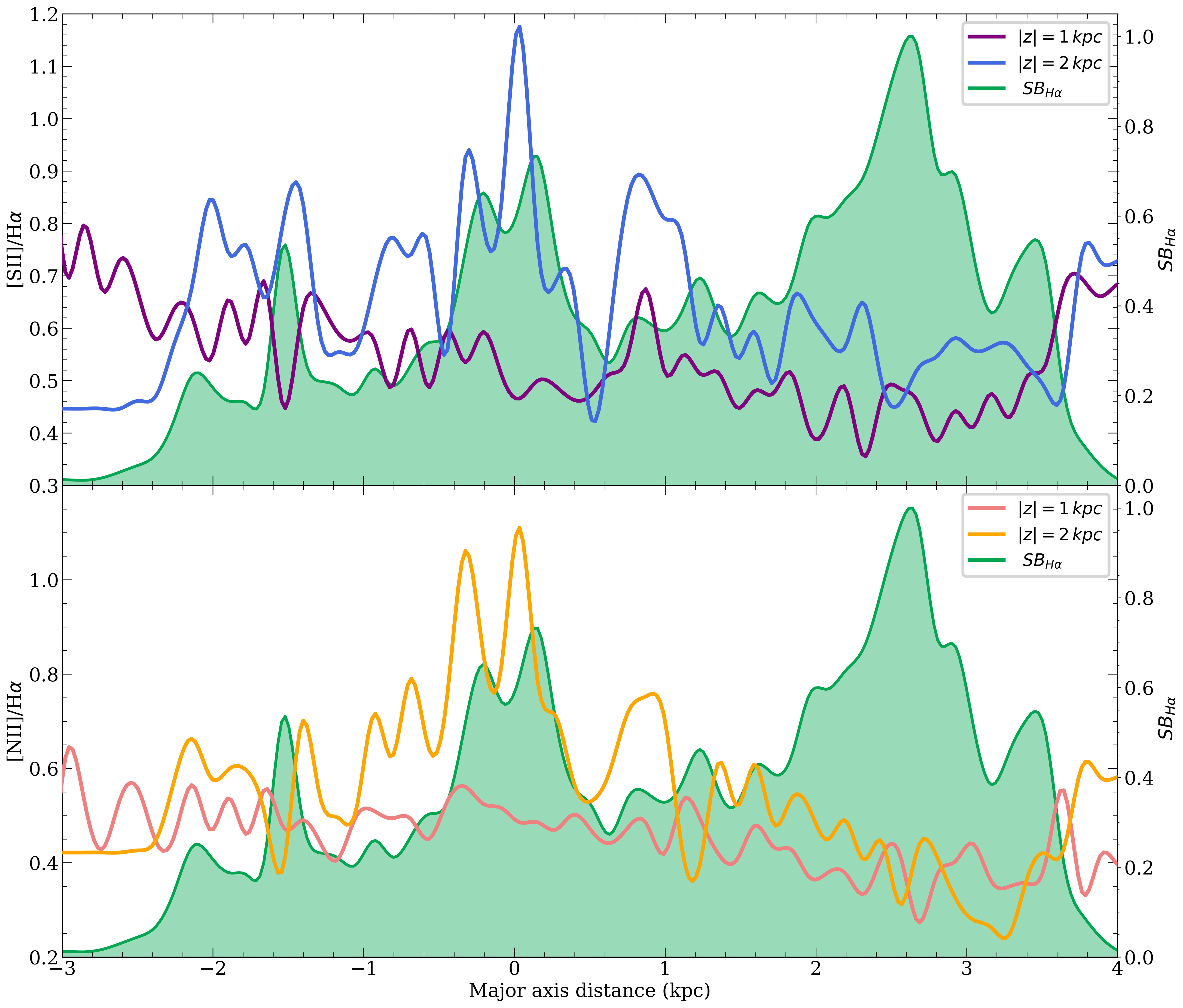}
        \caption{ESO157-49 MAD distribution for z = 1 and 2 kpc.}
        \label{fig:ESO157-49_MAD}
\end{figure}

ESO157-49 is the only galaxy in the sample to have nearby neighbours \citep{2017ApJ...843...16K}. It was observed that its closest neighbour, the dwarf galaxy ESO157-48 located southwest of the galaxy, has influence on its kinematics, that can result in an asymmetry in the properties of the extraplanar ionised gas \citep{2022A&A...659A.153R}. Figure \ref{fig:ESO157-49_maps} shows precisely this asymmetry. ESO157-49 presents a more uniform emission from the galactic plane, similar to ESO544-27 but with a slightly brighter emission observed between -1.5 kpc $\lesssim$ MAD $\lesssim$ 3 kpc. However, the extraplanar ionised structure at z > 0 is clearly different from that at z < 0, being the structure of the eDIG at z < 0 apparently tilted and shifted compared to the eDIG at z > 0. The line ratios height distributions (\ref{fig:ESO157-49_lines}) show that at z < 0 the \nii/\ha, \sii/\ha and \oi/\ha line ratios reach lower maximum values than at z > 0 (up to 0.2 on every ratio). Furthermore, at MAD = 2.5 kpc, where the \SBha is higher, the \nii/\ha and \sii/\ha line ratios are consistently remain lower, as observed for the rest of the sample (also seen in the MAD distributions; Figure \ref{fig:ESO157-49_MAD}).

\begin{figure}
    \raggedright
        \includegraphics[width=\columnwidth]{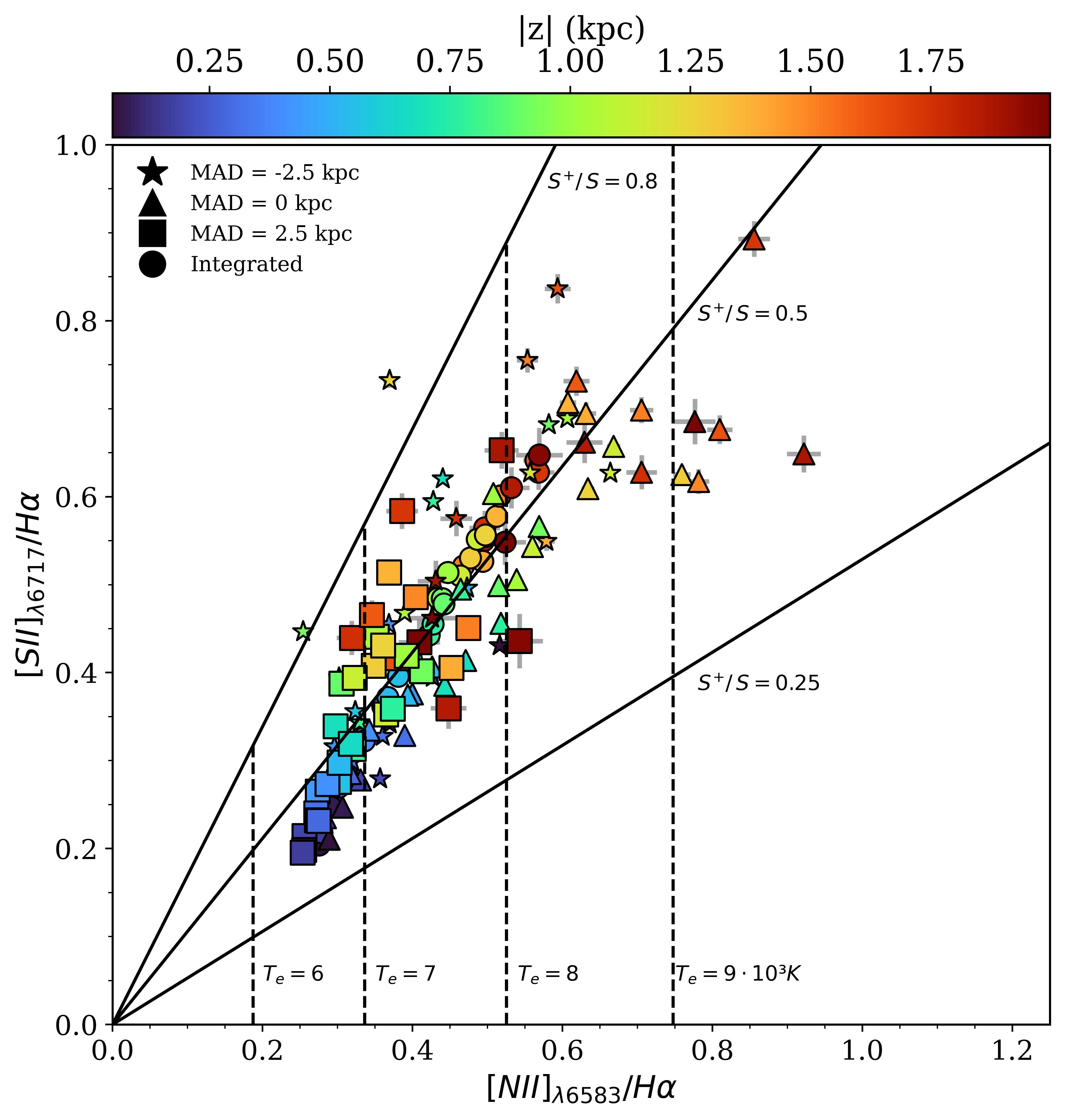}
        \caption{\nii$_{\lambda\,6583}$/\ha vs. \sii$_{\lambda\,6717}$/\ha for ESO157-49. Similarly to Figure \ref{fig:IC1553_T_S+S}, for MAD = -2.5 (stars), 0 (triangles) and 2.5 (squares) kpc.}
        \label{fig:ESO157-49_T_S+S}   
\end{figure}

The T$_e$ (see Figure \ref{fig:ESO157-49_T_S+S}) increases in height as usual, up to $\sim$9$\cdot$10$^3$ K, and the ionisation fraction S$^+$/S remains approximately constant at $\sim$0.5, dropping to 0.35 at the highest distances from the midplane (z $\simeq$ 2 kpc). In addition, at MAD = 2.5 kpc, where the \SBha is higher (corresponding to squares in Figure \ref{fig:ESO157-49_T_S+S}), the T$_e$ remains below $\sim$8$\cdot$10$^3$ K.

\begin{figure}
\centering
\includegraphics[width=\columnwidth]{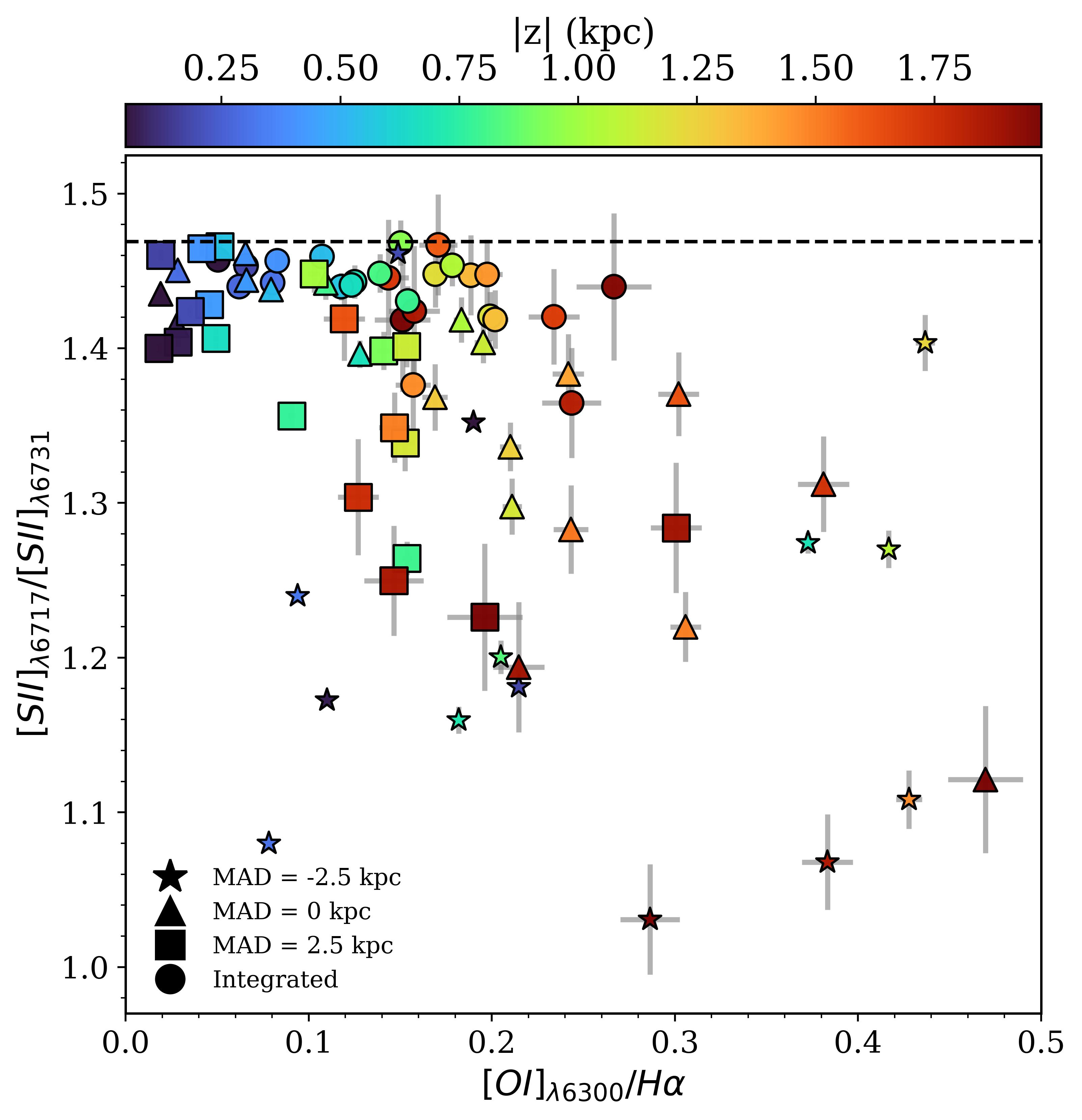} 
\caption{\sii$_{\lambda\,6717}$/\sii$_{\lambda\,6731}$ vs \oi/\ha for ESO157-49. Similar to Figure \ref{fig:IC1553_SII_OI}.}
\label{fig:ESO157-49_SII_OI}     
\end{figure}

\begin{figure*}[t!]
\centering
\includegraphics[width=0.9\textwidth]{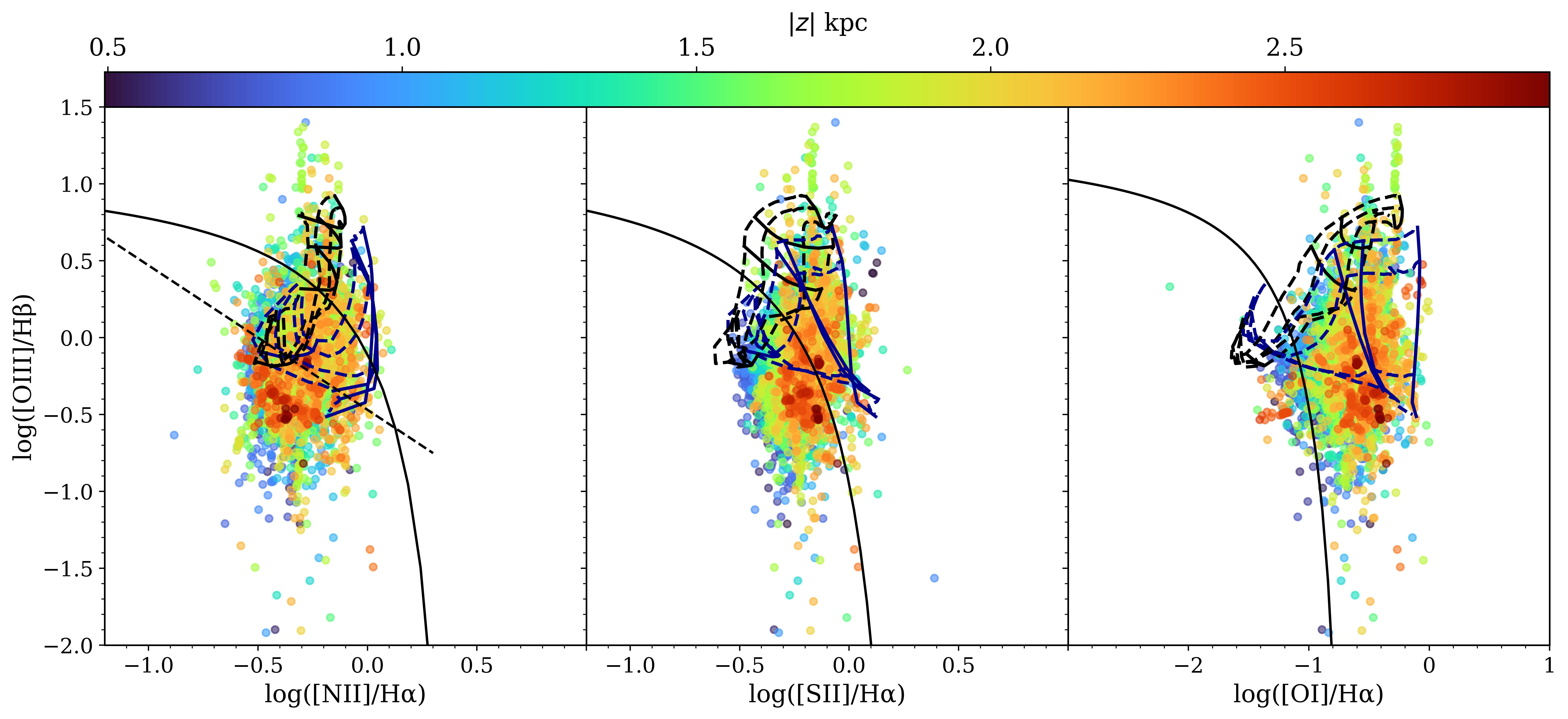} 
\caption{ESO157-49 BPT with hybrid models, similar to Figure \ref{fig:IC1553_BPT_total}. 50\% fast shocks and 50\% star formation with Z = 2Z$_\odot$ and q = $10^7$ cm/s.}
\label{fig:ESO157-49 BPT_1}     
\end{figure*}

\begin{figure*}
\centering
\includegraphics[width=0.8\textwidth]{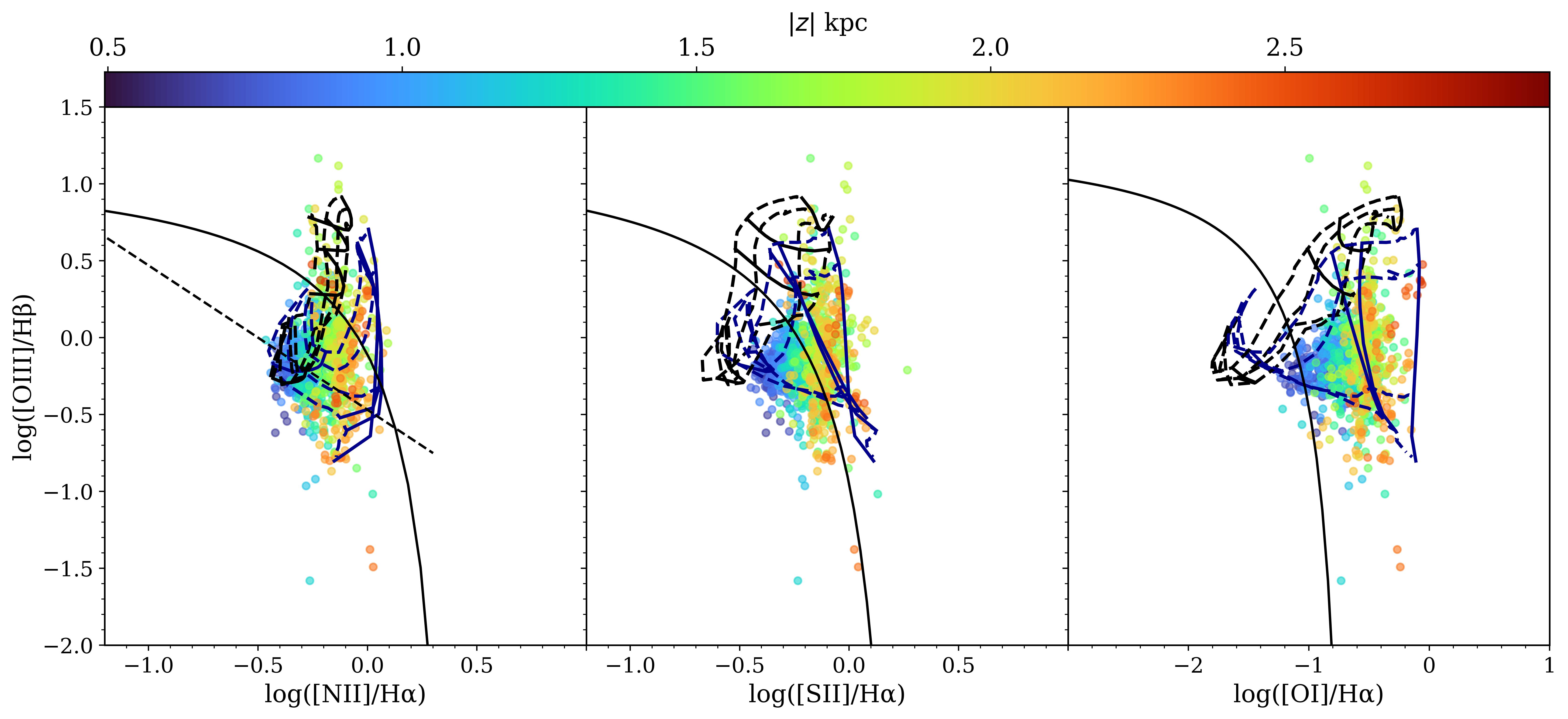} 
\includegraphics[width=0.8\textwidth]{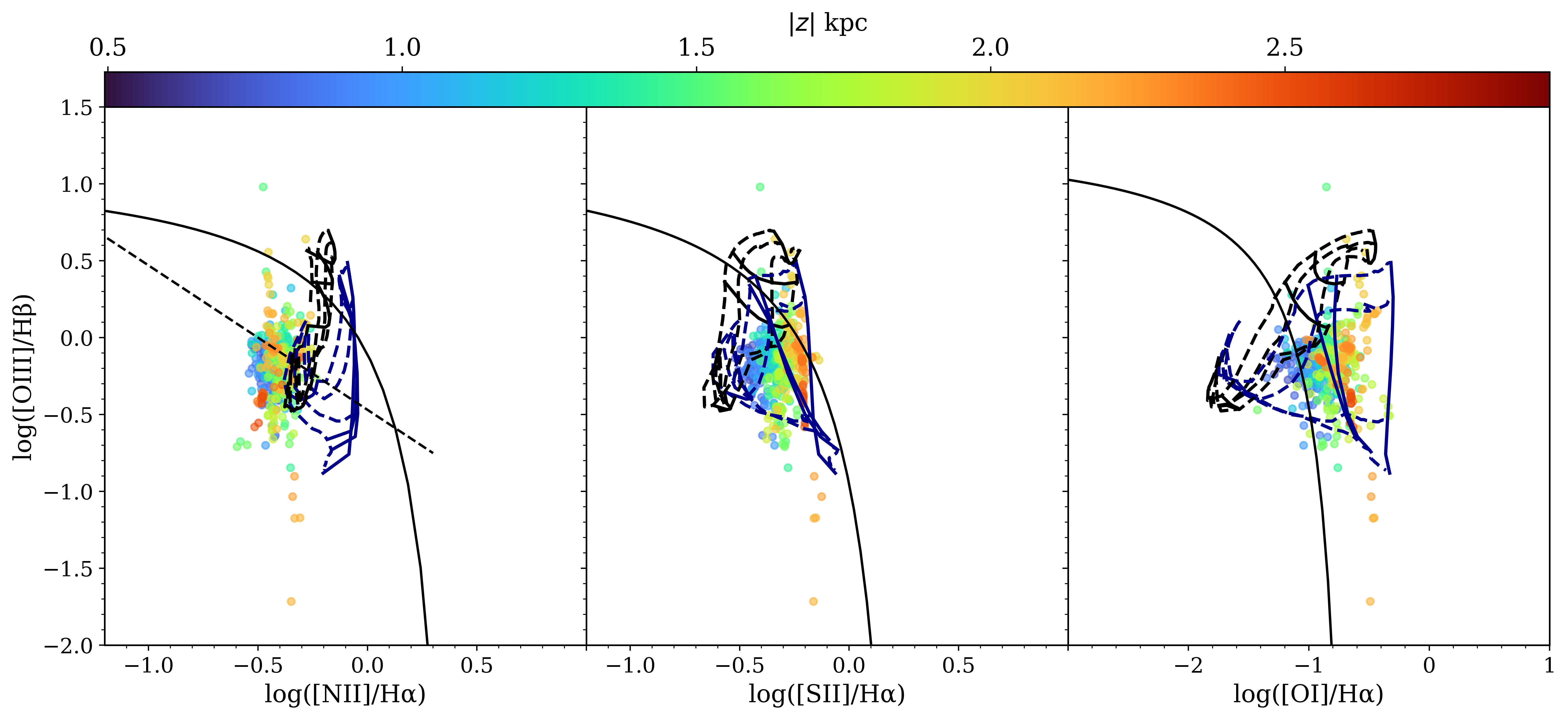} 
\caption{ESO157-49 BPT with hybrid models, similar to Figure \ref{fig:IC1553_BPT_MAD}. Up: BPT for bins between -0.5 kpc < MAD < 0.5 kpc. Hybrid models correspond to 50\% fast shocks and 50\% star formation with Z = 2Z$_\odot$ and q = $10^7$ cm/s. Down: BPT for bins between 2.25 kpc < MAD < 2.75 kpc. Hybrid models correspond to 30\% fast shocks and 70\% star formation with Z = 2Z$_\odot$ and q = $10^7$ cm/s.}
\label{fig:ESO157-49 BPT_2}     
\end{figure*}

Similar to ESO544-27, the significant uniformity in the emission from the galactic plane relative to the line of sight results in hybrid models that best fit the points on the BPT diagram (Figure \ref{fig:ESO157-49 BPT_1}) corresponding to 50\% ionisation due to fast shocks and 50\% due to star formation. However, at 2 kpc < MAD < 3 kpc, where the \SBha is higher, the contribution of fast shocks drops to 30\%, and at -0.5 kpc < MAD < 0.5 kpc, the contribution is still 50\% (see Figure \ref{fig:ESO157-49 BPT_2}). Additionally, in both cases, the dependence of the ionisation regime on height is evident, with bins corresponding to higher z being further from the demarcation of \citet{2001ApJ...556..121K}.

\FloatBarrier

\section{IC217} 
\label{sec:Ap_G}
\raggedbottom

\begin{figure}[ht!]
\centering
\includegraphics[width=\columnwidth]{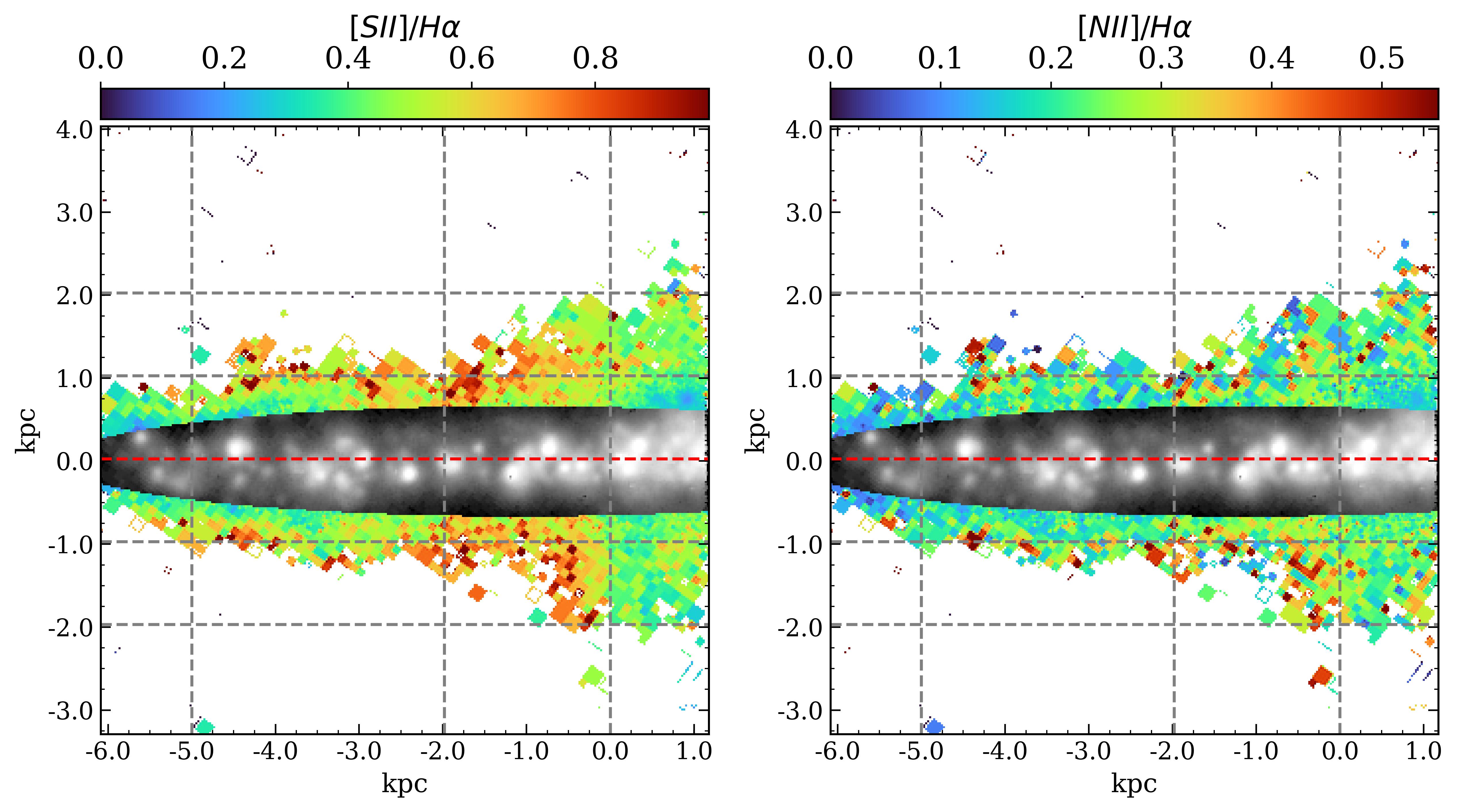}
\includegraphics[width=\columnwidth]{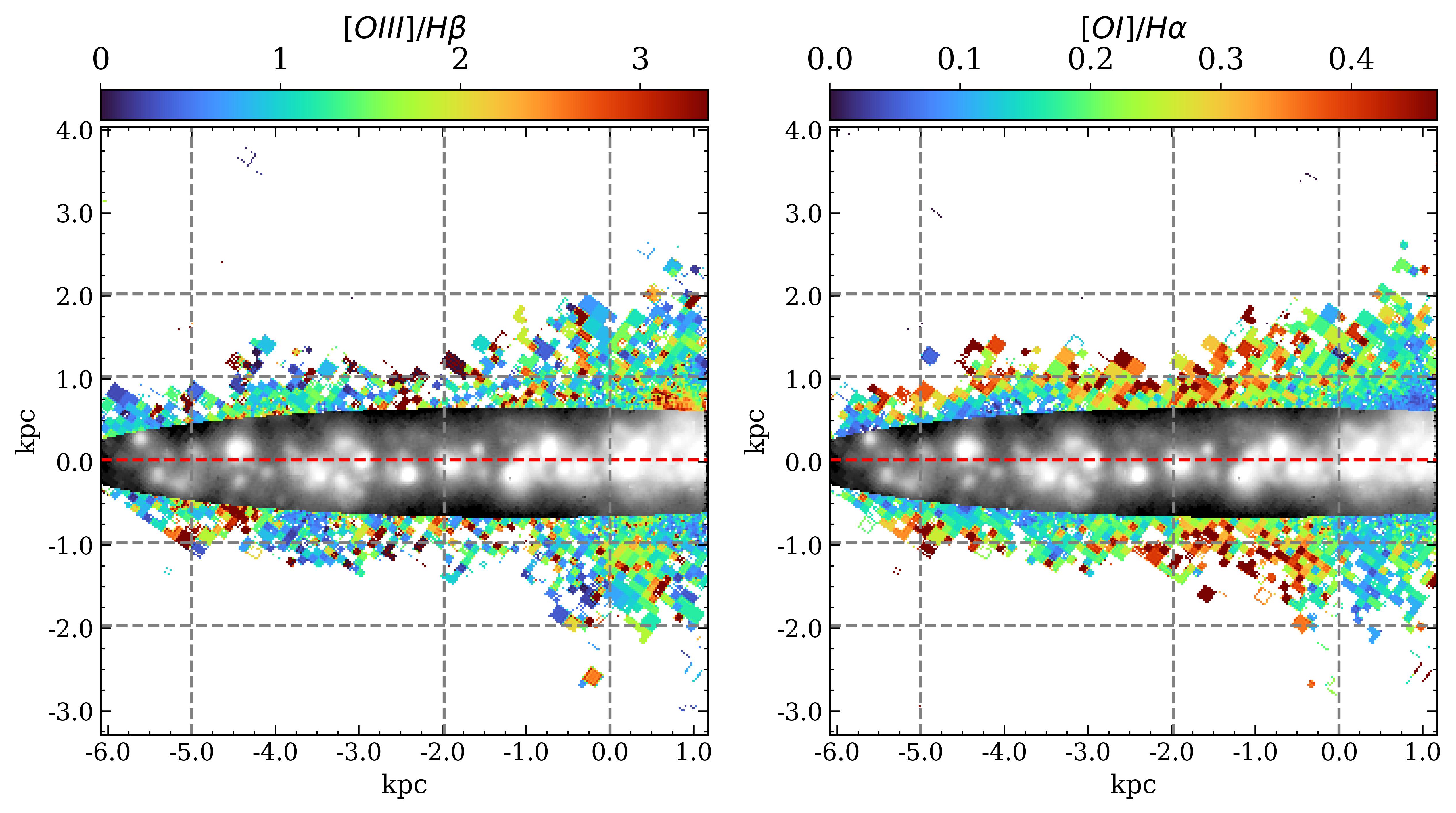} 
\captionof{figure}{IC217 line ratio maps, similar to Figure \ref{fig:IC1553_maps}. The grey dashed lines indicates
the heights with respect the midplane z = $\pm$1, $\pm$2 and major axis distances MAD = -5, -2, 0 kpc.}\label{fig:IC217_maps}
\end{figure}

\begin{figure}[t!]
\centering
\includegraphics[width=0.85\columnwidth]{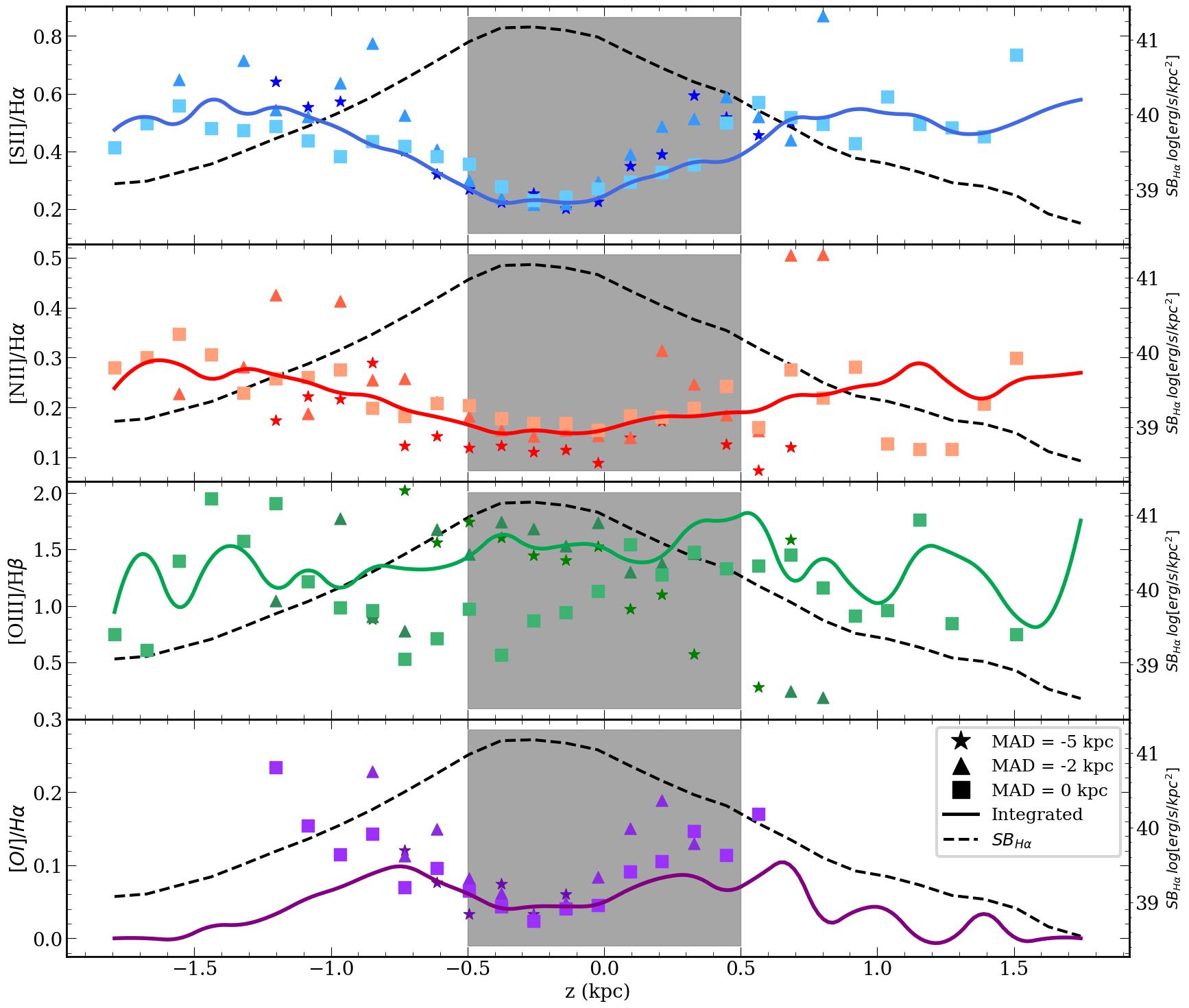}
\caption{IC217 line ratio distributions of the with respect the distance from the midplane. Similarly to Figure \ref{fig:IC1553_lines}, for MAD = -5 (stars), -2 (triangles) and 0 (squares) kpc.}
\label{fig:IC217_lines}     
\end{figure}

\begin{figure*}
\centering
\includegraphics[width=\textwidth]{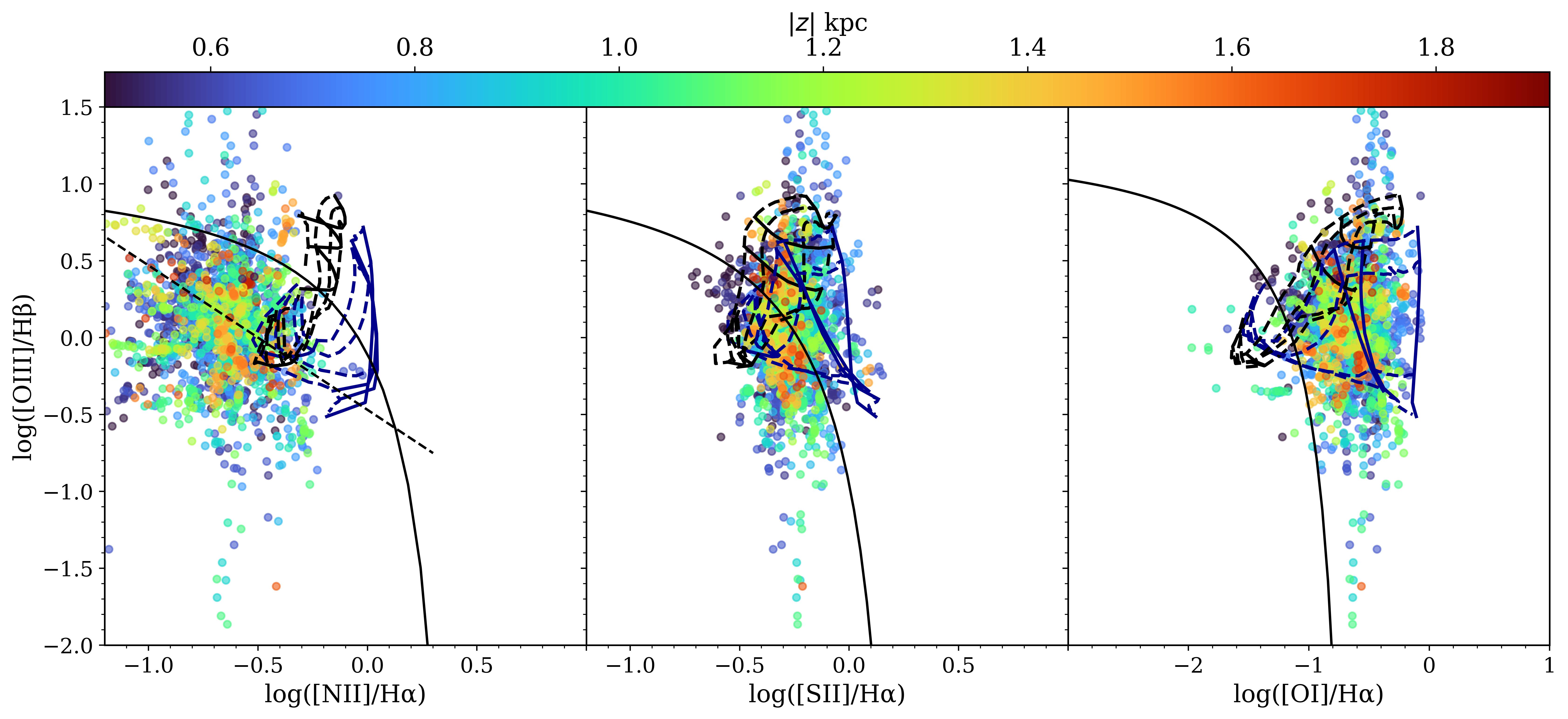} 
\caption{IC217 BPT with hybrid models, similar to Figure \ref{fig:IC1553_BPT_total}. 50\% fast shocks and 50\% star formation with Z = Z$_\odot$ and q = $10^7$ cm/s.}
\label{fig:IC217_BPT}     
\end{figure*}

IC217 has the lowest exposure time among the sample, only three exposures of 2624 seconds in contrast to the four exposures for the other galaxies. After excluding bins with S/N < 2 and relative errors exceeding 40\%, this results in the least comprehensive data sampling among the galaxies in the sample. However, Figure \ref{fig:IC217_maps} illustrates the variations in ionisation regimes at different MADs observed in the other galaxies. Between -3 kpc < MAD < -1 kpc, where the concentration of \hii regions with respect to the line of sight is lower, the line ratios in the eDIG exhibit higher values compared to the rest of the galaxy. Nevertheless, the level of data sampling is insufficient to conduct a more in-depth analysis of this galaxy. The line ratio distributions (Figure \ref{fig:IC217_lines}) and the BPT (Figure \ref{fig:IC217_BPT}) a mixed regime between different MADs, along with a general eDIG ionisation consisting of a 50\% contribution from fast shocks and 50\% of star formation.

\FloatBarrier

\section{IC1553}
\label{sec:Ap_F}

IC1553 is classified as an irregular galaxy \citep{1968TrSht..38....1V}. Despite its inclination of 78.6°, the irregular morphology of the disk is evident due to the highly asymmetrical distribution of \hii regions at 2 kpc $\lesssim$ MAD $\lesssim$ 
4 kpc observed with respect the line of sight.
For this reason, the differences in the ionisation conditions described in section \ref{sec:sec_3} are more evident. The column of eDIG just above the \hii regions exhibits a lower degree of ionisation for \nii and \sii compared to \ha than in the column of eDIG where the emission from the \hii regions is lower. In this latter scenario, the higher electron temperatures favour excitation due to collisions at a low-density regime \citep{2006agna.book.....O}. This translates to a more collisionally-ionised eDIG compared to the column above the \hii regions (with $f_{shock}$ = 0.4 and 0.2 respectively). However, at MAD $\simeq$ 0 kpc, the biconical structure observed in Figure \ref{fig:IC1553_maps} presents lower ionisation fraction (S$^+$/S $\simeq 0.35$) than the rest of the eDIG, but with an increasing electron temperature in height, reaching the highest values ($\sim$ 9$\cdot$10$^3$ K) of all the eDIG. This is consistent with the presence of gas accretion in the galaxy found by \citet{2022A&A...659A.153R} and \citet{2023A&A...678A..84D}.

The general behaviour of the \oiii/\hb and \oi/\ha ratios, as for \nii/\ha, \sii/\ha, is to increase with the height, also observed in previous studies \citep{1998ApJ...501..137R, 2009RvMP...81..969H,2016MNRAS.457.1257H, 2023A&A...678A..84D}. Nevertheless, in the BPT diagrams, there is no clear correlation between the location of the bins in the diagram and their distance with respect to the midplane. This lack of correlation is due to the two-dimensional dependence between the ionisation conditions and the galactic plane. For the \oi/\ha ratios, the emission line map is similar to the \nii/\ha and \sii/\ha maps. As the \oi/\ha ratio is related to the amount of H$^0$ relative to H$^+$, lower values of this ratio in the DIG will imply a higher degree of ionisation of the neutral hydrogen (\citealt{1998ApJ...501..137R, 2009RvMP...81..969H}; \citetalias{2024A&A...687A..20G}). As for the \nii/\ha and \sii/\ha ratios, this ratio is also lower in the eDIG between 2 kpc $\lesssim$ MAD $\lesssim$ 4 kpc than between -2 kpc $\lesssim$ MAD $\lesssim$ 2 kpc. 


\end{document}